\documentclass[aps,showpacs,twocolumn,amsmath,amssymb,superscriptaddress,prl,reprint]{revtex4-2}

\usepackage{graphicx}
\usepackage{bm}
\usepackage{color} 
\usepackage{comment}
\usepackage{physics}
\usepackage[T1]{fontenc}
\usepackage{hyperref}
\usepackage{tabularx}

\begin{document}

	\title{Weak Ferromagnetism in Altermagnets from Alternating $g$-Tensor Anisotropy}
	

	\author{Daegeun~Jo}
	\email{daegeun.jo@physics.uu.se}
	\affiliation{Department of Physics and Astronomy, Uppsala University, P.O. Box 516, SE-75120 Uppsala, Sweden}
	\affiliation{Wallenberg Initiative Materials Science for Sustainability, Uppsala University, SE-75120 Uppsala, Sweden}
	
	\author{Dongwook~Go}
	\affiliation{Institute of Physics, Johannes Gutenberg University Mainz, 55099 Mainz, Germany}
	
	\author{Yuriy~Mokrousov}
	\affiliation{Institute of Physics, Johannes Gutenberg University Mainz, 55099 Mainz, Germany}
	\affiliation{Peter Gr\"unberg Institut and Institute for Advanced Simulation, Forschungszentrum J\"ulich and JARA, 52425 J\"ulich, Germany \looseness=-1}

	\author{Peter~M.~Oppeneer}
	\affiliation{Department of Physics and Astronomy, Uppsala University, P.O. Box 516, SE-75120 Uppsala, Sweden}
	\affiliation{Wallenberg Initiative Materials Science for Sustainability, Uppsala University, SE-75120 Uppsala, Sweden}

	\author{Sang-Wook~Cheong}
	\email{sangc@physics.rutgers.edu}
	\affiliation{Rutgers Center for Emergent Materials and Department of Physics and Astronomy, Rutgers University, Piscataway, New Jersey 08854, USA}
	
	\author{Hyun-Woo~Lee}
	\email{hwl@postech.ac.kr}
	\affiliation{Department of Physics, Pohang University of Science and Technology, Pohang 37673, Korea}
	
	\begin{abstract}
		Altermagnets are magnetic materials with antiferromagnetic spin ordering but exhibit ferromagnetic properties. Understanding the microscopic origin of the latter is a central problem. Ferromagnet-like properties such as the anomalous Hall effect are linked with weak ferromagnetism, whose microscopic origin in altermagnets remains unclear however. We show theoretically that the alternating $g$-tensor anisotropy in altermagnets can induce weak ferromagnetism even when the Dzyaloshinskii-Moriya interaction is forbidden. We demonstrate this mechanism to explain weak ferromagnetism for both collinear and noncollinear spin altermagnets. Our findings provide new insights into the origin of weak ferromagnetism and suggest orbital-based ways for manipulating magnetic configurations in altermagnets.
	\end{abstract}
	
	\maketitle
	
	Magnetic materials have traditionally been categorized by spin ordering: ferromagnets (FMs) with parallel spins and antiferromagnets (AFMs) with antiparallel spins. However, recent studies~\cite{noda2016momentum, ahn2019antiferromagnetism, naka2019spin, hayami2019momentum, smejkal2020crystal, yuan2020giant, hayami2020bottom, reichlova2021macroscopic, yuan2021prediction, naka2021perovskite, hernandez2021efficient, yuan2021prediction, ma2021multifunctional, yuan2021strong, zhou2021crystal, egorov2021antiferromagnetism, mazin2021prediction, shao2021spin, smejkal2022giant, smejkal2022beyond, smejkal2022emerging} have highlighted the significant role of the interplay between spin ordering and local crystal structure surrounding magnetic atoms in characterizing magnetic materials. This insight has led to the classification of a new type of magnetism, dubbed altermagnetism~\cite{smejkal2022beyond, smejkal2022emerging, mazin2022editorial, chen2024emerging, bai2024altermagnetism}. Altermagnets (AMs) have antiferromagnetic spin ordering with alternating local crystal structures at sublattices. In view of the symmetry analysis based on the spin group theory, the alternating local crystal structure is crucial for the FM-like properties of the AMs. First of all, the electronic bands are spin-split~\cite{fedchenko2024observation, krempasky2024altermagnetic, zhu2024observation}. In the presence of spin-orbit coupling (SOC), they can exhibit further FM-like behavior such as the magneto-optical effect~\cite{zhou2021crystal, hariki2024xray} and the anomalous Hall effect~\cite{smejkal2020crystal, smejkal2022anomalous, feng2022anomalous}. Therefore, AMs offer potential for novel spintronic applications~\cite{bose2022tilted, bai2022observation, karube2022observation}, combining the benefits of both FMs~\cite{zutic2004spintronics, hirohata2020review} and AFMs~\cite{jungwirth2016antiferromagnetic, baltz2018antiferromagnetic}.

	Despite FM-like behavior being compatible with symmetry, a fundamental question still remains unanswered: What is a \emph{microscopic} defining feature of AMs that distinguishes them from conventional AFMs and gives rise to their FM-like behavior? Recently, ferroic ordering of magnetic octupoles has been proposed as a microscopic order parameter to describe $d$-wave AMs~\cite{bhowal2024ferroically}. According to a Landau theory augmented by spin-space symmetries, this magnetic octupole can couple linearly to the magnetic dipole through SOC~\cite{mcclarty2024landau}, enabling weak ferromagnetism (WFM) in AMs~\cite{mcclarty2024landau, autieri2023dzyaloshinskii, fernandes2024topological, antonenko2024mirror, milivojevic2024interplay, cheong2024altermagnetism, kluczyk2024coexistence, cheong2025altermagnetism} and their FM-like behavior. However, the microscopic origin of WFM in AMs remains unclear. Although the Dzyaloshinskii-Moriya interaction (DMI)~\cite{dzyaloshinsky1958thermodynamic, moriya1960anisotropic} can induce the WFM in some AMs such as $\alpha$-Fe$_2$O$_3$~\cite{moriya1960anisotropic} and ${\mathrm{La}}_{2}$${\mathrm{CuO}}_{4}$~\cite{cheong1989metamagnetism}, it cannot explain the substantial orbital magnetization of rutile-type~\cite{smejkal2020crystal, zhou2021crystal} and Mn$_3$Sn-type~\cite{sandratskii1996role, kubler2014noncollinear, chen2020manipulating} AMs predicted by first-principles calculations, necessitating deeper investigation into the origin of the WFM in AMs.
	
	In this Letter, we study a thus far unexplored origin of WFM in AMs: alternating $g$-tensor anisotropy. The $g$-tensor not only characterizes the spin exchange energy but also reflects the structural anisotropy, prompting us to examine its role in AMs. We demonstrate theoretically that the alternation of the $g$-tensor anisotropy among sublattices of AMs can naturally lead to WFM, primarily driven by the orbital moment, even when the DMI is forbidden. A wide range of AMs exhibit WFM either intrinsically (type-I AMs) or under strain (type-II AMs)~\cite{cheong2024altermagnetism}. This mechanism extends beyond conventional AMs with collinear spin ordering and straightforwardly applies to AMs with noncollinear spin ordering~\cite{cheong2024altermagnetism}. Our findings not only provide insights into FM-like properties of AMs but also offer a fresh perspective on previous studies of WFM in various compounds~\cite{dzyaloshinsky1958thermodynamic, moriya1960anisotropic, moriya1960theory, tomiyoshi1982magnetic, nagamiya1982triangular, cheong1989metamagnetism, ederer2005weak, kubler2014noncollinear, chen2020manipulating}.

	The spectroscopic $g$-tensor~\cite{kittel1949on} is defined as the ratio of the total magnetic moment to the spin angular momentum, which can be represented by a $3 \times 3$ matrix $\mathbf{g}$. In many high-symmetry systems, principal axes can be chosen such that the orbital and spin angular momenta are aligned with them. In this case, $\mathbf{g}$ can be expressed as a diagonal matrix with elements 
	%
	\begin{equation}\label{eq:g_diagonal}
		g_{\alpha} =  2 +    \frac{\langle  L_\alpha \rangle }{  \langle   S_\alpha  \rangle }    ,
	\end{equation}
	%
	where $\langle L_\alpha \rangle$ and $\langle S_\alpha \rangle$ are expectation values of orbital and spin angular momentum operators $\mathbf{L}$ and $\mathbf{S}$ along the principal axis $\alpha$, respectively. For example, let us consider a FM exhibiting the spin $S_0$ along the $\alpha$-axis in the absence of SOC, where the orbital is quenched. When weak SOC is taken into account, a finite $\langle L_\alpha \rangle$ is induced, which is linear in the SOC strength $\lambda_\mathrm{SO}$~\cite{bruno1989tight}. On the other hand, the spin-orbit correction to the spin is at least second-order~\cite{chen2020manipulating}, i.e., $\langle S_\alpha \rangle = S_0 + O(\lambda_\mathrm{SO}^2) $. Therefore, in the weak SOC regime, the SOC-induced magnetic moment originates primarily from the orbital moment, causing $g_{\alpha}$ to deviate from 2 by an amount proportional to the induced orbital moment or $\lambda_\mathrm{SO}$~\cite{moriya1960theory}.

\begin{figure}[t]
		\center\includegraphics[width=0.5\textwidth]{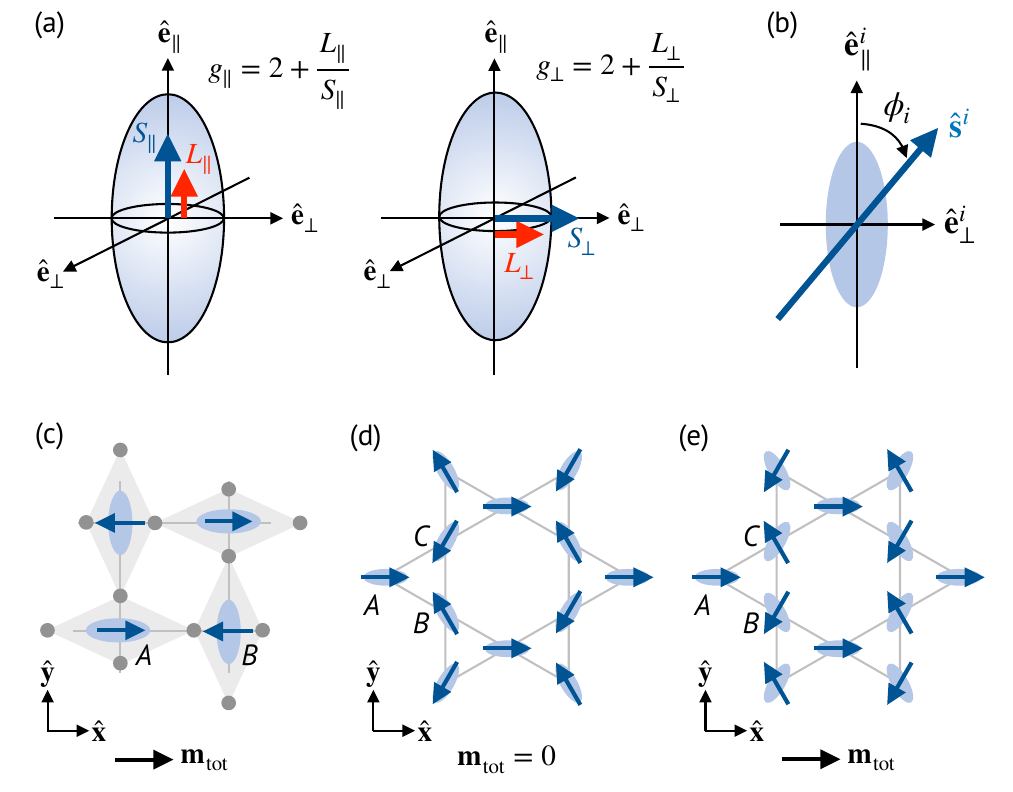}
		\caption{(a) Schematic illustration of the $g$-tensor anisotropy for a magnetic atom. The blue ellipsoids illustrate uniaxial anisotropy with principal axes $ \hat{\mathbf{e}}_{\parallel}$ and $\hat{\mathbf{e}}_{\perp}$. $S_{\parallel(\perp)}$ and $L_{\parallel(\perp)}$ are the spin and orbital angular momenta along $\hat{\mathbf{e}}_{\parallel(\perp)}$, respectively. (b) Example of the local structure of the magnetic sublattice $i$ in an AM. (c)-(e) Altermagnetic systems with alternating $g$-tensor anisotropy. Panels (c) and (e) exhibit the WFM (type-I AM), while panel (d) does not (type-II AM). 
		}
		\label{fig_illustration} 
	\end{figure}

	The anisotropic orbital moment is an essential ingredient of our discussion. If the crystal structure is anisotropic, the SOC-induced orbital moment depends on the spin exchange field direction $\hat{\mathbf{s}}$ due to the interplay between spin and crystal structure, which is also known to be responsible for magnetic anisotropy energy~\cite{bruno1989tight}. Here, we focus on its influence on the $g$-tensor. Consider a magnetic atom subject to an electrical potential featuring uniaxial anisotropy with principal axes $ \hat{\mathbf{e}}_{\parallel}$ and $\hat{\mathbf{e}}_{\perp}$ [Fig.~\ref{fig_illustration}(a)]. We compare two configurations where orbital and spin angular momenta, aligning with $\hat{\mathbf{s}}$, are oriented either along $ \hat{\mathbf{e}}_{\parallel}$ or $ \hat{\mathbf{e}}_{\perp}$. Based on the previous discussion, we expect that the orbital angular momenta $L_\parallel \neq L_\perp$ are linearly proportional to the SOC strength, while the spin angular momenta $S_\parallel$ and $S_\perp$ remain approximately equal to $S_0$. Consequently, the $g$-tensor can be expressed as $\mathbf{g} = \mathrm{diag} (g_\perp, g_\perp, g_\parallel)$, where $ g_{\parallel(\perp)} \approx 2 + L_{\parallel(\perp)} / S_0 $. This leads us to define the $g$-tensor anisotropy $\Delta g $ as:
	\begin{equation}
		\label{eq:g_ani}
		\Delta g \equiv g_\parallel - g_\perp \approx   \frac{L_\parallel  - L_\perp  }{S_0 } ,
	\end{equation}
	which can be experimentally evaluated from the ferromagnetic resonance spectroscopy~\cite{alahmed2022evidence}. Although we have assumed a ferromagnetic atom, our discussion can be generalized to altermagnetic systems by defining the $g$-tensor locally for each magnetic atom.

	Now, we show that $\Delta g$ combines with the alternating local crystal structure to generate the WFM of AMs. We consider a prototypical altermagnetic system. The $i$-th magnetic atom is surrounded by a crystalline potential with uniaxial anisotropy, having principal axes $\hat{\mathbf{e}}_{\parallel}^i$ and $\hat{\mathbf{e}}_{\perp}^i$ [Fig.~\ref{fig_illustration}(b)]. The local spin $\langle \mathbf{S}^i \rangle = S_0 \hat{\mathbf{s}}^i$ makes an angle $\phi_i$ with $\hat{\mathbf{e}}_{\parallel}^i$, satisfying $\sum_i^N \hat{\mathbf{s}}^i = 0$, where $N$ is the number of atoms in a unit cell. The local $g$-tensor $\mathbf{g}^i = \mathrm{diag} (g_\perp, g_\perp, g_\parallel)$ is defined in terms of the principal axes at each site. $g_\parallel$ and $g_\perp$ are independent of the sublattice, provided that the local crystal structures at each site are connected by rotational symmetry to each other [e.g., Figs.~\ref{fig_illustration}(c)-\ref{fig_illustration}(e)]. Thus, the total magnetic moment $ \mathbf{m}_\mathrm{tot} = -(\mu_\mathrm{B}/\hbar) \sum_i^N  \mathbf{g}^i \cdot  \langle  \mathbf{S}^i \rangle  $ is proportional to $\sum_i^N [g_\parallel \cos \phi_i \hat{\mathbf{e}}_\parallel^i + g_\perp (\hat{\mathbf{s}}^i -  \cos \phi_i \hat{\mathbf{e}}_\parallel^i)] $ and given by
	\begin{equation}\label{eq:m_tot}
		\mathbf{m}_\mathrm{tot}  = - \frac{\mu_\mathrm{B} }{\hbar} \Delta g  S_0 \mathbf{s}_\parallel, \qquad \mathbf{s}_\parallel = \sum_i^N \cos \phi_i \hat{\mathbf{e}}_\parallel^i,
	\end{equation}
	to first-order in SOC. Equation~\eqref{eq:m_tot} is the main result of our paper, which clearly demonstrates that the net magnetic moment of AMs is proportional to $\Delta g$ instead of the electron spin $g$-factor 2 for strong FMs. Our theory generalizes the single-ion anisotropy picture~\cite{moriya1960theory} by incorporating the orbital moment, providing a new perspective on WFM. Since $ \mathbf{m}_\mathrm{tot}  $ is dominated by the orbital contribution, it is naturally related to orbital-dominated anomalous Hall AFMs~\cite{ito2017anomalous, chen2020manipulating}. Notably, the direction of $ \mathbf{m}_\mathrm{tot}  $ is determined by $\mathbf{s}_\parallel$, representing the projection of spins with respect to local crystalline axes. This direction can differ from that of the DMI-induced net moment, and $ \mathbf{m}_\mathrm{tot}  $ can be nonzero even when the DMI is forbidden (see End Matter). Our mechanism based on $\Delta g$ is thus clearly distinct from the DMI-induced WFM. We further emphasize that our mechanism represents an \emph{intrinsic} property of AMs, since it arises from the alternating local crystal structure, which is a defining feature of AMs. By contrast, the origin of the DMI is not inherently related to the alternating local structure.

	Here, we examine AMs with collinear spins using Eq.~\eqref{eq:m_tot}. Figure~\ref{fig_illustration}(c) shows a two-dimensional square lattice consisting of two magnetic atoms $A$ and $B$ with opposite spins. The gray circles at the corners of diamonds represent the nonmagnetic atoms, which give rise to the anisotropic crystal potential depicted by blue ellipses. Similar types of structures are commonly found in rutile-type AMs, such as RuO$_2$~\cite{smejkal2020crystal}, MnF$_2$~\cite{yuan2020giant}, and NiF$_2$~\cite{moriya1960theory}. For an arbitrary in-plane spin configuration with $\phi_{A} = \phi$ and $\phi_{B} = \phi + \pi/2$, Eq.~\eqref{eq:m_tot} leads to
	\begin{equation}\label{eq:m_tot2}
		\mathbf{m}_\mathrm{tot}  = - \frac{\mu_\mathrm{B} }{\hbar} \Delta g  S_0  (\cos \phi \hat{\mathbf{x}} - \sin \phi \hat{\mathbf{y}}). 
	\end{equation}
	Interestingly, the predicted $\mathbf{m}_\mathrm{tot} \parallel \mathbf{s}_\parallel = \cos \phi \hat{\mathbf{x}} - \sin \phi \hat{\mathbf{y}}$ can be either parallel or perpendicular to the N\'eel vector $\hat{\mathbf{n}} =  \cos \phi \hat{\mathbf{x}} + \sin \phi \hat{\mathbf{y}}$ depending on $\phi$. This angular dependence is consistent with first-principles calculations for NiF$_2$ (see End Matter) and RuO$_2$~\cite{smejkal2020crystal, zhou2021crystal}, which cannot be attributed to the DMI. Such AMs exhibiting WFM have been categorized as type-I AMs~\cite{cheong2024altermagnetism}. Conversely, for the out-of-plane spin configuration with $\hat{\mathbf{s}}^i = \pm \hat{\mathbf{z}}$ (e.g., CoF$_2$~\cite{moriya1959piezomagnetism}), implying $\phi_{A} = \phi_{B}$, Eq.~\eqref{eq:m_tot} yields $\mathbf{m}_\mathrm{tot}=0$ since the relevant $g$-tensor anisotropy in this configuration is the difference between two $g_\perp$'s. Inducing difference between them by strain can induce net magnetization, which amounts to piezomagnetism~\cite{moriya1959piezomagnetism, disa2000polarizing}. This type of AMs, with zero net magnetization and piezomagnetism, has been categorized as type-II AMs~\cite{cheong2024altermagnetism}.

\begin{figure}[t]
		\center\includegraphics[width=0.5\textwidth]{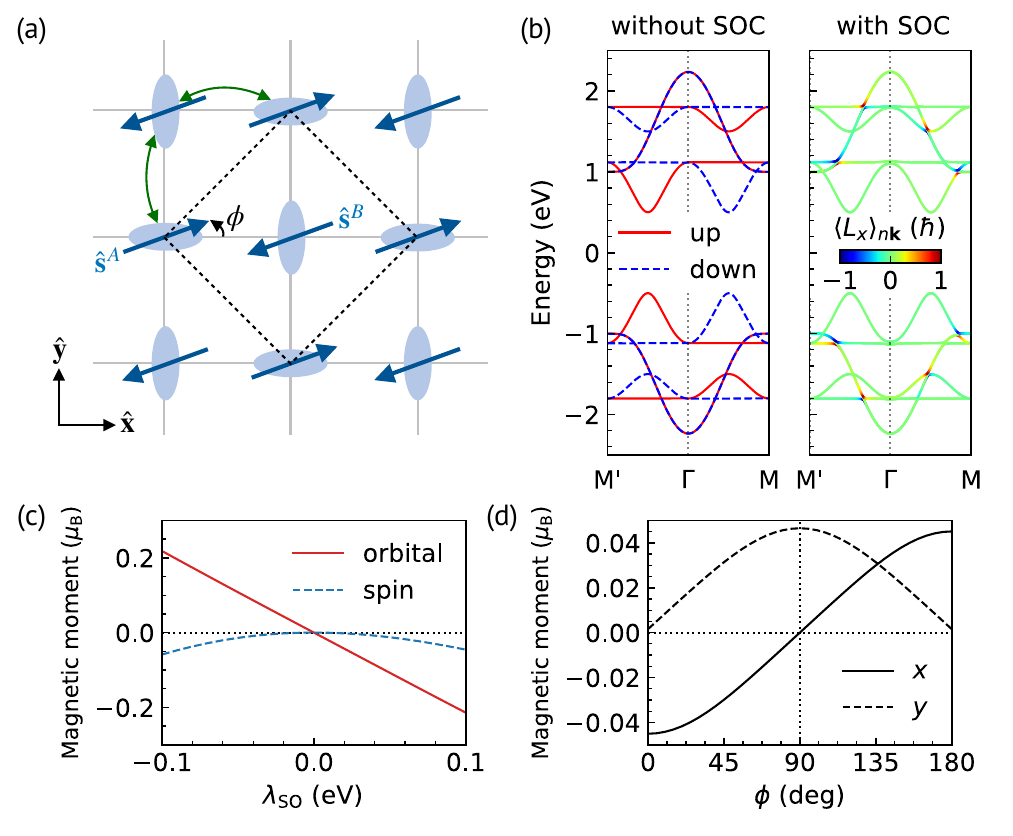}
		\caption{(a) Schematic illustration of the tight-binding model $\mathcal{H}$ for an AM with collinear spins. The dashed lines represent the unit cell, and the green arrows indicate the nearest-neighbor hoppings. The blue ellipses display the anisotropy. (b) Band structures without (left panel) and with (right panel) SOC. (c) The net orbital and spin magnetic moments at the Fermi energy $E_\mathrm{F} = -0.9$ eV as a function of the SOC strength $\lambda_\mathrm{SO}$. (d) $x$- and $y$-components of the total magnetic moment at $E_\mathrm{F} = -0.9$ eV depending on the spin angle $\phi$.
		}
		\label{fig_square} 
	\end{figure}

	Next, we test the above predictions by performing tight-binding calculations for a two-dimensional square lattice consisting of two magnetic atoms $A$ and $B$ with an antiferromagnetic spin ordering characterized by an angle $\phi$ [Fig.~\ref{fig_square}(a)]. Assuming the atomic orbital basis \{$ d_{xy}, d_{yz}, d_{zx} $\} for each atom, the Hamiltonian is written as (see Supplemental Material~\cite{supp} for details):
	\begin{align}\label{eq:h_tb}
		\mathcal{H}  = &	\sum_{\langle i,j \rangle  m n \sigma }	t_{ijmn} c_{im\sigma}^\dagger c_{jn\sigma} 
		- \frac{J_\mathrm{sd}}{\hbar}	\sum_{i  n \sigma \sigma' } 
		\hat{ \mathbf{s} }^i \cdot 	c_{in\sigma}^\dagger \mathbf{S}_{\sigma \sigma'} \; c_{in\sigma'} \nonumber \\
		&   +  \frac{\lambda_\mathrm{SO}}{\hbar^2}	\sum_{i  mn \sigma \sigma' } 	c_{im\sigma}^\dagger 	 \mathbf{L} _{mn} \cdot \mathbf{S}_{\sigma \sigma'} \; c_{in\sigma'} ,
	\end{align}
	where $c_{in\sigma}^\dagger$ ($c_{in\sigma}$) is the creation (annihilation) operator of an electron at site $i$ with orbital index $n$ and spin $\sigma = \uparrow,\downarrow$. The first term describes nearest-neighbor hopping, the second term represents the exchange interaction with the molecular field $(J_\mathrm{sd}/\hbar) \hat{\mathbf{s}}^i$, and the third term corresponds to the atomic SOC, where $\mathbf{L}$ is defined in terms of the atomic orbital basis. Note that this lattice does not contain nonmagnetic atoms in contrast to typical rutile-structure AMs [e.g., Fig.~\ref{fig_illustration}(c)]. Thus, Eq.~\eqref{eq:h_tb} reflects neither local anisotropy nor altermagnetism. To capture the alternating local crystal structure arising from the nonmagnetic atoms, we introduce the sublattice-dependent crystal field to each magnetic atom as follows: 
	\begin{equation}\label{eq:h_cf}
		\mathcal{H}_\mathrm{CF}   = \frac{\Delta_\mathrm{CF}}{2} \sum_{i  \sigma } 	\tau \,	(c_{i,yz,\sigma}^\dagger c_{i,yz,\sigma} - c_{i,zx,\sigma}^\dagger c_{i,zx,\sigma} ), 
	\end{equation}
	with the sublattice index $\tau$ being $+1(-1)$ for $A(B)$. This term leads to the alternating local anisotropy, as depicted by Fig.~\ref{fig_square}(a), thereby giving rise to altermagnetism. 
	
	Figure~\ref{fig_square}(b) shows the band structures without (left) and with (right) SOC for the total Hamiltonian $\mathcal{H} + \mathcal{H}_\mathrm{CF}$ with $J_\mathrm{sd}=1.0$ eV, $\phi=0$, $\lambda_\mathrm{SO}=0.02$ eV, and $\Delta_\mathrm{CF}=1.0$ eV. The nonrelativistic spin splitting alternates in sign between the $\mathbf{k}$ paths from $\Gamma$ to $\mathrm{M}=(0.5,0.5) $ and from $\Gamma$ to $\mathrm{M}'=(-0.5,0.5)$. With SOC, doubly degenerate $d_{xy}$ bands hybridize with $d_{zx}$ bands, inducing an orbital angular momentum $\langle L_x \rangle_{n\mathbf{k}}$ for each $\mathbf{k}$ point with band $n$. The anisotropy of $\langle L_x \rangle_{n\mathbf{k}}$ in $\mathbf{k}$ space results in the net orbital moment, which is linear in $\lambda_\mathrm{SO}$ [Fig.~\ref{fig_square}(c)] and dominant over the net spin moment that is quadratic in $\lambda_\mathrm{SO}$. These features agree with our expectations. Additionally, the $\phi$ dependence of $\mathbf{m}_\mathrm{tot}$ [Fig.~\ref{fig_square}(d)] is consistent with Eq.~\eqref{eq:m_tot2}.

\begin{figure}[t]
		\center\includegraphics[width=0.5\textwidth]{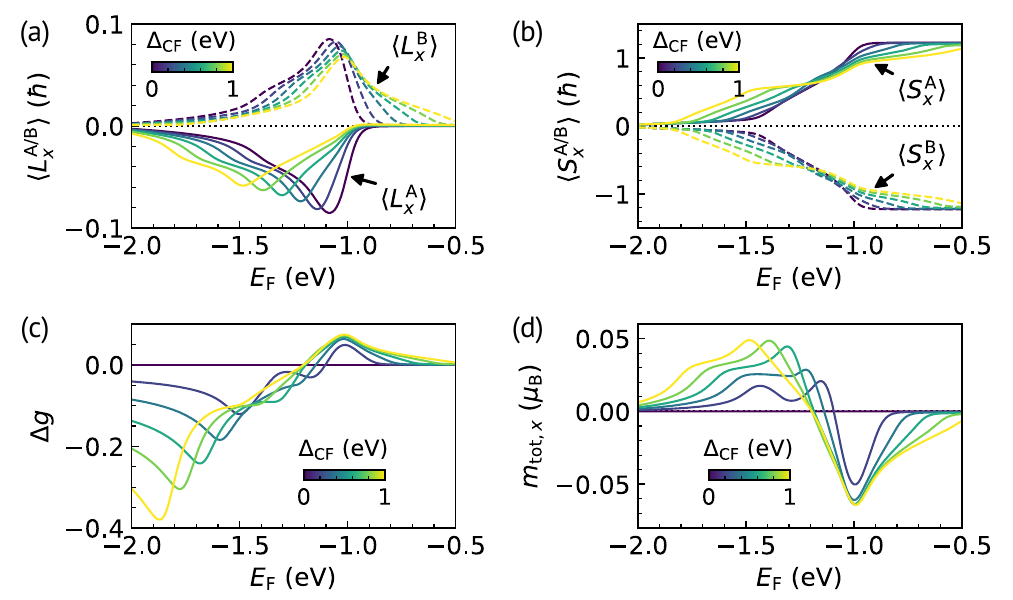}
		\caption{(a),(b) Fermi energy ($E_\mathrm{F}$) dependences of (a) orbital and (b) spin angular momenta for atom $A$ (solid lines) and atom $B$ (dashed lines) with varying crystal field parameter $\Delta_\mathrm{CF}$. (c),(d) $E_\mathrm{F}$ dependences of (c) $g$-tensor anisotropy $\Delta g$ and (d) total magnetic moment $m_{\mathrm{tot},x}$ with varying $\Delta_\mathrm{CF}$.
		}
		\label{fig_square2} 
	\end{figure}

	Next, we examine the connection between the $g$-tensor anisotropy and net magnetic moment. For this, we first calculate the orbital (spin) angular momentum projected onto atoms $A$ and $B$, denoted by $\langle L_x^{A} \rangle$ and $\langle L_x^{B} \rangle$ ($\langle S_x^{A} \rangle$ and $\langle S_x^{B} \rangle$), respectively, as a function of the Fermi energy $E_\mathrm{F}$ and as a function of the alternating crystal field parameter $\Delta_\mathrm{CF}$ in $\mathcal{H}_\mathrm{CF}$. Figure~\ref{fig_square2}(a) demonstrates that $\langle L_x^{A} \rangle + \langle L_x^{B} \rangle$ is zero for  $\Delta_\mathrm{CF}=0$ and becomes larger in magnitude as $\Delta_\mathrm{CF}$ increases. In contrast, $\langle S_x^{A} \rangle + \langle S_x^{B} \rangle$ remains almost zero regardless of $\Delta_\mathrm{CF}$ [Fig.~\ref{fig_square2}(b)]. Then, the $g$-tensor anisotropy, evaluated by $\Delta g = \langle L_x^{A} \rangle/\langle S_x^{A} \rangle - \langle L_x^{B} \rangle / \langle S_x^{B} \rangle$, gradually increases with $\Delta_\mathrm{CF}$ [Fig.~\ref{fig_square2}(c)]. Note that the total magnetic moment [Fig.~\ref{fig_square2}(d)] exhibits a clear correlation with $\Delta g$. Moreover, we find that the product $\Delta g (\vert \langle S_x^{A} \rangle \vert + \vert \langle S_x^{B} \rangle \vert)/2$ exhibits the essentially identical variation as $m_{\mathrm{tot},x}$ with respect to $E_\mathrm{F}$ or $\Delta_\mathrm{CF}$ change, providing further support to Eq.~\eqref{eq:m_tot}.

	It is worthwhile to compare the magnitude of the WFM arising from two mechanisms: $\Delta g$ and the DMI. Spin moments canted by an angle $\psi$ due to the DMI produce a net transverse spin moment of $m_\mathrm{DMI} = -(\mu_\mathrm{B}/\hbar) S_0 \sin \psi$. Given the similarity between this relation and Eq.~\eqref{eq:m_tot2}, the comparison between the two effects reduces to that between $\sin \psi$ and $\Delta g$. In prototypical DMI antiferromagnets~\cite{thoma2021revealing}, $\psi$ typically reaches the order of $1^\circ$, leading to $\sin \psi \approx 0.02$. On the other hand, $\Delta g$ can reach the order of 0.1, corresponding to an angle of $\sin^{-1} (\Delta g) = 5.7^\circ$, in our model or rutile AMs (see End Matter) in the metallic phase. This illustrates that the WFM arising from $\Delta g$ can be much larger than that induced by the DMI.

	Next, we extend our theory to AMs with noncollinear spin structures~\cite{cheong2024altermagnetism}. For example, Fig.~\ref{fig_illustration}(d) displays a kagome lattice with a triangular spin structure where the spins of sublattices $A$, $B$, and $C$ are each rotated by $120^\circ$ relative to one another and the local $g$-tensor of each sublattice is anisotropic. In this normal triangular spin structure, the angles between spin and $g$-tensor's principal axis are identical for all sublattices, i.e., $\phi_{A} = \phi_{B} = \phi_{C}$, which makes $\mathbf{m}_\mathrm{tot}$ in Eq.~\eqref{eq:m_tot} zero (type-II AM). On the other hand, for an inverse triangular spin structure [Fig.~\ref{fig_illustration}(e)], where the spins are rotated by $-120^\circ$, Eq.~\eqref{eq:m_tot} predicts a nonvanishing $\mathbf{m}_\mathrm{tot}$ (type-I AM). For arbitrary angles, $\phi_{A} = \phi$, $\phi_{B} = \phi +2\pi/3$, and $\phi_{C} = \phi +4\pi/3$, $\mathbf{m}_\mathrm{tot}$ follows Eq.~\eqref{eq:m_tot2}~\cite{ito2017anomalous}. We note that in both normal [Fig.~\ref{fig_illustration}(d)] and inverse [Fig.~\ref{fig_illustration}(e)] triangular spin structures, the DMI does not induce any net spin moment~\cite{tomiyoshi1982magnetic} since these spin configurations with complete spin compensation minimize the DMI energy.

\begin{figure}[t]
		\center\includegraphics[width=0.5\textwidth]{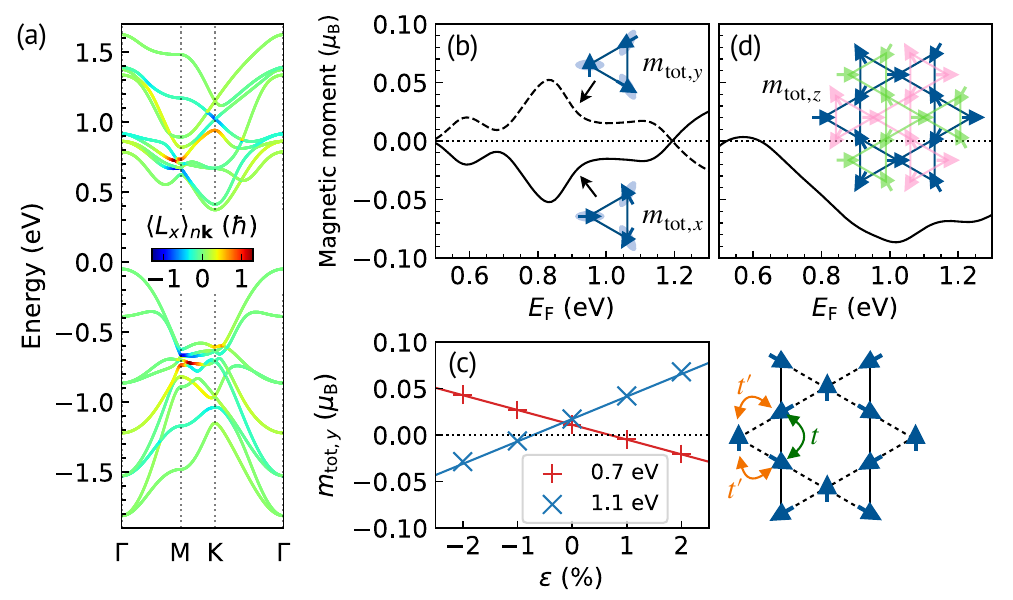}
		\caption{(a) Band structure of the kagome lattice having an inverse triangular spin structure with $\phi=0$. (b) Fermi energy ($E_\mathrm{F}$) dependences of the net magnetic moment $m_{\mathrm{tot},x}$ for $\phi=0$ (solid line)  and $m_{\mathrm{tot},y}$ for $\phi=\pi/2$ (dashed line). The insets display the spin configuration for the three sublattices. (c) $m_{\mathrm{tot},y}$ of the distorted kagome lattice (right panel) at $E_\mathrm{F} = 0.7$ eV (red $+$ symbols) and $E_\mathrm{F} = 1.1$ eV (blue $\times$ symbols) with varying strain parameter $\varepsilon$. Solid lines indicate a linear relation between $\varepsilon$ and $m_{\mathrm{tot},y}$. (d) $E_\mathrm{F}$ dependence of $m_{\mathrm{tot},z}$ for the bulk system with ABC-stacked kagome planes. The inset shows the top view of the structure, where blue, pink, and green layers are stacked in order. 
		}
		\label{fig_kagome} 
	\end{figure}

	We also examine these predictions numerically by applying $\mathcal{H}$ to the kagome lattices (see Supplemental Material~\cite{supp} for details and parameters used). Since the magnetic atoms themselves in the kagome lattices give rise to the alternating local crystal structure, we consider in our calculation only $\mathcal{H}$ and ignore $\mathcal{H}_\mathrm{CF}$. For each magnetic atom, the orbital basis \{$d_{3z^2-r^2}, d_{yz}$, $d_{zx}$\} is taken into account. The angle between the spin at sublattice $A$ and the $x$-axis is given by $\phi$. We first consider the inverse triangular spin structure [Fig.~\ref{fig_illustration}(e)]. Its band structure for $\phi=0$ [Fig.~\ref{fig_kagome}(a)] shows band-resolved $\langle L_x \rangle_{n\mathbf{k}}$ summed over three sublattices $A$, $B$, and $C$. Analogous to the collinear spin case, the orbital is the main contribution to $\mathbf{m}_\mathrm{tot}$. In Fig.~\ref{fig_kagome}(b), the solid and dashed lines present the $x$- and $y$-components of $\mathbf{m}_\mathrm{tot}$ for $\phi=0$ (e.g., Mn$_3$Ge and Mn$_3$Ga) and $\phi=\pi/2$ (e.g, Mn$_3$Sn), respectively. Note that they follow the predicted relation $m_{\mathrm{tot},x}(\phi) = - m_{\mathrm{tot},y}(\phi + \pi/2)$ [Eq.~\eqref{eq:m_tot2}].

	Motivated by an experiment~\cite{ikhlas2022piezomagnetic} on the piezomagnetic properties of the kagome lattice with the inverse triangular spin structure, we examine the strain effect on $\mathbf{m}_\mathrm{tot}$. The right panel of Fig.~\ref{fig_kagome}(c) shows a distorted kagome lattice, with solid and dashed lines indicating different bond lengths $r$ and $r'$, respectively. Suppose $r'= (1+\varepsilon) r$ due to strain with the parameter $\varepsilon$. The original hopping parameter $t$ is then modified to $t'$ under strain, which can be approximated as $t' = t/ (1+\varepsilon)^2$~\cite{harrison2012electronic}. We incorporate this effect into the tight-binding model while preserving the lattice structure~\cite{supp}. The left panel in Fig.~\ref{fig_kagome}(c) presents $m_{\mathrm{tot},y}$ at $E_\mathrm{F} = 0.7$ eV and $E_\mathrm{F} = 1.1$ eV, demonstrating a piezomagnetic effect~\cite{moriya1959piezomagnetism}: magnetization is induced by strain linearly. As $\mathbf{m}_\mathrm{tot}$ for $\varepsilon = 0$ is tiny, $\mathbf{m}_\mathrm{tot}$ can be easily switched by the piezomagnetic effect, in qualitative agreement with the experiment~\cite{ikhlas2022piezomagnetic}.

	For the normal triangular spin structure [Fig.~\ref{fig_illustration}(d)], our numerical calculation finds $\mathbf{m}_\mathrm{tot} = 0$, as predicted by Eq.~\eqref{eq:m_tot}. However, WFM can be induced via symmetry-breaking stacking. For instance, the structures of Mn$_3$Rh, Mn$_3$Ir, and Mn$_3$Pt consist of kagome planes with ABC-type stacking along the (111) direction~\cite{chen2014anomalous, zhang2017strong}, which breaks the $C_{2x}$ symmetry. To examine these systems, we apply $\mathcal{H}$ to the three-dimensional bulk structure composed of ABC-stacked kagome lattices with the normal triangular spin structure [inset of Fig.~\ref{fig_kagome}(d)]~\cite{supp}. In this system, the off-diagonal $g$-tensor element $g_{zx} =   \langle L_z  + 2 S_z \rangle / \langle S_x \rangle   $ becomes nonzero due to the broken $C_{2x}$ symmetry. As a result, out-of-plane magnetization is induced, as shown in Fig.~\ref{fig_kagome}(d). Our analysis based on the $g$-tensor explains previous studies on kagome materials~\cite{tomiyoshi1982magnetic, nagamiya1982triangular, nakatsuji2015large, chen2014anomalous, kubler2014noncollinear, chen2020manipulating} with a fresh mechanism in terms of altermagnetism.

	In conclusion, we have demonstrated that WFM can be induced in AMs by the alternating $g$-tensor anisotropy $\Delta g$, which is a distinguishing feature of AMs that sets them apart from conventional AFMs. This mechanism applies to AMs with either collinear or noncollinear spin structures. The concept of $\Delta g$ is useful for classifying type-I and type-II AMs. Our theory provides new understanding of the WFM in general and proposes ways to manipulating magnetic configurations in various AMs. We suggest further investigations on its role in novel AMs with twisted~\cite{he2023nonrelativistic, sheoran2024nonrelativistic}, Janus~\cite{mazin2023induced, sheoran2024nonrelativistic, zhu2024multipiezo}, or supercell~\cite{jaeschke2024supercell} structures. Lastly, our findings suggest that AMs could act as orbital ferromagnets, where emergent orbital effects are potentially linked to $\Delta g$. This opens a promising research direction toward orbital engineering, e.g., orbitronics~\cite{go2021orbitronics, jo2024spintronics} utilizing AMs.

	\begin{acknowledgments}
		D.J. and P.M.O. were supported by the Swedish Research Council (VR), the Knut and Alice Wallenberg Foundation (Grants No. 2022.0079 and 2023.0336), and the Wallenberg Initiative Materials Science for Sustainability (WISE) funded by the Knut and Allice Wallenberg Foundation. D.G., Y.M., and P.M.O. acknowledge funding from the European Union’s HORIZON EUROPE, under the grant agreement No 101129641, and D.G. and Y.M. acknowledge financial support by the Deutsche Forschungsgemeinschaft (DFG, German Research Foundation)$-$TRR 288$-$422213477 (project B06), and TRR 173/3$-$268565370 (project A11). S.-W.C. was supported by the DOE under Grant No. DOE: DE-FG02-07ER46382. H.-W.L. was supported by the National Research Foundation of Korea (NRF) (No. RS-2024-00356270, RS-2024-00410027). The calculations were supported by resources provided by the National Academic Infrastructure for Supercomputing in Sweden (NAISS) at NSC Linköping, partially funded by VR through Grant No. 2022-06725. 
	\end{acknowledgments}

	\bibliography{ref.bib}

\begin{thebibliography}{85}%
\makeatletter
\providecommand \@ifxundefined [1]{%
 \@ifx{#1\undefined}
}%
\providecommand \@ifnum [1]{%
 \ifnum #1\expandafter \@firstoftwo
 \else \expandafter \@secondoftwo
 \fi
}%
\providecommand \@ifx [1]{%
 \ifx #1\expandafter \@firstoftwo
 \else \expandafter \@secondoftwo
 \fi
}%
\providecommand \natexlab [1]{#1}%
\providecommand \enquote  [1]{``#1''}%
\providecommand \bibnamefont  [1]{#1}%
\providecommand \bibfnamefont [1]{#1}%
\providecommand \citenamefont [1]{#1}%
\providecommand \href@noop [0]{\@secondoftwo}%
\providecommand \href [0]{\begingroup \@sanitize@url \@href}%
\providecommand \@href[1]{\@@startlink{#1}\@@href}%
\providecommand \@@href[1]{\endgroup#1\@@endlink}%
\providecommand \@sanitize@url [0]{\catcode `\\12\catcode `\$12\catcode
  `\&12\catcode `\#12\catcode `\^12\catcode `\_12\catcode `\%12\relax}%
\providecommand \@@startlink[1]{}%
\providecommand \@@endlink[0]{}%
\providecommand \url  [0]{\begingroup\@sanitize@url \@url }%
\providecommand \@url [1]{\endgroup\@href {#1}{\urlprefix }}%
\providecommand \urlprefix  [0]{URL }%
\providecommand \Eprint [0]{\href }%
\providecommand \doibase [0]{https://doi.org/}%
\providecommand \selectlanguage [0]{\@gobble}%
\providecommand \bibinfo  [0]{\@secondoftwo}%
\providecommand \bibfield  [0]{\@secondoftwo}%
\providecommand \translation [1]{[#1]}%
\providecommand \BibitemOpen [0]{}%
\providecommand \bibitemStop [0]{}%
\providecommand \bibitemNoStop [0]{.\EOS\space}%
\providecommand \EOS [0]{\spacefactor3000\relax}%
\providecommand \BibitemShut  [1]{\csname bibitem#1\endcsname}%
\let\auto@bib@innerbib\@empty
\bibitem [{\citenamefont {Noda}\ \emph {et~al.}(2016)\citenamefont {Noda},
  \citenamefont {Ohno},\ and\ \citenamefont {Nakamura}}]{noda2016momentum}%
  \BibitemOpen
  \bibfield  {author} {\bibinfo {author} {\bibfnamefont {Y.}~\bibnamefont
  {Noda}}, \bibinfo {author} {\bibfnamefont {K.}~\bibnamefont {Ohno}},\ and\
  \bibinfo {author} {\bibfnamefont {S.}~\bibnamefont {Nakamura}},\ }\bibfield
  {title} {\bibinfo {title} {{Momentum-dependent band spin splitting in
  semiconducting MnO$_2$: a density functional calculation}},\ }\href
  {https://doi.org/10.1039/C5CP07806G} {\bibfield  {journal} {\bibinfo
  {journal} {Physical Chemistry Chemical Physics}\ }\textbf {\bibinfo {volume}
  {18}},\ \bibinfo {pages} {13294} (\bibinfo {year} {2016})}\BibitemShut
  {NoStop}%
\bibitem [{\citenamefont {Ahn}\ \emph {et~al.}(2019)\citenamefont {Ahn},
  \citenamefont {Hariki}, \citenamefont {Lee},\ and\ \citenamefont
  {Kune\ifmmode~\check{s}\else \v{s}\fi{}}}]{ahn2019antiferromagnetism}%
  \BibitemOpen
  \bibfield  {author} {\bibinfo {author} {\bibfnamefont {K.-H.}\ \bibnamefont
  {Ahn}}, \bibinfo {author} {\bibfnamefont {A.}~\bibnamefont {Hariki}},
  \bibinfo {author} {\bibfnamefont {K.-W.}\ \bibnamefont {Lee}},\ and\ \bibinfo
  {author} {\bibfnamefont {J.}~\bibnamefont {Kune\ifmmode~\check{s}\else
  \v{s}\fi{}}},\ }\bibfield  {title} {\bibinfo {title} {{Antiferromagnetism in
  ${\mathrm{RuO}}_{2}$ as $d$-wave Pomeranchuk instability}},\ }\href
  {https://doi.org/10.1103/PhysRevB.99.184432} {\bibfield  {journal} {\bibinfo
  {journal} {Phys. Rev. B}\ }\textbf {\bibinfo {volume} {99}},\ \bibinfo
  {pages} {184432} (\bibinfo {year} {2019})}\BibitemShut {NoStop}%
\bibitem [{\citenamefont {Naka}\ \emph {et~al.}(2019)\citenamefont {Naka},
  \citenamefont {Hayami}, \citenamefont {Kusunose}, \citenamefont {Yanagi},
  \citenamefont {Motome},\ and\ \citenamefont {Seo}}]{naka2019spin}%
  \BibitemOpen
  \bibfield  {author} {\bibinfo {author} {\bibfnamefont {M.}~\bibnamefont
  {Naka}}, \bibinfo {author} {\bibfnamefont {S.}~\bibnamefont {Hayami}},
  \bibinfo {author} {\bibfnamefont {H.}~\bibnamefont {Kusunose}}, \bibinfo
  {author} {\bibfnamefont {Y.}~\bibnamefont {Yanagi}}, \bibinfo {author}
  {\bibfnamefont {Y.}~\bibnamefont {Motome}},\ and\ \bibinfo {author}
  {\bibfnamefont {H.}~\bibnamefont {Seo}},\ }\bibfield  {title} {\bibinfo
  {title} {{Spin current generation in organic antiferromagnets}},\ }\href
  {https://doi.org/10.1038/s41467-019-12229-y} {\bibfield  {journal} {\bibinfo
  {journal} {Nature Communications}\ }\textbf {\bibinfo {volume} {10}},\
  \bibinfo {pages} {4305} (\bibinfo {year} {2019})}\BibitemShut {NoStop}%
\bibitem [{\citenamefont {Hayami}\ \emph {et~al.}(2019)\citenamefont {Hayami},
  \citenamefont {Yanagi},\ and\ \citenamefont {Kusunose}}]{hayami2019momentum}%
  \BibitemOpen
  \bibfield  {author} {\bibinfo {author} {\bibfnamefont {S.}~\bibnamefont
  {Hayami}}, \bibinfo {author} {\bibfnamefont {Y.}~\bibnamefont {Yanagi}},\
  and\ \bibinfo {author} {\bibfnamefont {H.}~\bibnamefont {Kusunose}},\
  }\bibfield  {title} {\bibinfo {title} {{Momentum-Dependent Spin Splitting by
  Collinear Antiferromagnetic Ordering}},\ }\href
  {https://doi.org/10.7566/JPSJ.88.123702} {\bibfield  {journal} {\bibinfo
  {journal} {Journal of the Physical Society of Japan}\ }\textbf {\bibinfo
  {volume} {88}},\ \bibinfo {pages} {123702} (\bibinfo {year}
  {2019})}\BibitemShut {NoStop}%
\bibitem [{\citenamefont {Šmejkal}\ \emph {et~al.}(2020)\citenamefont
  {Šmejkal}, \citenamefont {González-Hernández}, \citenamefont {Jungwirth},\
  and\ \citenamefont {Sinova}}]{smejkal2020crystal}%
  \BibitemOpen
  \bibfield  {author} {\bibinfo {author} {\bibfnamefont {L.}~\bibnamefont
  {Šmejkal}}, \bibinfo {author} {\bibfnamefont {R.}~\bibnamefont
  {González-Hernández}}, \bibinfo {author} {\bibfnamefont {T.}~\bibnamefont
  {Jungwirth}},\ and\ \bibinfo {author} {\bibfnamefont {J.}~\bibnamefont
  {Sinova}},\ }\bibfield  {title} {\bibinfo {title} {{Crystal time-reversal
  symmetry breaking and spontaneous Hall effect in collinear
  antiferromagnets}},\ }\href {https://doi.org/10.1126/sciadv.aaz8809}
  {\bibfield  {journal} {\bibinfo  {journal} {Science Advances}\ }\textbf
  {\bibinfo {volume} {6}},\ \bibinfo {pages} {eaaz8809} (\bibinfo {year}
  {2020})}\BibitemShut {NoStop}%
\bibitem [{\citenamefont {Yuan}\ \emph {et~al.}(2020)\citenamefont {Yuan},
  \citenamefont {Wang}, \citenamefont {Luo}, \citenamefont {Rashba},\ and\
  \citenamefont {Zunger}}]{yuan2020giant}%
  \BibitemOpen
  \bibfield  {author} {\bibinfo {author} {\bibfnamefont {L.-D.}\ \bibnamefont
  {Yuan}}, \bibinfo {author} {\bibfnamefont {Z.}~\bibnamefont {Wang}}, \bibinfo
  {author} {\bibfnamefont {J.-W.}\ \bibnamefont {Luo}}, \bibinfo {author}
  {\bibfnamefont {E.~I.}\ \bibnamefont {Rashba}},\ and\ \bibinfo {author}
  {\bibfnamefont {A.}~\bibnamefont {Zunger}},\ }\bibfield  {title} {\bibinfo
  {title} {{Giant momentum-dependent spin splitting in centrosymmetric low-$Z$
  antiferromagnets}},\ }\href {https://doi.org/10.1103/PhysRevB.102.014422}
  {\bibfield  {journal} {\bibinfo  {journal} {Phys. Rev. B}\ }\textbf {\bibinfo
  {volume} {102}},\ \bibinfo {pages} {014422} (\bibinfo {year}
  {2020})}\BibitemShut {NoStop}%
\bibitem [{\citenamefont {Hayami}\ \emph {et~al.}(2020)\citenamefont {Hayami},
  \citenamefont {Yanagi},\ and\ \citenamefont {Kusunose}}]{hayami2020bottom}%
  \BibitemOpen
  \bibfield  {author} {\bibinfo {author} {\bibfnamefont {S.}~\bibnamefont
  {Hayami}}, \bibinfo {author} {\bibfnamefont {Y.}~\bibnamefont {Yanagi}},\
  and\ \bibinfo {author} {\bibfnamefont {H.}~\bibnamefont {Kusunose}},\
  }\bibfield  {title} {\bibinfo {title} {{Bottom-up design of spin-split and
  reshaped electronic band structures in antiferromagnets without spin-orbit
  coupling: Procedure on the basis of augmented multipoles}},\ }\href
  {https://doi.org/10.1103/PhysRevB.102.144441} {\bibfield  {journal} {\bibinfo
   {journal} {Phys. Rev. B}\ }\textbf {\bibinfo {volume} {102}},\ \bibinfo
  {pages} {144441} (\bibinfo {year} {2020})}\BibitemShut {NoStop}%
\bibitem [{\citenamefont {Reichlová}\ \emph {et~al.}(2021)\citenamefont
  {Reichlová}, \citenamefont {Seeger}, \citenamefont {González-Hernández},
  \citenamefont {Kounta}, \citenamefont {Schlitz}, \citenamefont {Kriegner},
  \citenamefont {Ritzinger}, \citenamefont {Lammel}, \citenamefont {Leiviskä},
  \citenamefont {Petříček}, \citenamefont {Doležal}, \citenamefont
  {Schmoranzerová}, \citenamefont {Bad'ura}, \citenamefont {Thomas},
  \citenamefont {Baltz}, \citenamefont {Michez}, \citenamefont {Sinova},
  \citenamefont {Goennenwein}, \citenamefont {Jungwirth},\ and\ \citenamefont
  {Šmejkal}}]{reichlova2021macroscopic}%
  \BibitemOpen
  \bibfield  {author} {\bibinfo {author} {\bibfnamefont {H.}~\bibnamefont
  {Reichlová}}, \bibinfo {author} {\bibfnamefont {R.~L.}\ \bibnamefont
  {Seeger}}, \bibinfo {author} {\bibfnamefont {R.}~\bibnamefont
  {González-Hernández}}, \bibinfo {author} {\bibfnamefont {I.}~\bibnamefont
  {Kounta}}, \bibinfo {author} {\bibfnamefont {R.}~\bibnamefont {Schlitz}},
  \bibinfo {author} {\bibfnamefont {D.}~\bibnamefont {Kriegner}}, \bibinfo
  {author} {\bibfnamefont {P.}~\bibnamefont {Ritzinger}}, \bibinfo {author}
  {\bibfnamefont {M.}~\bibnamefont {Lammel}}, \bibinfo {author} {\bibfnamefont
  {M.}~\bibnamefont {Leiviskä}}, \bibinfo {author} {\bibfnamefont
  {V.}~\bibnamefont {Petříček}}, \bibinfo {author} {\bibfnamefont
  {P.}~\bibnamefont {Doležal}}, \bibinfo {author} {\bibfnamefont
  {E.}~\bibnamefont {Schmoranzerová}}, \bibinfo {author} {\bibfnamefont
  {A.}~\bibnamefont {Bad'ura}}, \bibinfo {author} {\bibfnamefont
  {A.}~\bibnamefont {Thomas}}, \bibinfo {author} {\bibfnamefont
  {V.}~\bibnamefont {Baltz}}, \bibinfo {author} {\bibfnamefont
  {L.}~\bibnamefont {Michez}}, \bibinfo {author} {\bibfnamefont
  {J.}~\bibnamefont {Sinova}}, \bibinfo {author} {\bibfnamefont {S.~T.~B.}\
  \bibnamefont {Goennenwein}}, \bibinfo {author} {\bibfnamefont
  {T.}~\bibnamefont {Jungwirth}},\ and\ \bibinfo {author} {\bibfnamefont
  {L.}~\bibnamefont {Šmejkal}},\ }\href@noop {} {\bibinfo {title}
  {{Macroscopic time reversal symmetry breaking by staggered spin-momentum
  interaction}}} (\bibinfo {year} {2021}),\ \Eprint
  {https://arxiv.org/abs/2012.15651} {arXiv:2012.15651 [cond-mat.mes-hall]}
  \BibitemShut {NoStop}%
\bibitem [{\citenamefont {Yuan}\ \emph
  {et~al.}(2021{\natexlab{a}})\citenamefont {Yuan}, \citenamefont {Wang},
  \citenamefont {Luo},\ and\ \citenamefont {Zunger}}]{yuan2021prediction}%
  \BibitemOpen
  \bibfield  {author} {\bibinfo {author} {\bibfnamefont {L.-D.}\ \bibnamefont
  {Yuan}}, \bibinfo {author} {\bibfnamefont {Z.}~\bibnamefont {Wang}}, \bibinfo
  {author} {\bibfnamefont {J.-W.}\ \bibnamefont {Luo}},\ and\ \bibinfo {author}
  {\bibfnamefont {A.}~\bibnamefont {Zunger}},\ }\bibfield  {title} {\bibinfo
  {title} {{Prediction of low-Z collinear and noncollinear antiferromagnetic
  compounds having momentum-dependent spin splitting even without spin-orbit
  coupling}},\ }\href {https://doi.org/10.1103/PhysRevMaterials.5.014409}
  {\bibfield  {journal} {\bibinfo  {journal} {Phys. Rev. Mater.}\ }\textbf
  {\bibinfo {volume} {5}},\ \bibinfo {pages} {014409} (\bibinfo {year}
  {2021}{\natexlab{a}})}\BibitemShut {NoStop}%
\bibitem [{\citenamefont {Naka}\ \emph {et~al.}(2021)\citenamefont {Naka},
  \citenamefont {Motome},\ and\ \citenamefont {Seo}}]{naka2021perovskite}%
  \BibitemOpen
  \bibfield  {author} {\bibinfo {author} {\bibfnamefont {M.}~\bibnamefont
  {Naka}}, \bibinfo {author} {\bibfnamefont {Y.}~\bibnamefont {Motome}},\ and\
  \bibinfo {author} {\bibfnamefont {H.}~\bibnamefont {Seo}},\ }\bibfield
  {title} {\bibinfo {title} {{Perovskite as a spin current generator}},\ }\href
  {https://doi.org/10.1103/PhysRevB.103.125114} {\bibfield  {journal} {\bibinfo
   {journal} {Phys. Rev. B}\ }\textbf {\bibinfo {volume} {103}},\ \bibinfo
  {pages} {125114} (\bibinfo {year} {2021})}\BibitemShut {NoStop}%
\bibitem [{\citenamefont {Gonz\'alez-Hern\'andez}\ \emph
  {et~al.}(2021)\citenamefont {Gonz\'alez-Hern\'andez}, \citenamefont
  {\ifmmode~\check{S}\else \v{S}\fi{}mejkal}, \citenamefont {V\'yborn\'y},
  \citenamefont {Yahagi}, \citenamefont {Sinova}, \citenamefont {Jungwirth},\
  and\ \citenamefont {\ifmmode~\check{Z}\else
  \v{Z}\fi{}elezn\'y}}]{hernandez2021efficient}%
  \BibitemOpen
  \bibfield  {author} {\bibinfo {author} {\bibfnamefont {R.}~\bibnamefont
  {Gonz\'alez-Hern\'andez}}, \bibinfo {author} {\bibfnamefont {L.}~\bibnamefont
  {\ifmmode~\check{S}\else \v{S}\fi{}mejkal}}, \bibinfo {author} {\bibfnamefont
  {K.}~\bibnamefont {V\'yborn\'y}}, \bibinfo {author} {\bibfnamefont
  {Y.}~\bibnamefont {Yahagi}}, \bibinfo {author} {\bibfnamefont
  {J.}~\bibnamefont {Sinova}}, \bibinfo {author} {\bibfnamefont
  {T.}~\bibnamefont {Jungwirth}},\ and\ \bibinfo {author} {\bibfnamefont
  {J.}~\bibnamefont {\ifmmode~\check{Z}\else \v{Z}\fi{}elezn\'y}},\ }\bibfield
  {title} {\bibinfo {title} {{Efficient Electrical Spin Splitter Based on
  Nonrelativistic Collinear Antiferromagnetism}},\ }\href
  {https://doi.org/10.1103/PhysRevLett.126.127701} {\bibfield  {journal}
  {\bibinfo  {journal} {Phys. Rev. Lett.}\ }\textbf {\bibinfo {volume} {126}},\
  \bibinfo {pages} {127701} (\bibinfo {year} {2021})}\BibitemShut {NoStop}%
\bibitem [{\citenamefont {Ma}\ \emph {et~al.}(2021)\citenamefont {Ma},
  \citenamefont {Hu}, \citenamefont {Li}, \citenamefont {Liu}, \citenamefont
  {Yao}, \citenamefont {Jia},\ and\ \citenamefont
  {Liu}}]{ma2021multifunctional}%
  \BibitemOpen
  \bibfield  {author} {\bibinfo {author} {\bibfnamefont {H.-Y.}\ \bibnamefont
  {Ma}}, \bibinfo {author} {\bibfnamefont {M.}~\bibnamefont {Hu}}, \bibinfo
  {author} {\bibfnamefont {N.}~\bibnamefont {Li}}, \bibinfo {author}
  {\bibfnamefont {J.}~\bibnamefont {Liu}}, \bibinfo {author} {\bibfnamefont
  {W.}~\bibnamefont {Yao}}, \bibinfo {author} {\bibfnamefont {J.-F.}\
  \bibnamefont {Jia}},\ and\ \bibinfo {author} {\bibfnamefont {J.}~\bibnamefont
  {Liu}},\ }\bibfield  {title} {\bibinfo {title} {Multifunctional
  antiferromagnetic materials with giant piezomagnetism and noncollinear spin
  current},\ }\href {https://doi.org/10.1038/s41467-021-23127-7} {\bibfield
  {journal} {\bibinfo  {journal} {Nature Communications}\ }\textbf {\bibinfo
  {volume} {12}},\ \bibinfo {pages} {2846} (\bibinfo {year}
  {2021})}\BibitemShut {NoStop}%
\bibitem [{\citenamefont {Yuan}\ \emph
  {et~al.}(2021{\natexlab{b}})\citenamefont {Yuan}, \citenamefont {Wang},
  \citenamefont {Luo},\ and\ \citenamefont {Zunger}}]{yuan2021strong}%
  \BibitemOpen
  \bibfield  {author} {\bibinfo {author} {\bibfnamefont {L.-D.}\ \bibnamefont
  {Yuan}}, \bibinfo {author} {\bibfnamefont {Z.}~\bibnamefont {Wang}}, \bibinfo
  {author} {\bibfnamefont {J.-W.}\ \bibnamefont {Luo}},\ and\ \bibinfo {author}
  {\bibfnamefont {A.}~\bibnamefont {Zunger}},\ }\bibfield  {title} {\bibinfo
  {title} {Strong influence of nonmagnetic ligands on the momentum-dependent
  spin splitting in antiferromagnets},\ }\href
  {https://doi.org/10.1103/PhysRevB.103.224410} {\bibfield  {journal} {\bibinfo
   {journal} {Phys. Rev. B}\ }\textbf {\bibinfo {volume} {103}},\ \bibinfo
  {pages} {224410} (\bibinfo {year} {2021}{\natexlab{b}})}\BibitemShut
  {NoStop}%
\bibitem [{\citenamefont {Zhou}\ \emph {et~al.}(2021)\citenamefont {Zhou},
  \citenamefont {Feng}, \citenamefont {Yang}, \citenamefont {Guo},\ and\
  \citenamefont {Yao}}]{zhou2021crystal}%
  \BibitemOpen
  \bibfield  {author} {\bibinfo {author} {\bibfnamefont {X.}~\bibnamefont
  {Zhou}}, \bibinfo {author} {\bibfnamefont {W.}~\bibnamefont {Feng}}, \bibinfo
  {author} {\bibfnamefont {X.}~\bibnamefont {Yang}}, \bibinfo {author}
  {\bibfnamefont {G.-Y.}\ \bibnamefont {Guo}},\ and\ \bibinfo {author}
  {\bibfnamefont {Y.}~\bibnamefont {Yao}},\ }\bibfield  {title} {\bibinfo
  {title} {{Crystal chirality magneto-optical effects in collinear
  antiferromagnets}},\ }\href {https://doi.org/10.1103/PhysRevB.104.024401}
  {\bibfield  {journal} {\bibinfo  {journal} {Phys. Rev. B}\ }\textbf {\bibinfo
  {volume} {104}},\ \bibinfo {pages} {024401} (\bibinfo {year}
  {2021})}\BibitemShut {NoStop}%
\bibitem [{\citenamefont {Egorov}\ \emph {et~al.}(2021)\citenamefont {Egorov},
  \citenamefont {Litvin},\ and\ \citenamefont
  {Evarestov}}]{egorov2021antiferromagnetism}%
  \BibitemOpen
  \bibfield  {author} {\bibinfo {author} {\bibfnamefont {S.~A.}\ \bibnamefont
  {Egorov}}, \bibinfo {author} {\bibfnamefont {D.~B.}\ \bibnamefont {Litvin}},\
  and\ \bibinfo {author} {\bibfnamefont {R.~A.}\ \bibnamefont {Evarestov}},\
  }\bibfield  {title} {\bibinfo {title} {{Antiferromagnetism-Induced Spin
  Splitting in Systems Described by Magnetic Layer Groups}},\ }\href
  {https://doi.org/10.1021/acs.jpcc.1c02653} {\bibfield  {journal} {\bibinfo
  {journal} {The Journal of Physical Chemistry C}\ }\textbf {\bibinfo {volume}
  {125}},\ \bibinfo {pages} {16147} (\bibinfo {year} {2021})}\BibitemShut
  {NoStop}%
\bibitem [{\citenamefont {Mazin}\ \emph {et~al.}(2021)\citenamefont {Mazin},
  \citenamefont {Koepernik}, \citenamefont {Johannes}, \citenamefont
  {González-Hernández},\ and\ \citenamefont
  {Šmejkal}}]{mazin2021prediction}%
  \BibitemOpen
  \bibfield  {author} {\bibinfo {author} {\bibfnamefont {I.~I.}\ \bibnamefont
  {Mazin}}, \bibinfo {author} {\bibfnamefont {K.}~\bibnamefont {Koepernik}},
  \bibinfo {author} {\bibfnamefont {M.~D.}\ \bibnamefont {Johannes}}, \bibinfo
  {author} {\bibfnamefont {R.}~\bibnamefont {González-Hernández}},\ and\
  \bibinfo {author} {\bibfnamefont {L.}~\bibnamefont {Šmejkal}},\ }\bibfield
  {title} {\bibinfo {title} {{Prediction of unconventional magnetism in doped
  FeSb$_2$}},\ }\href {https://doi.org/10.1073/pnas.2108924118} {\bibfield
  {journal} {\bibinfo  {journal} {Proceedings of the National Academy of
  Sciences}\ }\textbf {\bibinfo {volume} {118}},\ \bibinfo {pages}
  {e2108924118} (\bibinfo {year} {2021})}\BibitemShut {NoStop}%
\bibitem [{\citenamefont {Shao}\ \emph {et~al.}(2021)\citenamefont {Shao},
  \citenamefont {Zhang}, \citenamefont {Li}, \citenamefont {Eom},\ and\
  \citenamefont {Tsymbal}}]{shao2021spin}%
  \BibitemOpen
  \bibfield  {author} {\bibinfo {author} {\bibfnamefont {D.-F.}\ \bibnamefont
  {Shao}}, \bibinfo {author} {\bibfnamefont {S.-H.}\ \bibnamefont {Zhang}},
  \bibinfo {author} {\bibfnamefont {M.}~\bibnamefont {Li}}, \bibinfo {author}
  {\bibfnamefont {C.-B.}\ \bibnamefont {Eom}},\ and\ \bibinfo {author}
  {\bibfnamefont {E.~Y.}\ \bibnamefont {Tsymbal}},\ }\bibfield  {title}
  {\bibinfo {title} {{Spin-neutral currents for spintronics}},\ }\href
  {https://doi.org/10.1038/s41467-021-26915-3} {\bibfield  {journal} {\bibinfo
  {journal} {Nature Communications}\ }\textbf {\bibinfo {volume} {12}},\
  \bibinfo {pages} {7061} (\bibinfo {year} {2021})}\BibitemShut {NoStop}%
\bibitem [{\citenamefont {\ifmmode~\check{S}\else \v{S}\fi{}mejkal}\ \emph
  {et~al.}(2022{\natexlab{a}})\citenamefont {\ifmmode~\check{S}\else
  \v{S}\fi{}mejkal}, \citenamefont {Hellenes}, \citenamefont
  {Gonz\'alez-Hern\'andez}, \citenamefont {Sinova},\ and\ \citenamefont
  {Jungwirth}}]{smejkal2022giant}%
  \BibitemOpen
  \bibfield  {author} {\bibinfo {author} {\bibfnamefont {L.}~\bibnamefont
  {\ifmmode~\check{S}\else \v{S}\fi{}mejkal}}, \bibinfo {author} {\bibfnamefont
  {A.~B.}\ \bibnamefont {Hellenes}}, \bibinfo {author} {\bibfnamefont
  {R.}~\bibnamefont {Gonz\'alez-Hern\'andez}}, \bibinfo {author} {\bibfnamefont
  {J.}~\bibnamefont {Sinova}},\ and\ \bibinfo {author} {\bibfnamefont
  {T.}~\bibnamefont {Jungwirth}},\ }\bibfield  {title} {\bibinfo {title}
  {{Giant and Tunneling Magnetoresistance in Unconventional Collinear
  Antiferromagnets with Nonrelativistic Spin-Momentum Coupling}},\ }\href
  {https://doi.org/10.1103/PhysRevX.12.011028} {\bibfield  {journal} {\bibinfo
  {journal} {Phys. Rev. X}\ }\textbf {\bibinfo {volume} {12}},\ \bibinfo
  {pages} {011028} (\bibinfo {year} {2022}{\natexlab{a}})}\BibitemShut
  {NoStop}%
\bibitem [{\citenamefont {\ifmmode~\check{S}\else \v{S}\fi{}mejkal}\ \emph
  {et~al.}(2022{\natexlab{b}})\citenamefont {\ifmmode~\check{S}\else
  \v{S}\fi{}mejkal}, \citenamefont {Sinova},\ and\ \citenamefont
  {Jungwirth}}]{smejkal2022beyond}%
  \BibitemOpen
  \bibfield  {author} {\bibinfo {author} {\bibfnamefont {L.}~\bibnamefont
  {\ifmmode~\check{S}\else \v{S}\fi{}mejkal}}, \bibinfo {author} {\bibfnamefont
  {J.}~\bibnamefont {Sinova}},\ and\ \bibinfo {author} {\bibfnamefont
  {T.}~\bibnamefont {Jungwirth}},\ }\bibfield  {title} {\bibinfo {title}
  {{Beyond Conventional Ferromagnetism and Antiferromagnetism: A Phase with
  Nonrelativistic Spin and Crystal Rotation Symmetry}},\ }\href
  {https://doi.org/10.1103/PhysRevX.12.031042} {\bibfield  {journal} {\bibinfo
  {journal} {Phys. Rev. X}\ }\textbf {\bibinfo {volume} {12}},\ \bibinfo
  {pages} {031042} (\bibinfo {year} {2022}{\natexlab{b}})}\BibitemShut
  {NoStop}%
\bibitem [{\citenamefont {\ifmmode~\check{S}\else \v{S}\fi{}mejkal}\ \emph
  {et~al.}(2022{\natexlab{c}})\citenamefont {\ifmmode~\check{S}\else
  \v{S}\fi{}mejkal}, \citenamefont {Sinova},\ and\ \citenamefont
  {Jungwirth}}]{smejkal2022emerging}%
  \BibitemOpen
  \bibfield  {author} {\bibinfo {author} {\bibfnamefont {L.}~\bibnamefont
  {\ifmmode~\check{S}\else \v{S}\fi{}mejkal}}, \bibinfo {author} {\bibfnamefont
  {J.}~\bibnamefont {Sinova}},\ and\ \bibinfo {author} {\bibfnamefont
  {T.}~\bibnamefont {Jungwirth}},\ }\bibfield  {title} {\bibinfo {title}
  {{Emerging Research Landscape of Altermagnetism}},\ }\href
  {https://doi.org/10.1103/PhysRevX.12.040501} {\bibfield  {journal} {\bibinfo
  {journal} {Phys. Rev. X}\ }\textbf {\bibinfo {volume} {12}},\ \bibinfo
  {pages} {040501} (\bibinfo {year} {2022}{\natexlab{c}})}\BibitemShut
  {NoStop}%
\bibitem [{\citenamefont {Mazin}(2022)}]{mazin2022editorial}%
  \BibitemOpen
  \bibfield  {author} {\bibinfo {author} {\bibfnamefont {I.}~\bibnamefont
  {Mazin}} (\bibinfo {collaboration} {The PRX Editors}),\ }\bibfield  {title}
  {\bibinfo {title} {{Editorial: Altermagnetism---A New Punch Line of
  Fundamental Magnetism}},\ }\href {https://doi.org/10.1103/PhysRevX.12.040002}
  {\bibfield  {journal} {\bibinfo  {journal} {Phys. Rev. X}\ }\textbf {\bibinfo
  {volume} {12}},\ \bibinfo {pages} {040002} (\bibinfo {year}
  {2022})}\BibitemShut {NoStop}%
\bibitem [{\citenamefont {Chen}\ \emph {et~al.}(2024)\citenamefont {Chen},
  \citenamefont {Liu}, \citenamefont {Zhou}, \citenamefont {Meng},
  \citenamefont {Wang}, \citenamefont {Duan}, \citenamefont {Zhao},
  \citenamefont {Yan}, \citenamefont {Qin},\ and\ \citenamefont
  {Liu}}]{chen2024emerging}%
  \BibitemOpen
  \bibfield  {author} {\bibinfo {author} {\bibfnamefont {H.}~\bibnamefont
  {Chen}}, \bibinfo {author} {\bibfnamefont {L.}~\bibnamefont {Liu}}, \bibinfo
  {author} {\bibfnamefont {X.}~\bibnamefont {Zhou}}, \bibinfo {author}
  {\bibfnamefont {Z.}~\bibnamefont {Meng}}, \bibinfo {author} {\bibfnamefont
  {X.}~\bibnamefont {Wang}}, \bibinfo {author} {\bibfnamefont {Z.}~\bibnamefont
  {Duan}}, \bibinfo {author} {\bibfnamefont {G.}~\bibnamefont {Zhao}}, \bibinfo
  {author} {\bibfnamefont {H.}~\bibnamefont {Yan}}, \bibinfo {author}
  {\bibfnamefont {P.}~\bibnamefont {Qin}},\ and\ \bibinfo {author}
  {\bibfnamefont {Z.}~\bibnamefont {Liu}},\ }\bibfield  {title} {\bibinfo
  {title} {{Emerging Antiferromagnets for Spintronics}},\ }\href
  {https://doi.org/https://doi.org/10.1002/adma.202310379} {\bibfield
  {journal} {\bibinfo  {journal} {Advanced Materials}\ }\textbf {\bibinfo
  {volume} {36}},\ \bibinfo {pages} {2310379} (\bibinfo {year}
  {2024})}\BibitemShut {NoStop}%
\bibitem [{\citenamefont {Bai}\ \emph {et~al.}(2024)\citenamefont {Bai},
  \citenamefont {Feng}, \citenamefont {Liu}, \citenamefont {Šmejkal},
  \citenamefont {Mokrousov},\ and\ \citenamefont
  {Yao}}]{bai2024altermagnetism}%
  \BibitemOpen
  \bibfield  {author} {\bibinfo {author} {\bibfnamefont {L.}~\bibnamefont
  {Bai}}, \bibinfo {author} {\bibfnamefont {W.}~\bibnamefont {Feng}}, \bibinfo
  {author} {\bibfnamefont {S.}~\bibnamefont {Liu}}, \bibinfo {author}
  {\bibfnamefont {L.}~\bibnamefont {Šmejkal}}, \bibinfo {author}
  {\bibfnamefont {Y.}~\bibnamefont {Mokrousov}},\ and\ \bibinfo {author}
  {\bibfnamefont {Y.}~\bibnamefont {Yao}},\ }\href@noop {} {\bibinfo {title}
  {{Altermagnetism: Exploring New Frontiers in Magnetism and Spintronics}}}
  (\bibinfo {year} {2024}),\ \Eprint {https://arxiv.org/abs/2406.02123}
  {arXiv:2406.02123 [cond-mat.mtrl-sci]} \BibitemShut {NoStop}%
\bibitem [{\citenamefont {Fedchenko}\ \emph {et~al.}(2024)\citenamefont
  {Fedchenko}, \citenamefont {Minár}, \citenamefont {Akashdeep}, \citenamefont
  {D’Souza}, \citenamefont {Vasilyev}, \citenamefont {Tkach}, \citenamefont
  {Odenbreit}, \citenamefont {Nguyen}, \citenamefont {Kutnyakhov},
  \citenamefont {Wind}, \citenamefont {Wenthaus}, \citenamefont {Scholz},
  \citenamefont {Rossnagel}, \citenamefont {Hoesch}, \citenamefont
  {Aeschlimann}, \citenamefont {Stadtmüller}, \citenamefont {Kläui},
  \citenamefont {Schönhense}, \citenamefont {Jungwirth}, \citenamefont
  {Hellenes}, \citenamefont {Jakob}, \citenamefont {Šmejkal}, \citenamefont
  {Sinova},\ and\ \citenamefont {Elmers}}]{fedchenko2024observation}%
  \BibitemOpen
  \bibfield  {author} {\bibinfo {author} {\bibfnamefont {O.}~\bibnamefont
  {Fedchenko}}, \bibinfo {author} {\bibfnamefont {J.}~\bibnamefont {Minár}},
  \bibinfo {author} {\bibfnamefont {A.}~\bibnamefont {Akashdeep}}, \bibinfo
  {author} {\bibfnamefont {S.~W.}\ \bibnamefont {D’Souza}}, \bibinfo {author}
  {\bibfnamefont {D.}~\bibnamefont {Vasilyev}}, \bibinfo {author}
  {\bibfnamefont {O.}~\bibnamefont {Tkach}}, \bibinfo {author} {\bibfnamefont
  {L.}~\bibnamefont {Odenbreit}}, \bibinfo {author} {\bibfnamefont
  {Q.}~\bibnamefont {Nguyen}}, \bibinfo {author} {\bibfnamefont
  {D.}~\bibnamefont {Kutnyakhov}}, \bibinfo {author} {\bibfnamefont
  {N.}~\bibnamefont {Wind}}, \bibinfo {author} {\bibfnamefont {L.}~\bibnamefont
  {Wenthaus}}, \bibinfo {author} {\bibfnamefont {M.}~\bibnamefont {Scholz}},
  \bibinfo {author} {\bibfnamefont {K.}~\bibnamefont {Rossnagel}}, \bibinfo
  {author} {\bibfnamefont {M.}~\bibnamefont {Hoesch}}, \bibinfo {author}
  {\bibfnamefont {M.}~\bibnamefont {Aeschlimann}}, \bibinfo {author}
  {\bibfnamefont {B.}~\bibnamefont {Stadtmüller}}, \bibinfo {author}
  {\bibfnamefont {M.}~\bibnamefont {Kläui}}, \bibinfo {author} {\bibfnamefont
  {G.}~\bibnamefont {Schönhense}}, \bibinfo {author} {\bibfnamefont
  {T.}~\bibnamefont {Jungwirth}}, \bibinfo {author} {\bibfnamefont {A.~B.}\
  \bibnamefont {Hellenes}}, \bibinfo {author} {\bibfnamefont {G.}~\bibnamefont
  {Jakob}}, \bibinfo {author} {\bibfnamefont {L.}~\bibnamefont {Šmejkal}},
  \bibinfo {author} {\bibfnamefont {J.}~\bibnamefont {Sinova}},\ and\ \bibinfo
  {author} {\bibfnamefont {H.-J.}\ \bibnamefont {Elmers}},\ }\bibfield  {title}
  {\bibinfo {title} {{Observation of time-reversal symmetry breaking in the
  band structure of altermagnetic ${\mathrm{RuO}}_{2}$}},\ }\href
  {https://doi.org/10.1126/sciadv.adj4883} {\bibfield  {journal} {\bibinfo
  {journal} {Science Advances}\ }\textbf {\bibinfo {volume} {10}},\ \bibinfo
  {pages} {eadj4883} (\bibinfo {year} {2024})}\BibitemShut {NoStop}%
\bibitem [{\citenamefont {Krempask{\'y}}\ \emph {et~al.}(2024)\citenamefont
  {Krempask{\'y}}, \citenamefont {{\v S}mejkal}, \citenamefont {D'Souza},
  \citenamefont {Hajlaoui}, \citenamefont {Springholz}, \citenamefont
  {Uhl{\'\i}{\v r}ov{\'a}}, \citenamefont {Alarab}, \citenamefont
  {Constantinou}, \citenamefont {Strocov}, \citenamefont {Usanov},
  \citenamefont {Pudelko}, \citenamefont {Gonz{\'a}lez-Hern{\'a}ndez},
  \citenamefont {Birk~Hellenes}, \citenamefont {Jansa}, \citenamefont
  {Reichlov{\'a}}, \citenamefont {{\v S}ob{\'a}{\v n}}, \citenamefont
  {Gonzalez~Betancourt}, \citenamefont {Wadley}, \citenamefont {Sinova},
  \citenamefont {Kriegner}, \citenamefont {Min{\'a}r}, \citenamefont {Dil},\
  and\ \citenamefont {Jungwirth}}]{krempasky2024altermagnetic}%
  \BibitemOpen
  \bibfield  {author} {\bibinfo {author} {\bibfnamefont {J.}~\bibnamefont
  {Krempask{\'y}}}, \bibinfo {author} {\bibfnamefont {L.}~\bibnamefont {{\v
  S}mejkal}}, \bibinfo {author} {\bibfnamefont {S.~W.}\ \bibnamefont
  {D'Souza}}, \bibinfo {author} {\bibfnamefont {M.}~\bibnamefont {Hajlaoui}},
  \bibinfo {author} {\bibfnamefont {G.}~\bibnamefont {Springholz}}, \bibinfo
  {author} {\bibfnamefont {K.}~\bibnamefont {Uhl{\'\i}{\v r}ov{\'a}}}, \bibinfo
  {author} {\bibfnamefont {F.}~\bibnamefont {Alarab}}, \bibinfo {author}
  {\bibfnamefont {P.~C.}\ \bibnamefont {Constantinou}}, \bibinfo {author}
  {\bibfnamefont {V.}~\bibnamefont {Strocov}}, \bibinfo {author} {\bibfnamefont
  {D.}~\bibnamefont {Usanov}}, \bibinfo {author} {\bibfnamefont {W.~R.}\
  \bibnamefont {Pudelko}}, \bibinfo {author} {\bibfnamefont {R.}~\bibnamefont
  {Gonz{\'a}lez-Hern{\'a}ndez}}, \bibinfo {author} {\bibfnamefont
  {A.}~\bibnamefont {Birk~Hellenes}}, \bibinfo {author} {\bibfnamefont
  {Z.}~\bibnamefont {Jansa}}, \bibinfo {author} {\bibfnamefont
  {H.}~\bibnamefont {Reichlov{\'a}}}, \bibinfo {author} {\bibfnamefont
  {Z.}~\bibnamefont {{\v S}ob{\'a}{\v n}}}, \bibinfo {author} {\bibfnamefont
  {R.~D.}\ \bibnamefont {Gonzalez~Betancourt}}, \bibinfo {author}
  {\bibfnamefont {P.}~\bibnamefont {Wadley}}, \bibinfo {author} {\bibfnamefont
  {J.}~\bibnamefont {Sinova}}, \bibinfo {author} {\bibfnamefont
  {D.}~\bibnamefont {Kriegner}}, \bibinfo {author} {\bibfnamefont
  {J.}~\bibnamefont {Min{\'a}r}}, \bibinfo {author} {\bibfnamefont {J.~H.}\
  \bibnamefont {Dil}},\ and\ \bibinfo {author} {\bibfnamefont {T.}~\bibnamefont
  {Jungwirth}},\ }\bibfield  {title} {\bibinfo {title} {{Altermagnetic lifting
  of Kramers spin degeneracy}},\ }\href
  {https://doi.org/10.1038/s41586-023-06907-7} {\bibfield  {journal} {\bibinfo
  {journal} {Nature}\ }\textbf {\bibinfo {volume} {626}},\ \bibinfo {pages}
  {517} (\bibinfo {year} {2024})}\BibitemShut {NoStop}%
\bibitem [{\citenamefont {Zhu}\ \emph {et~al.}(2024{\natexlab{a}})\citenamefont
  {Zhu}, \citenamefont {Chen}, \citenamefont {Liu}, \citenamefont {Liu},
  \citenamefont {Liu}, \citenamefont {Zha}, \citenamefont {Qu}, \citenamefont
  {Hong}, \citenamefont {Li}, \citenamefont {Jiang}, \citenamefont {Ma},
  \citenamefont {Hao}, \citenamefont {Zhu}, \citenamefont {Liu}, \citenamefont
  {Zeng}, \citenamefont {Jayaram}, \citenamefont {Lenger}, \citenamefont
  {Ding}, \citenamefont {Mo}, \citenamefont {Tanaka}, \citenamefont {Arita},
  \citenamefont {Liu}, \citenamefont {Ye}, \citenamefont {Shen}, \citenamefont
  {Wrachtrup}, \citenamefont {Huang}, \citenamefont {He}, \citenamefont {Qiao},
  \citenamefont {Liu},\ and\ \citenamefont {Liu}}]{zhu2024observation}%
  \BibitemOpen
  \bibfield  {author} {\bibinfo {author} {\bibfnamefont {Y.-P.}\ \bibnamefont
  {Zhu}}, \bibinfo {author} {\bibfnamefont {X.}~\bibnamefont {Chen}}, \bibinfo
  {author} {\bibfnamefont {X.-R.}\ \bibnamefont {Liu}}, \bibinfo {author}
  {\bibfnamefont {Y.}~\bibnamefont {Liu}}, \bibinfo {author} {\bibfnamefont
  {P.}~\bibnamefont {Liu}}, \bibinfo {author} {\bibfnamefont {H.}~\bibnamefont
  {Zha}}, \bibinfo {author} {\bibfnamefont {G.}~\bibnamefont {Qu}}, \bibinfo
  {author} {\bibfnamefont {C.}~\bibnamefont {Hong}}, \bibinfo {author}
  {\bibfnamefont {J.}~\bibnamefont {Li}}, \bibinfo {author} {\bibfnamefont
  {Z.}~\bibnamefont {Jiang}}, \bibinfo {author} {\bibfnamefont {X.-M.}\
  \bibnamefont {Ma}}, \bibinfo {author} {\bibfnamefont {Y.-J.}\ \bibnamefont
  {Hao}}, \bibinfo {author} {\bibfnamefont {M.-Y.}\ \bibnamefont {Zhu}},
  \bibinfo {author} {\bibfnamefont {W.}~\bibnamefont {Liu}}, \bibinfo {author}
  {\bibfnamefont {M.}~\bibnamefont {Zeng}}, \bibinfo {author} {\bibfnamefont
  {S.}~\bibnamefont {Jayaram}}, \bibinfo {author} {\bibfnamefont
  {M.}~\bibnamefont {Lenger}}, \bibinfo {author} {\bibfnamefont
  {J.}~\bibnamefont {Ding}}, \bibinfo {author} {\bibfnamefont {S.}~\bibnamefont
  {Mo}}, \bibinfo {author} {\bibfnamefont {K.}~\bibnamefont {Tanaka}}, \bibinfo
  {author} {\bibfnamefont {M.}~\bibnamefont {Arita}}, \bibinfo {author}
  {\bibfnamefont {Z.}~\bibnamefont {Liu}}, \bibinfo {author} {\bibfnamefont
  {M.}~\bibnamefont {Ye}}, \bibinfo {author} {\bibfnamefont {D.}~\bibnamefont
  {Shen}}, \bibinfo {author} {\bibfnamefont {J.}~\bibnamefont {Wrachtrup}},
  \bibinfo {author} {\bibfnamefont {Y.}~\bibnamefont {Huang}}, \bibinfo
  {author} {\bibfnamefont {R.-H.}\ \bibnamefont {He}}, \bibinfo {author}
  {\bibfnamefont {S.}~\bibnamefont {Qiao}}, \bibinfo {author} {\bibfnamefont
  {Q.}~\bibnamefont {Liu}},\ and\ \bibinfo {author} {\bibfnamefont
  {C.}~\bibnamefont {Liu}},\ }\bibfield  {title} {\bibinfo {title}
  {{Observation of plaid-like spin splitting in a noncoplanar
  antiferromagnet}},\ }\href {https://doi.org/10.1038/s41586-024-07023-w}
  {\bibfield  {journal} {\bibinfo  {journal} {Nature}\ }\textbf {\bibinfo
  {volume} {626}},\ \bibinfo {pages} {523} (\bibinfo {year}
  {2024}{\natexlab{a}})}\BibitemShut {NoStop}%
\bibitem [{\citenamefont {Hariki}\ \emph {et~al.}(2024)\citenamefont {Hariki},
  \citenamefont {Dal~Din}, \citenamefont {Amin}, \citenamefont {Yamaguchi},
  \citenamefont {Badura}, \citenamefont {Kriegner}, \citenamefont {Edmonds},
  \citenamefont {Campion}, \citenamefont {Wadley}, \citenamefont {Backes},
  \citenamefont {Veiga}, \citenamefont {Dhesi}, \citenamefont {Springholz},
  \citenamefont {\ifmmode~\check{S}\else \v{S}\fi{}mejkal}, \citenamefont
  {V\'yborn\'y}, \citenamefont {Jungwirth},\ and\ \citenamefont
  {Kune\ifmmode~\check{s}\else \v{s}\fi{}}}]{hariki2024xray}%
  \BibitemOpen
  \bibfield  {author} {\bibinfo {author} {\bibfnamefont {A.}~\bibnamefont
  {Hariki}}, \bibinfo {author} {\bibfnamefont {A.}~\bibnamefont {Dal~Din}},
  \bibinfo {author} {\bibfnamefont {O.~J.}\ \bibnamefont {Amin}}, \bibinfo
  {author} {\bibfnamefont {T.}~\bibnamefont {Yamaguchi}}, \bibinfo {author}
  {\bibfnamefont {A.}~\bibnamefont {Badura}}, \bibinfo {author} {\bibfnamefont
  {D.}~\bibnamefont {Kriegner}}, \bibinfo {author} {\bibfnamefont {K.~W.}\
  \bibnamefont {Edmonds}}, \bibinfo {author} {\bibfnamefont {R.~P.}\
  \bibnamefont {Campion}}, \bibinfo {author} {\bibfnamefont {P.}~\bibnamefont
  {Wadley}}, \bibinfo {author} {\bibfnamefont {D.}~\bibnamefont {Backes}},
  \bibinfo {author} {\bibfnamefont {L.~S.~I.}\ \bibnamefont {Veiga}}, \bibinfo
  {author} {\bibfnamefont {S.~S.}\ \bibnamefont {Dhesi}}, \bibinfo {author}
  {\bibfnamefont {G.}~\bibnamefont {Springholz}}, \bibinfo {author}
  {\bibfnamefont {L.}~\bibnamefont {\ifmmode~\check{S}\else \v{S}\fi{}mejkal}},
  \bibinfo {author} {\bibfnamefont {K.}~\bibnamefont {V\'yborn\'y}}, \bibinfo
  {author} {\bibfnamefont {T.}~\bibnamefont {Jungwirth}},\ and\ \bibinfo
  {author} {\bibfnamefont {J.}~\bibnamefont {Kune\ifmmode~\check{s}\else
  \v{s}\fi{}}},\ }\bibfield  {title} {\bibinfo {title} {{X-Ray Magnetic
  Circular Dichroism in Altermagnetic $\ensuremath{\alpha}$-MnTe}},\ }\href
  {https://doi.org/10.1103/PhysRevLett.132.176701} {\bibfield  {journal}
  {\bibinfo  {journal} {Phys. Rev. Lett.}\ }\textbf {\bibinfo {volume} {132}},\
  \bibinfo {pages} {176701} (\bibinfo {year} {2024})}\BibitemShut {NoStop}%
\bibitem [{\citenamefont {{\v S}mejkal}\ \emph {et~al.}(2022)\citenamefont {{\v
  S}mejkal}, \citenamefont {MacDonald}, \citenamefont {Sinova}, \citenamefont
  {Nakatsuji},\ and\ \citenamefont {Jungwirth}}]{smejkal2022anomalous}%
  \BibitemOpen
  \bibfield  {author} {\bibinfo {author} {\bibfnamefont {L.}~\bibnamefont {{\v
  S}mejkal}}, \bibinfo {author} {\bibfnamefont {A.~H.}\ \bibnamefont
  {MacDonald}}, \bibinfo {author} {\bibfnamefont {J.}~\bibnamefont {Sinova}},
  \bibinfo {author} {\bibfnamefont {S.}~\bibnamefont {Nakatsuji}},\ and\
  \bibinfo {author} {\bibfnamefont {T.}~\bibnamefont {Jungwirth}},\ }\bibfield
  {title} {\bibinfo {title} {{Anomalous Hall antiferromagnets}},\ }\href
  {https://doi.org/10.1038/s41578-022-00430-3} {\bibfield  {journal} {\bibinfo
  {journal} {Nature Reviews Materials}\ }\textbf {\bibinfo {volume} {7}},\
  \bibinfo {pages} {482} (\bibinfo {year} {2022})}\BibitemShut {NoStop}%
\bibitem [{\citenamefont {Feng}\ \emph {et~al.}(2022)\citenamefont {Feng},
  \citenamefont {Zhou}, \citenamefont {{\v S}mejkal}, \citenamefont {Wu},
  \citenamefont {Zhu}, \citenamefont {Guo}, \citenamefont
  {Gonz{\'a}lez-Hern{\'a}ndez}, \citenamefont {Wang}, \citenamefont {Yan},
  \citenamefont {Qin}, \citenamefont {Zhang}, \citenamefont {Wu}, \citenamefont
  {Chen}, \citenamefont {Meng}, \citenamefont {Liu}, \citenamefont {Xia},
  \citenamefont {Sinova}, \citenamefont {Jungwirth},\ and\ \citenamefont
  {Liu}}]{feng2022anomalous}%
  \BibitemOpen
  \bibfield  {author} {\bibinfo {author} {\bibfnamefont {Z.}~\bibnamefont
  {Feng}}, \bibinfo {author} {\bibfnamefont {X.}~\bibnamefont {Zhou}}, \bibinfo
  {author} {\bibfnamefont {L.}~\bibnamefont {{\v S}mejkal}}, \bibinfo {author}
  {\bibfnamefont {L.}~\bibnamefont {Wu}}, \bibinfo {author} {\bibfnamefont
  {Z.}~\bibnamefont {Zhu}}, \bibinfo {author} {\bibfnamefont {H.}~\bibnamefont
  {Guo}}, \bibinfo {author} {\bibfnamefont {R.}~\bibnamefont
  {Gonz{\'a}lez-Hern{\'a}ndez}}, \bibinfo {author} {\bibfnamefont
  {X.}~\bibnamefont {Wang}}, \bibinfo {author} {\bibfnamefont {H.}~\bibnamefont
  {Yan}}, \bibinfo {author} {\bibfnamefont {P.}~\bibnamefont {Qin}}, \bibinfo
  {author} {\bibfnamefont {X.}~\bibnamefont {Zhang}}, \bibinfo {author}
  {\bibfnamefont {H.}~\bibnamefont {Wu}}, \bibinfo {author} {\bibfnamefont
  {H.}~\bibnamefont {Chen}}, \bibinfo {author} {\bibfnamefont {Z.}~\bibnamefont
  {Meng}}, \bibinfo {author} {\bibfnamefont {L.}~\bibnamefont {Liu}}, \bibinfo
  {author} {\bibfnamefont {Z.}~\bibnamefont {Xia}}, \bibinfo {author}
  {\bibfnamefont {J.}~\bibnamefont {Sinova}}, \bibinfo {author} {\bibfnamefont
  {T.}~\bibnamefont {Jungwirth}},\ and\ \bibinfo {author} {\bibfnamefont
  {Z.}~\bibnamefont {Liu}},\ }\bibfield  {title} {\bibinfo {title} {{An
  anomalous Hall effect in altermagnetic ruthenium dioxide}},\ }\href
  {https://doi.org/10.1038/s41928-022-00866-z} {\bibfield  {journal} {\bibinfo
  {journal} {Nature Electronics}\ }\textbf {\bibinfo {volume} {5}},\ \bibinfo
  {pages} {735} (\bibinfo {year} {2022})}\BibitemShut {NoStop}%
\bibitem [{\citenamefont {Bose}\ \emph {et~al.}(2022)\citenamefont {Bose},
  \citenamefont {Schreiber}, \citenamefont {Jain}, \citenamefont {Shao},
  \citenamefont {Nair}, \citenamefont {Sun}, \citenamefont {Zhang},
  \citenamefont {Muller}, \citenamefont {Tsymbal}, \citenamefont {Schlom},\
  and\ \citenamefont {Ralph}}]{bose2022tilted}%
  \BibitemOpen
  \bibfield  {author} {\bibinfo {author} {\bibfnamefont {A.}~\bibnamefont
  {Bose}}, \bibinfo {author} {\bibfnamefont {N.~J.}\ \bibnamefont {Schreiber}},
  \bibinfo {author} {\bibfnamefont {R.}~\bibnamefont {Jain}}, \bibinfo {author}
  {\bibfnamefont {D.-F.}\ \bibnamefont {Shao}}, \bibinfo {author}
  {\bibfnamefont {H.~P.}\ \bibnamefont {Nair}}, \bibinfo {author}
  {\bibfnamefont {J.}~\bibnamefont {Sun}}, \bibinfo {author} {\bibfnamefont
  {X.~S.}\ \bibnamefont {Zhang}}, \bibinfo {author} {\bibfnamefont {D.~A.}\
  \bibnamefont {Muller}}, \bibinfo {author} {\bibfnamefont {E.~Y.}\
  \bibnamefont {Tsymbal}}, \bibinfo {author} {\bibfnamefont {D.~G.}\
  \bibnamefont {Schlom}},\ and\ \bibinfo {author} {\bibfnamefont {D.~C.}\
  \bibnamefont {Ralph}},\ }\bibfield  {title} {\bibinfo {title} {{Tilted spin
  current generated by the collinear antiferromagnet ruthenium dioxide}},\
  }\href {https://doi.org/10.1038/s41928-022-00744-8} {\bibfield  {journal}
  {\bibinfo  {journal} {Nature Electronics}\ }\textbf {\bibinfo {volume} {5}},\
  \bibinfo {pages} {267} (\bibinfo {year} {2022})}\BibitemShut {NoStop}%
\bibitem [{\citenamefont {Bai}\ \emph {et~al.}(2022)\citenamefont {Bai},
  \citenamefont {Han}, \citenamefont {Feng}, \citenamefont {Zhou},
  \citenamefont {Su}, \citenamefont {Wang}, \citenamefont {Liao}, \citenamefont
  {Zhu}, \citenamefont {Chen}, \citenamefont {Pan}, \citenamefont {Fan},\ and\
  \citenamefont {Song}}]{bai2022observation}%
  \BibitemOpen
  \bibfield  {author} {\bibinfo {author} {\bibfnamefont {H.}~\bibnamefont
  {Bai}}, \bibinfo {author} {\bibfnamefont {L.}~\bibnamefont {Han}}, \bibinfo
  {author} {\bibfnamefont {X.~Y.}\ \bibnamefont {Feng}}, \bibinfo {author}
  {\bibfnamefont {Y.~J.}\ \bibnamefont {Zhou}}, \bibinfo {author}
  {\bibfnamefont {R.~X.}\ \bibnamefont {Su}}, \bibinfo {author} {\bibfnamefont
  {Q.}~\bibnamefont {Wang}}, \bibinfo {author} {\bibfnamefont {L.~Y.}\
  \bibnamefont {Liao}}, \bibinfo {author} {\bibfnamefont {W.~X.}\ \bibnamefont
  {Zhu}}, \bibinfo {author} {\bibfnamefont {X.~Z.}\ \bibnamefont {Chen}},
  \bibinfo {author} {\bibfnamefont {F.}~\bibnamefont {Pan}}, \bibinfo {author}
  {\bibfnamefont {X.~L.}\ \bibnamefont {Fan}},\ and\ \bibinfo {author}
  {\bibfnamefont {C.}~\bibnamefont {Song}},\ }\bibfield  {title} {\bibinfo
  {title} {{Observation of Spin Splitting Torque in a Collinear Antiferromagnet
  ${\mathrm{RuO}}_{2}$}},\ }\href
  {https://doi.org/10.1103/PhysRevLett.128.197202} {\bibfield  {journal}
  {\bibinfo  {journal} {Phys. Rev. Lett.}\ }\textbf {\bibinfo {volume} {128}},\
  \bibinfo {pages} {197202} (\bibinfo {year} {2022})}\BibitemShut {NoStop}%
\bibitem [{\citenamefont {Karube}\ \emph {et~al.}(2022)\citenamefont {Karube},
  \citenamefont {Tanaka}, \citenamefont {Sugawara}, \citenamefont {Kadoguchi},
  \citenamefont {Kohda},\ and\ \citenamefont {Nitta}}]{karube2022observation}%
  \BibitemOpen
  \bibfield  {author} {\bibinfo {author} {\bibfnamefont {S.}~\bibnamefont
  {Karube}}, \bibinfo {author} {\bibfnamefont {T.}~\bibnamefont {Tanaka}},
  \bibinfo {author} {\bibfnamefont {D.}~\bibnamefont {Sugawara}}, \bibinfo
  {author} {\bibfnamefont {N.}~\bibnamefont {Kadoguchi}}, \bibinfo {author}
  {\bibfnamefont {M.}~\bibnamefont {Kohda}},\ and\ \bibinfo {author}
  {\bibfnamefont {J.}~\bibnamefont {Nitta}},\ }\bibfield  {title} {\bibinfo
  {title} {{Observation of Spin-Splitter Torque in Collinear Antiferromagnetic
  ${\mathrm{RuO}}_{2}$}},\ }\href
  {https://doi.org/10.1103/PhysRevLett.129.137201} {\bibfield  {journal}
  {\bibinfo  {journal} {Phys. Rev. Lett.}\ }\textbf {\bibinfo {volume} {129}},\
  \bibinfo {pages} {137201} (\bibinfo {year} {2022})}\BibitemShut {NoStop}%
\bibitem [{\citenamefont {\ifmmode \check{Z}\else
  \v{Z}\fi{}uti\ifmmode~\acute{c}\else \'{c}\fi{}}\ \emph
  {et~al.}(2004)\citenamefont {\ifmmode \check{Z}\else
  \v{Z}\fi{}uti\ifmmode~\acute{c}\else \'{c}\fi{}}, \citenamefont {Fabian},\
  and\ \citenamefont {Das~Sarma}}]{zutic2004spintronics}%
  \BibitemOpen
  \bibfield  {author} {\bibinfo {author} {\bibfnamefont {I.}~\bibnamefont
  {\ifmmode \check{Z}\else \v{Z}\fi{}uti\ifmmode~\acute{c}\else \'{c}\fi{}}},
  \bibinfo {author} {\bibfnamefont {J.}~\bibnamefont {Fabian}},\ and\ \bibinfo
  {author} {\bibfnamefont {S.}~\bibnamefont {Das~Sarma}},\ }\bibfield  {title}
  {\bibinfo {title} {{Spintronics: Fundamentals and applications}},\ }\href
  {https://doi.org/10.1103/RevModPhys.76.323} {\bibfield  {journal} {\bibinfo
  {journal} {Rev. Mod. Phys.}\ }\textbf {\bibinfo {volume} {76}},\ \bibinfo
  {pages} {323} (\bibinfo {year} {2004})}\BibitemShut {NoStop}%
\bibitem [{\citenamefont {Hirohata}\ \emph {et~al.}(2020)\citenamefont
  {Hirohata}, \citenamefont {Yamada}, \citenamefont {Nakatani}, \citenamefont
  {Prejbeanu}, \citenamefont {Diény}, \citenamefont {Pirro},\ and\
  \citenamefont {Hillebrands}}]{hirohata2020review}%
  \BibitemOpen
  \bibfield  {author} {\bibinfo {author} {\bibfnamefont {A.}~\bibnamefont
  {Hirohata}}, \bibinfo {author} {\bibfnamefont {K.}~\bibnamefont {Yamada}},
  \bibinfo {author} {\bibfnamefont {Y.}~\bibnamefont {Nakatani}}, \bibinfo
  {author} {\bibfnamefont {I.-L.}\ \bibnamefont {Prejbeanu}}, \bibinfo {author}
  {\bibfnamefont {B.}~\bibnamefont {Diény}}, \bibinfo {author} {\bibfnamefont
  {P.}~\bibnamefont {Pirro}},\ and\ \bibinfo {author} {\bibfnamefont
  {B.}~\bibnamefont {Hillebrands}},\ }\bibfield  {title} {\bibinfo {title}
  {{Review on spintronics: Principles and device applications}},\ }\href
  {https://doi.org/https://doi.org/10.1016/j.jmmm.2020.166711} {\bibfield
  {journal} {\bibinfo  {journal} {Journal of Magnetism and Magnetic Materials}\
  }\textbf {\bibinfo {volume} {509}},\ \bibinfo {pages} {166711} (\bibinfo
  {year} {2020})}\BibitemShut {NoStop}%
\bibitem [{\citenamefont {Jungwirth}\ \emph {et~al.}(2016)\citenamefont
  {Jungwirth}, \citenamefont {Marti}, \citenamefont {Wadley},\ and\
  \citenamefont {Wunderlich}}]{jungwirth2016antiferromagnetic}%
  \BibitemOpen
  \bibfield  {author} {\bibinfo {author} {\bibfnamefont {T.}~\bibnamefont
  {Jungwirth}}, \bibinfo {author} {\bibfnamefont {X.}~\bibnamefont {Marti}},
  \bibinfo {author} {\bibfnamefont {P.}~\bibnamefont {Wadley}},\ and\ \bibinfo
  {author} {\bibfnamefont {J.}~\bibnamefont {Wunderlich}},\ }\bibfield  {title}
  {\bibinfo {title} {{Antiferromagnetic spintronics}},\ }\href
  {https://doi.org/10.1038/nnano.2016.18} {\bibfield  {journal} {\bibinfo
  {journal} {Nature Nanotechnology}\ }\textbf {\bibinfo {volume} {11}},\
  \bibinfo {pages} {231} (\bibinfo {year} {2016})}\BibitemShut {NoStop}%
\bibitem [{\citenamefont {Baltz}\ \emph {et~al.}(2018)\citenamefont {Baltz},
  \citenamefont {Manchon}, \citenamefont {Tsoi}, \citenamefont {Moriyama},
  \citenamefont {Ono},\ and\ \citenamefont
  {Tserkovnyak}}]{baltz2018antiferromagnetic}%
  \BibitemOpen
  \bibfield  {author} {\bibinfo {author} {\bibfnamefont {V.}~\bibnamefont
  {Baltz}}, \bibinfo {author} {\bibfnamefont {A.}~\bibnamefont {Manchon}},
  \bibinfo {author} {\bibfnamefont {M.}~\bibnamefont {Tsoi}}, \bibinfo {author}
  {\bibfnamefont {T.}~\bibnamefont {Moriyama}}, \bibinfo {author}
  {\bibfnamefont {T.}~\bibnamefont {Ono}},\ and\ \bibinfo {author}
  {\bibfnamefont {Y.}~\bibnamefont {Tserkovnyak}},\ }\bibfield  {title}
  {\bibinfo {title} {{Antiferromagnetic spintronics}},\ }\href
  {https://doi.org/10.1103/RevModPhys.90.015005} {\bibfield  {journal}
  {\bibinfo  {journal} {Rev. Mod. Phys.}\ }\textbf {\bibinfo {volume} {90}},\
  \bibinfo {pages} {015005} (\bibinfo {year} {2018})}\BibitemShut {NoStop}%
\bibitem [{\citenamefont {Bhowal}\ and\ \citenamefont
  {Spaldin}(2024)}]{bhowal2024ferroically}%
  \BibitemOpen
  \bibfield  {author} {\bibinfo {author} {\bibfnamefont {S.}~\bibnamefont
  {Bhowal}}\ and\ \bibinfo {author} {\bibfnamefont {N.~A.}\ \bibnamefont
  {Spaldin}},\ }\bibfield  {title} {\bibinfo {title} {{Ferroically Ordered
  Magnetic Octupoles in $d$-Wave Altermagnets}},\ }\href
  {https://doi.org/10.1103/PhysRevX.14.011019} {\bibfield  {journal} {\bibinfo
  {journal} {Phys. Rev. X}\ }\textbf {\bibinfo {volume} {14}},\ \bibinfo
  {pages} {011019} (\bibinfo {year} {2024})}\BibitemShut {NoStop}%
\bibitem [{\citenamefont {McClarty}\ and\ \citenamefont
  {Rau}(2024)}]{mcclarty2024landau}%
  \BibitemOpen
  \bibfield  {author} {\bibinfo {author} {\bibfnamefont {P.~A.}\ \bibnamefont
  {McClarty}}\ and\ \bibinfo {author} {\bibfnamefont {J.~G.}\ \bibnamefont
  {Rau}},\ }\bibfield  {title} {\bibinfo {title} {Landau theory of
  altermagnetism},\ }\href {https://doi.org/10.1103/PhysRevLett.132.176702}
  {\bibfield  {journal} {\bibinfo  {journal} {Phys. Rev. Lett.}\ }\textbf
  {\bibinfo {volume} {132}},\ \bibinfo {pages} {176702} (\bibinfo {year}
  {2024})}\BibitemShut {NoStop}%
\bibitem [{\citenamefont {Autieri}\ \emph {et~al.}(2023)\citenamefont
  {Autieri}, \citenamefont {Sattigeri}, \citenamefont {Cuono},\ and\
  \citenamefont {Fakhredine}}]{autieri2023dzyaloshinskii}%
  \BibitemOpen
  \bibfield  {author} {\bibinfo {author} {\bibfnamefont {C.}~\bibnamefont
  {Autieri}}, \bibinfo {author} {\bibfnamefont {R.~M.}\ \bibnamefont
  {Sattigeri}}, \bibinfo {author} {\bibfnamefont {G.}~\bibnamefont {Cuono}},\
  and\ \bibinfo {author} {\bibfnamefont {A.}~\bibnamefont {Fakhredine}},\
  }\href@noop {} {\bibinfo {title} {{Dzyaloshinskii-Moriya interaction inducing
  weak ferromagnetism in centrosymmetric altermagnets and weak ferrimagnetism
  in noncentrosymmetric altermagnets}}} (\bibinfo {year} {2023}),\ \Eprint
  {https://arxiv.org/abs/2312.07678} {arXiv:2312.07678 [cond-mat.mtrl-sci]}
  \BibitemShut {NoStop}%
\bibitem [{\citenamefont {Fernandes}\ \emph {et~al.}(2024)\citenamefont
  {Fernandes}, \citenamefont {de~Carvalho}, \citenamefont {Birol},\ and\
  \citenamefont {Pereira}}]{fernandes2024topological}%
  \BibitemOpen
  \bibfield  {author} {\bibinfo {author} {\bibfnamefont {R.~M.}\ \bibnamefont
  {Fernandes}}, \bibinfo {author} {\bibfnamefont {V.~S.}\ \bibnamefont
  {de~Carvalho}}, \bibinfo {author} {\bibfnamefont {T.}~\bibnamefont {Birol}},\
  and\ \bibinfo {author} {\bibfnamefont {R.~G.}\ \bibnamefont {Pereira}},\
  }\bibfield  {title} {\bibinfo {title} {{Topological transition from nodal to
  nodeless Zeeman splitting in altermagnets}},\ }\href
  {https://doi.org/10.1103/PhysRevB.109.024404} {\bibfield  {journal} {\bibinfo
   {journal} {Phys. Rev. B}\ }\textbf {\bibinfo {volume} {109}},\ \bibinfo
  {pages} {024404} (\bibinfo {year} {2024})}\BibitemShut {NoStop}%
\bibitem [{\citenamefont {Antonenko}\ \emph {et~al.}(2024)\citenamefont
  {Antonenko}, \citenamefont {Fernandes},\ and\ \citenamefont
  {Venderbos}}]{antonenko2024mirror}%
  \BibitemOpen
  \bibfield  {author} {\bibinfo {author} {\bibfnamefont {D.~S.}\ \bibnamefont
  {Antonenko}}, \bibinfo {author} {\bibfnamefont {R.~M.}\ \bibnamefont
  {Fernandes}},\ and\ \bibinfo {author} {\bibfnamefont {J.~W.~F.}\ \bibnamefont
  {Venderbos}},\ }\href@noop {} {\bibinfo {title} {{Mirror Chern Bands and Weyl
  Nodal Loops in Altermagnets}}} (\bibinfo {year} {2024}),\ \Eprint
  {https://arxiv.org/abs/2402.10201} {arXiv:2402.10201 [cond-mat.mes-hall]}
  \BibitemShut {NoStop}%
\bibitem [{\citenamefont {Milivojević}\ \emph {et~al.}(2024)\citenamefont
  {Milivojević}, \citenamefont {Orozović}, \citenamefont {Picozzi},
  \citenamefont {Gmitra},\ and\ \citenamefont
  {Stavrić}}]{milivojevic2024interplay}%
  \BibitemOpen
  \bibfield  {author} {\bibinfo {author} {\bibfnamefont {M.}~\bibnamefont
  {Milivojević}}, \bibinfo {author} {\bibfnamefont {M.}~\bibnamefont
  {Orozović}}, \bibinfo {author} {\bibfnamefont {S.}~\bibnamefont {Picozzi}},
  \bibinfo {author} {\bibfnamefont {M.}~\bibnamefont {Gmitra}},\ and\ \bibinfo
  {author} {\bibfnamefont {S.}~\bibnamefont {Stavrić}},\ }\bibfield  {title}
  {\bibinfo {title} {{Interplay of altermagnetism and weak ferromagnetism in
  two-dimensional RuF4}},\ }\href {https://doi.org/10.1088/2053-1583/ad4c73}
  {\bibfield  {journal} {\bibinfo  {journal} {2D Materials}\ }\textbf {\bibinfo
  {volume} {11}},\ \bibinfo {pages} {035025} (\bibinfo {year}
  {2024})}\BibitemShut {NoStop}%
\bibitem [{\citenamefont {Cheong}\ and\ \citenamefont
  {Huang}(2024)}]{cheong2024altermagnetism}%
  \BibitemOpen
  \bibfield  {author} {\bibinfo {author} {\bibfnamefont {S.-W.}\ \bibnamefont
  {Cheong}}\ and\ \bibinfo {author} {\bibfnamefont {F.-T.}\ \bibnamefont
  {Huang}},\ }\bibfield  {title} {\bibinfo {title} {{Altermagnetism with
  non-collinear spins}},\ }\href {https://doi.org/10.1038/s41535-024-00626-6}
  {\bibfield  {journal} {\bibinfo  {journal} {npj Quantum Materials}\ }\textbf
  {\bibinfo {volume} {9}},\ \bibinfo {pages} {13} (\bibinfo {year}
  {2024})}\BibitemShut {NoStop}%
\bibitem [{\citenamefont {Kluczyk}\ \emph {et~al.}(2024)\citenamefont
  {Kluczyk}, \citenamefont {Gas}, \citenamefont {Grzybowski}, \citenamefont
  {Skupi\ifmmode~\acute{n}\else \'{n}\fi{}ski}, \citenamefont {Borysiewicz},
  \citenamefont {F\k{a}s}, \citenamefont {Suffczy\ifmmode~\acute{n}\else
  \'{n}\fi{}ski}, \citenamefont {Domagala}, \citenamefont {Grasza},
  \citenamefont {Mycielski}, \citenamefont {Baj}, \citenamefont {Ahn},
  \citenamefont {V\'yborn\'y}, \citenamefont {Sawicki},\ and\ \citenamefont
  {Gryglas-Borysiewicz}}]{kluczyk2024coexistence}%
  \BibitemOpen
  \bibfield  {author} {\bibinfo {author} {\bibfnamefont {K.~P.}\ \bibnamefont
  {Kluczyk}}, \bibinfo {author} {\bibfnamefont {K.}~\bibnamefont {Gas}},
  \bibinfo {author} {\bibfnamefont {M.~J.}\ \bibnamefont {Grzybowski}},
  \bibinfo {author} {\bibfnamefont {P.}~\bibnamefont
  {Skupi\ifmmode~\acute{n}\else \'{n}\fi{}ski}}, \bibinfo {author}
  {\bibfnamefont {M.~A.}\ \bibnamefont {Borysiewicz}}, \bibinfo {author}
  {\bibfnamefont {T.}~\bibnamefont {F\k{a}s}}, \bibinfo {author} {\bibfnamefont
  {J.}~\bibnamefont {Suffczy\ifmmode~\acute{n}\else \'{n}\fi{}ski}}, \bibinfo
  {author} {\bibfnamefont {J.~Z.}\ \bibnamefont {Domagala}}, \bibinfo {author}
  {\bibfnamefont {K.}~\bibnamefont {Grasza}}, \bibinfo {author} {\bibfnamefont
  {A.}~\bibnamefont {Mycielski}}, \bibinfo {author} {\bibfnamefont
  {M.}~\bibnamefont {Baj}}, \bibinfo {author} {\bibfnamefont {K.~H.}\
  \bibnamefont {Ahn}}, \bibinfo {author} {\bibfnamefont {K.}~\bibnamefont
  {V\'yborn\'y}}, \bibinfo {author} {\bibfnamefont {M.}~\bibnamefont
  {Sawicki}},\ and\ \bibinfo {author} {\bibfnamefont {M.}~\bibnamefont
  {Gryglas-Borysiewicz}},\ }\bibfield  {title} {\bibinfo {title} {{Coexistence
  of anomalous Hall effect and weak magnetization in a nominally collinear
  antiferromagnet MnTe}},\ }\href {https://doi.org/10.1103/PhysRevB.110.155201}
  {\bibfield  {journal} {\bibinfo  {journal} {Phys. Rev. B}\ }\textbf {\bibinfo
  {volume} {110}},\ \bibinfo {pages} {155201} (\bibinfo {year}
  {2024})}\BibitemShut {NoStop}%
\bibitem [{\citenamefont {Cheong}\ and\ \citenamefont
  {Huang}(2025)}]{cheong2025altermagnetism}%
  \BibitemOpen
  \bibfield  {author} {\bibinfo {author} {\bibfnamefont {S.-W.}\ \bibnamefont
  {Cheong}}\ and\ \bibinfo {author} {\bibfnamefont {F.-T.}\ \bibnamefont
  {Huang}},\ }\href@noop {} {\bibinfo {title} {{Altermagnetism
  Classification}}} (\bibinfo {year} {2025}),\ \Eprint
  {https://arxiv.org/abs/2409.20456} {arXiv:2409.20456 [cond-mat.mtrl-sci]}
  \BibitemShut {NoStop}%
\bibitem [{\citenamefont
  {Dzyaloshinsky}(1958)}]{dzyaloshinsky1958thermodynamic}%
  \BibitemOpen
  \bibfield  {author} {\bibinfo {author} {\bibfnamefont {I.}~\bibnamefont
  {Dzyaloshinsky}},\ }\bibfield  {title} {\bibinfo {title} {{A thermodynamic
  theory of “weak” ferromagnetism of antiferromagnetics}},\ }\href
  {https://doi.org/https://doi.org/10.1016/0022-3697(58)90076-3} {\bibfield
  {journal} {\bibinfo  {journal} {Journal of Physics and Chemistry of Solids}\
  }\textbf {\bibinfo {volume} {4}},\ \bibinfo {pages} {241} (\bibinfo {year}
  {1958})}\BibitemShut {NoStop}%
\bibitem [{\citenamefont {Moriya}(1960{\natexlab{a}})}]{moriya1960anisotropic}%
  \BibitemOpen
  \bibfield  {author} {\bibinfo {author} {\bibfnamefont {T.}~\bibnamefont
  {Moriya}},\ }\bibfield  {title} {\bibinfo {title} {{Anisotropic Superexchange
  Interaction and Weak Ferromagnetism}},\ }\href
  {https://doi.org/10.1103/PhysRev.120.91} {\bibfield  {journal} {\bibinfo
  {journal} {Phys. Rev.}\ }\textbf {\bibinfo {volume} {120}},\ \bibinfo {pages}
  {91} (\bibinfo {year} {1960}{\natexlab{a}})}\BibitemShut {NoStop}%
\bibitem [{\citenamefont {Cheong}\ \emph {et~al.}(1989)\citenamefont {Cheong},
  \citenamefont {Thompson},\ and\ \citenamefont
  {Fisk}}]{cheong1989metamagnetism}%
  \BibitemOpen
  \bibfield  {author} {\bibinfo {author} {\bibfnamefont {S.-W.}\ \bibnamefont
  {Cheong}}, \bibinfo {author} {\bibfnamefont {J.~D.}\ \bibnamefont
  {Thompson}},\ and\ \bibinfo {author} {\bibfnamefont {Z.}~\bibnamefont
  {Fisk}},\ }\bibfield  {title} {\bibinfo {title} {{Metamagnetism in
  ${\mathrm{La}}_{2}$${\mathrm{CuO}}_{4}$}},\ }\href
  {https://doi.org/10.1103/PhysRevB.39.4395} {\bibfield  {journal} {\bibinfo
  {journal} {Phys. Rev. B}\ }\textbf {\bibinfo {volume} {39}},\ \bibinfo
  {pages} {4395} (\bibinfo {year} {1989})}\BibitemShut {NoStop}%
\bibitem [{\citenamefont {Sandratskii}\ and\ \citenamefont
  {K\"ubler}(1996)}]{sandratskii1996role}%
  \BibitemOpen
  \bibfield  {author} {\bibinfo {author} {\bibfnamefont {L.~M.}\ \bibnamefont
  {Sandratskii}}\ and\ \bibinfo {author} {\bibfnamefont {J.}~\bibnamefont
  {K\"ubler}},\ }\bibfield  {title} {\bibinfo {title} {{Role of Orbital
  Polarization in Weak Ferromagnetism}},\ }\href
  {https://doi.org/10.1103/PhysRevLett.76.4963} {\bibfield  {journal} {\bibinfo
   {journal} {Phys. Rev. Lett.}\ }\textbf {\bibinfo {volume} {76}},\ \bibinfo
  {pages} {4963} (\bibinfo {year} {1996})}\BibitemShut {NoStop}%
\bibitem [{\citenamefont {Kübler}\ and\ \citenamefont
  {Felser}(2014)}]{kubler2014noncollinear}%
  \BibitemOpen
  \bibfield  {author} {\bibinfo {author} {\bibfnamefont {J.}~\bibnamefont
  {Kübler}}\ and\ \bibinfo {author} {\bibfnamefont {C.}~\bibnamefont
  {Felser}},\ }\bibfield  {title} {\bibinfo {title} {{Non-collinear
  antiferromagnets and the anomalous Hall effect}},\ }\href
  {https://doi.org/10.1209/0295-5075/108/67001} {\bibfield  {journal} {\bibinfo
   {journal} {Europhysics Letters}\ }\textbf {\bibinfo {volume} {108}},\
  \bibinfo {pages} {67001} (\bibinfo {year} {2014})}\BibitemShut {NoStop}%
\bibitem [{\citenamefont {Chen}\ \emph {et~al.}(2020)\citenamefont {Chen},
  \citenamefont {Wang}, \citenamefont {Xiao}, \citenamefont {Guo},
  \citenamefont {Niu},\ and\ \citenamefont {MacDonald}}]{chen2020manipulating}%
  \BibitemOpen
  \bibfield  {author} {\bibinfo {author} {\bibfnamefont {H.}~\bibnamefont
  {Chen}}, \bibinfo {author} {\bibfnamefont {T.-C.}\ \bibnamefont {Wang}},
  \bibinfo {author} {\bibfnamefont {D.}~\bibnamefont {Xiao}}, \bibinfo {author}
  {\bibfnamefont {G.-Y.}\ \bibnamefont {Guo}}, \bibinfo {author} {\bibfnamefont
  {Q.}~\bibnamefont {Niu}},\ and\ \bibinfo {author} {\bibfnamefont {A.~H.}\
  \bibnamefont {MacDonald}},\ }\bibfield  {title} {\bibinfo {title}
  {{Manipulating anomalous Hall antiferromagnets with magnetic fields}},\
  }\href {https://doi.org/10.1103/PhysRevB.101.104418} {\bibfield  {journal}
  {\bibinfo  {journal} {Phys. Rev. B}\ }\textbf {\bibinfo {volume} {101}},\
  \bibinfo {pages} {104418} (\bibinfo {year} {2020})}\BibitemShut {NoStop}%
\bibitem [{\citenamefont {Moriya}(1960{\natexlab{b}})}]{moriya1960theory}%
  \BibitemOpen
  \bibfield  {author} {\bibinfo {author} {\bibfnamefont {T.}~\bibnamefont
  {Moriya}},\ }\bibfield  {title} {\bibinfo {title} {{Theory of Magnetism of
  Ni${\mathrm{F}}_{2}$}},\ }\href {https://doi.org/10.1103/PhysRev.117.635}
  {\bibfield  {journal} {\bibinfo  {journal} {Phys. Rev.}\ }\textbf {\bibinfo
  {volume} {117}},\ \bibinfo {pages} {635} (\bibinfo {year}
  {1960}{\natexlab{b}})}\BibitemShut {NoStop}%
\bibitem [{\citenamefont {Tomiyoshi}\ and\ \citenamefont
  {Yamaguchi}(1982)}]{tomiyoshi1982magnetic}%
  \BibitemOpen
  \bibfield  {author} {\bibinfo {author} {\bibfnamefont {S.}~\bibnamefont
  {Tomiyoshi}}\ and\ \bibinfo {author} {\bibfnamefont {Y.}~\bibnamefont
  {Yamaguchi}},\ }\bibfield  {title} {\bibinfo {title} {{Magnetic Structure and
  Weak Ferromagnetism of Mn$_3$Sn Studied by Polarized Neutron Diffraction}},\
  }\href {https://doi.org/10.1143/JPSJ.51.2478} {\bibfield  {journal} {\bibinfo
   {journal} {Journal of the Physical Society of Japan}\ }\textbf {\bibinfo
  {volume} {51}},\ \bibinfo {pages} {2478} (\bibinfo {year}
  {1982})}\BibitemShut {NoStop}%
\bibitem [{\citenamefont {Nagamiya}\ \emph {et~al.}(1982)\citenamefont
  {Nagamiya}, \citenamefont {Tomiyoshi},\ and\ \citenamefont
  {Yamaguchi}}]{nagamiya1982triangular}%
  \BibitemOpen
  \bibfield  {author} {\bibinfo {author} {\bibfnamefont {T.}~\bibnamefont
  {Nagamiya}}, \bibinfo {author} {\bibfnamefont {S.}~\bibnamefont
  {Tomiyoshi}},\ and\ \bibinfo {author} {\bibfnamefont {Y.}~\bibnamefont
  {Yamaguchi}},\ }\bibfield  {title} {\bibinfo {title} {{Triangular spin
  configuration and weak ferromagnetism of Mn$_3$Sn and Mn$_3$Ge}},\ }\href
  {https://doi.org/https://doi.org/10.1016/0038-1098(82)90159-4} {\bibfield
  {journal} {\bibinfo  {journal} {Solid State Communications}\ }\textbf
  {\bibinfo {volume} {42}},\ \bibinfo {pages} {385} (\bibinfo {year}
  {1982})}\BibitemShut {NoStop}%
\bibitem [{\citenamefont {Ederer}\ and\ \citenamefont
  {Spaldin}(2005)}]{ederer2005weak}%
  \BibitemOpen
  \bibfield  {author} {\bibinfo {author} {\bibfnamefont {C.}~\bibnamefont
  {Ederer}}\ and\ \bibinfo {author} {\bibfnamefont {N.~A.}\ \bibnamefont
  {Spaldin}},\ }\bibfield  {title} {\bibinfo {title} {{Weak ferromagnetism and
  magnetoelectric coupling in bismuth ferrite}},\ }\href
  {https://doi.org/10.1103/PhysRevB.71.060401} {\bibfield  {journal} {\bibinfo
  {journal} {Phys. Rev. B}\ }\textbf {\bibinfo {volume} {71}},\ \bibinfo
  {pages} {060401(R)} (\bibinfo {year} {2005})}\BibitemShut {NoStop}%
\bibitem [{\citenamefont {Kittel}(1949)}]{kittel1949on}%
  \BibitemOpen
  \bibfield  {author} {\bibinfo {author} {\bibfnamefont {C.}~\bibnamefont
  {Kittel}},\ }\bibfield  {title} {\bibinfo {title} {{On the Gyromagnetic Ratio
  and Spectroscopic Splitting Factor of Ferromagnetic Substances}},\ }\href
  {https://doi.org/10.1103/PhysRev.76.743} {\bibfield  {journal} {\bibinfo
  {journal} {Phys. Rev.}\ }\textbf {\bibinfo {volume} {76}},\ \bibinfo {pages}
  {743} (\bibinfo {year} {1949})}\BibitemShut {NoStop}%
\bibitem [{\citenamefont {Bruno}(1989)}]{bruno1989tight}%
  \BibitemOpen
  \bibfield  {author} {\bibinfo {author} {\bibfnamefont {P.}~\bibnamefont
  {Bruno}},\ }\bibfield  {title} {\bibinfo {title} {{Tight-binding approach to
  the orbital magnetic moment and magnetocrystalline anisotropy of
  transition-metal monolayers}},\ }\href
  {https://doi.org/10.1103/PhysRevB.39.865} {\bibfield  {journal} {\bibinfo
  {journal} {Phys. Rev. B}\ }\textbf {\bibinfo {volume} {39}},\ \bibinfo
  {pages} {865} (\bibinfo {year} {1989})}\BibitemShut {NoStop}%
\bibitem [{\citenamefont {Alahmed}\ \emph {et~al.}(2022)\citenamefont
  {Alahmed}, \citenamefont {Zhang}, \citenamefont {Wen}, \citenamefont {Xiong},
  \citenamefont {Li}, \citenamefont {Zhang}, \citenamefont {Lux}, \citenamefont
  {Freimuth}, \citenamefont {Mahdi}, \citenamefont {Mokrousov}, \citenamefont
  {Novosad}, \citenamefont {Kwok}, \citenamefont {Yu}, \citenamefont {Zhang},
  \citenamefont {Lee},\ and\ \citenamefont {Li}}]{alahmed2022evidence}%
  \BibitemOpen
  \bibfield  {author} {\bibinfo {author} {\bibfnamefont {L.}~\bibnamefont
  {Alahmed}}, \bibinfo {author} {\bibfnamefont {X.}~\bibnamefont {Zhang}},
  \bibinfo {author} {\bibfnamefont {J.}~\bibnamefont {Wen}}, \bibinfo {author}
  {\bibfnamefont {Y.}~\bibnamefont {Xiong}}, \bibinfo {author} {\bibfnamefont
  {Y.}~\bibnamefont {Li}}, \bibinfo {author} {\bibfnamefont {L.-c.}\
  \bibnamefont {Zhang}}, \bibinfo {author} {\bibfnamefont {F.}~\bibnamefont
  {Lux}}, \bibinfo {author} {\bibfnamefont {F.}~\bibnamefont {Freimuth}},
  \bibinfo {author} {\bibfnamefont {M.}~\bibnamefont {Mahdi}}, \bibinfo
  {author} {\bibfnamefont {Y.}~\bibnamefont {Mokrousov}}, \bibinfo {author}
  {\bibfnamefont {V.}~\bibnamefont {Novosad}}, \bibinfo {author} {\bibfnamefont
  {W.-K.}\ \bibnamefont {Kwok}}, \bibinfo {author} {\bibfnamefont
  {D.}~\bibnamefont {Yu}}, \bibinfo {author} {\bibfnamefont {W.}~\bibnamefont
  {Zhang}}, \bibinfo {author} {\bibfnamefont {Y.~S.}\ \bibnamefont {Lee}},\
  and\ \bibinfo {author} {\bibfnamefont {P.}~\bibnamefont {Li}},\ }\bibfield
  {title} {\bibinfo {title} {{Evidence of Magnon-Mediated Orbital Magnetism in
  a Quasi-2D Topological Magnon Insulator}},\ }\href
  {https://doi.org/10.1021/acs.nanolett.2c00562} {\bibfield  {journal}
  {\bibinfo  {journal} {Nano Letters}\ }\textbf {\bibinfo {volume} {22}},\
  \bibinfo {pages} {5114} (\bibinfo {year} {2022})}\BibitemShut {NoStop}%
\bibitem [{\citenamefont {Ito}\ and\ \citenamefont
  {Nomura}(2017)}]{ito2017anomalous}%
  \BibitemOpen
  \bibfield  {author} {\bibinfo {author} {\bibfnamefont {N.}~\bibnamefont
  {Ito}}\ and\ \bibinfo {author} {\bibfnamefont {K.}~\bibnamefont {Nomura}},\
  }\bibfield  {title} {\bibinfo {title} {{Anomalous Hall Effect and Spontaneous
  Orbital Magnetization in Antiferromagnetic Weyl Metal}},\ }\href
  {https://doi.org/10.7566/JPSJ.86.063703} {\bibfield  {journal} {\bibinfo
  {journal} {Journal of the Physical Society of Japan}\ }\textbf {\bibinfo
  {volume} {86}},\ \bibinfo {pages} {063703} (\bibinfo {year}
  {2017})}\BibitemShut {NoStop}%
\bibitem [{\citenamefont {Moriya}(1959)}]{moriya1959piezomagnetism}%
  \BibitemOpen
  \bibfield  {author} {\bibinfo {author} {\bibfnamefont {T.}~\bibnamefont
  {Moriya}},\ }\bibfield  {title} {\bibinfo {title} {{Piezomagnetism in
  CoF2}},\ }\href
  {https://doi.org/https://doi.org/10.1016/0022-3697(59)90043-5} {\bibfield
  {journal} {\bibinfo  {journal} {Journal of Physics and Chemistry of Solids}\
  }\textbf {\bibinfo {volume} {11}},\ \bibinfo {pages} {73} (\bibinfo {year}
  {1959})}\BibitemShut {NoStop}%
\bibitem [{\citenamefont {Disa}\ \emph {et~al.}(2020)\citenamefont {Disa},
  \citenamefont {Fechner}, \citenamefont {Nova}, \citenamefont {Liu},
  \citenamefont {F{\"o}rst}, \citenamefont {Prabhakaran}, \citenamefont
  {Radaelli},\ and\ \citenamefont {Cavalleri}}]{disa2000polarizing}%
  \BibitemOpen
  \bibfield  {author} {\bibinfo {author} {\bibfnamefont {A.~S.}\ \bibnamefont
  {Disa}}, \bibinfo {author} {\bibfnamefont {M.}~\bibnamefont {Fechner}},
  \bibinfo {author} {\bibfnamefont {T.~F.}\ \bibnamefont {Nova}}, \bibinfo
  {author} {\bibfnamefont {B.}~\bibnamefont {Liu}}, \bibinfo {author}
  {\bibfnamefont {M.}~\bibnamefont {F{\"o}rst}}, \bibinfo {author}
  {\bibfnamefont {D.}~\bibnamefont {Prabhakaran}}, \bibinfo {author}
  {\bibfnamefont {P.~G.}\ \bibnamefont {Radaelli}},\ and\ \bibinfo {author}
  {\bibfnamefont {A.}~\bibnamefont {Cavalleri}},\ }\bibfield  {title} {\bibinfo
  {title} {{Polarizing an antiferromagnet by optical engineering of the crystal
  field}},\ }\href {https://doi.org/10.1038/s41567-020-0936-3} {\bibfield
  {journal} {\bibinfo  {journal} {Nature Physics}\ }\textbf {\bibinfo {volume}
  {16}},\ \bibinfo {pages} {937} (\bibinfo {year} {2020})}\BibitemShut
  {NoStop}%
\bibitem [{sup()}]{supp}%
  \BibitemOpen
  \href@noop {} {}\bibinfo {note} {See Supplemental Material for details of the
  tight-binding models and the first-principle calculation, which includes
  Refs.~\cite{slater1954simplified, fleurWeb, fleurCode, wimmer1981full,
  perdew1981self, li1990magnetic, moreira2000ab, monkhorst1976special,
  pizzi2020wannier90}}\BibitemShut {NoStop}%
\bibitem [{\citenamefont {Thoma}\ \emph {et~al.}(2021)\citenamefont {Thoma},
  \citenamefont {Hutanu}, \citenamefont {Deng}, \citenamefont {Dmitrienko},
  \citenamefont {Brown}, \citenamefont {Gukasov}, \citenamefont {Roth},\ and\
  \citenamefont {Angst}}]{thoma2021revealing}%
  \BibitemOpen
  \bibfield  {author} {\bibinfo {author} {\bibfnamefont {H.}~\bibnamefont
  {Thoma}}, \bibinfo {author} {\bibfnamefont {V.}~\bibnamefont {Hutanu}},
  \bibinfo {author} {\bibfnamefont {H.}~\bibnamefont {Deng}}, \bibinfo {author}
  {\bibfnamefont {V.~E.}\ \bibnamefont {Dmitrienko}}, \bibinfo {author}
  {\bibfnamefont {P.~J.}\ \bibnamefont {Brown}}, \bibinfo {author}
  {\bibfnamefont {A.}~\bibnamefont {Gukasov}}, \bibinfo {author} {\bibfnamefont
  {G.}~\bibnamefont {Roth}},\ and\ \bibinfo {author} {\bibfnamefont
  {M.}~\bibnamefont {Angst}},\ }\bibfield  {title} {\bibinfo {title}
  {{Revealing the Absolute Direction of the Dzyaloshinskii-Moriya Interaction
  in Prototypical Weak Ferromagnets by Polarized Neutrons}},\ }\href
  {https://doi.org/10.1103/PhysRevX.11.011060} {\bibfield  {journal} {\bibinfo
  {journal} {Phys. Rev. X}\ }\textbf {\bibinfo {volume} {11}},\ \bibinfo
  {pages} {011060} (\bibinfo {year} {2021})}\BibitemShut {NoStop}%
\bibitem [{\citenamefont {Ikhlas}\ \emph {et~al.}(2022)\citenamefont {Ikhlas},
  \citenamefont {Dasgupta}, \citenamefont {Theuss}, \citenamefont {Higo},
  \citenamefont {Kittaka}, \citenamefont {Ramshaw}, \citenamefont
  {Tchernyshyov}, \citenamefont {Hicks},\ and\ \citenamefont
  {Nakatsuji}}]{ikhlas2022piezomagnetic}%
  \BibitemOpen
  \bibfield  {author} {\bibinfo {author} {\bibfnamefont {M.}~\bibnamefont
  {Ikhlas}}, \bibinfo {author} {\bibfnamefont {S.}~\bibnamefont {Dasgupta}},
  \bibinfo {author} {\bibfnamefont {F.}~\bibnamefont {Theuss}}, \bibinfo
  {author} {\bibfnamefont {T.}~\bibnamefont {Higo}}, \bibinfo {author}
  {\bibfnamefont {S.}~\bibnamefont {Kittaka}}, \bibinfo {author} {\bibfnamefont
  {B.~J.}\ \bibnamefont {Ramshaw}}, \bibinfo {author} {\bibfnamefont
  {O.}~\bibnamefont {Tchernyshyov}}, \bibinfo {author} {\bibfnamefont {C.~W.}\
  \bibnamefont {Hicks}},\ and\ \bibinfo {author} {\bibfnamefont
  {S.}~\bibnamefont {Nakatsuji}},\ }\bibfield  {title} {\bibinfo {title}
  {{Piezomagnetic switching of the anomalous Hall effect in an antiferromagnet
  at room temperature}},\ }\href {https://doi.org/10.1038/s41567-022-01645-5}
  {\bibfield  {journal} {\bibinfo  {journal} {Nature Physics}\ }\textbf
  {\bibinfo {volume} {18}},\ \bibinfo {pages} {1086} (\bibinfo {year}
  {2022})}\BibitemShut {NoStop}%
\bibitem [{\citenamefont {Harrison}(1989)}]{harrison2012electronic}%
  \BibitemOpen
  \bibfield  {author} {\bibinfo {author} {\bibfnamefont {W.~A.}\ \bibnamefont
  {Harrison}},\ }\href@noop {} {\emph {\bibinfo {title} {{Electronic structure
  and the properties of solids}}}}\ (\bibinfo  {publisher} {Dover},\ \bibinfo
  {address} {New York},\ \bibinfo {year} {1989})\BibitemShut {NoStop}%
\bibitem [{\citenamefont {Chen}\ \emph {et~al.}(2014)\citenamefont {Chen},
  \citenamefont {Niu},\ and\ \citenamefont {MacDonald}}]{chen2014anomalous}%
  \BibitemOpen
  \bibfield  {author} {\bibinfo {author} {\bibfnamefont {H.}~\bibnamefont
  {Chen}}, \bibinfo {author} {\bibfnamefont {Q.}~\bibnamefont {Niu}},\ and\
  \bibinfo {author} {\bibfnamefont {A.~H.}\ \bibnamefont {MacDonald}},\
  }\bibfield  {title} {\bibinfo {title} {{Anomalous Hall Effect Arising from
  Noncollinear Antiferromagnetism}},\ }\href
  {https://doi.org/10.1103/PhysRevLett.112.017205} {\bibfield  {journal}
  {\bibinfo  {journal} {Phys. Rev. Lett.}\ }\textbf {\bibinfo {volume} {112}},\
  \bibinfo {pages} {017205} (\bibinfo {year} {2014})}\BibitemShut {NoStop}%
\bibitem [{\citenamefont {Zhang}\ \emph {et~al.}(2017)\citenamefont {Zhang},
  \citenamefont {Sun}, \citenamefont {Yang}, \citenamefont
  {\ifmmode~\check{Z}\else \v{Z}\fi{}elezn\'y}, \citenamefont {Parkin},
  \citenamefont {Felser},\ and\ \citenamefont {Yan}}]{zhang2017strong}%
  \BibitemOpen
  \bibfield  {author} {\bibinfo {author} {\bibfnamefont {Y.}~\bibnamefont
  {Zhang}}, \bibinfo {author} {\bibfnamefont {Y.}~\bibnamefont {Sun}}, \bibinfo
  {author} {\bibfnamefont {H.}~\bibnamefont {Yang}}, \bibinfo {author}
  {\bibfnamefont {J.}~\bibnamefont {\ifmmode~\check{Z}\else
  \v{Z}\fi{}elezn\'y}}, \bibinfo {author} {\bibfnamefont {S.~P.~P.}\
  \bibnamefont {Parkin}}, \bibinfo {author} {\bibfnamefont {C.}~\bibnamefont
  {Felser}},\ and\ \bibinfo {author} {\bibfnamefont {B.}~\bibnamefont {Yan}},\
  }\bibfield  {title} {\bibinfo {title} {{Strong anisotropic anomalous Hall
  effect and spin Hall effect in the chiral antiferromagnetic compounds
  ${\mathrm{Mn}}_{3}X$ ($X=\mathrm{Ge}$, Sn, Ga, Ir, Rh, and Pt)}},\ }\href
  {https://doi.org/10.1103/PhysRevB.95.075128} {\bibfield  {journal} {\bibinfo
  {journal} {Phys. Rev. B}\ }\textbf {\bibinfo {volume} {95}},\ \bibinfo
  {pages} {075128} (\bibinfo {year} {2017})}\BibitemShut {NoStop}%
\bibitem [{\citenamefont {Nakatsuji}\ \emph {et~al.}(2015)\citenamefont
  {Nakatsuji}, \citenamefont {Kiyohara},\ and\ \citenamefont
  {Higo}}]{nakatsuji2015large}%
  \BibitemOpen
  \bibfield  {author} {\bibinfo {author} {\bibfnamefont {S.}~\bibnamefont
  {Nakatsuji}}, \bibinfo {author} {\bibfnamefont {N.}~\bibnamefont
  {Kiyohara}},\ and\ \bibinfo {author} {\bibfnamefont {T.}~\bibnamefont
  {Higo}},\ }\bibfield  {title} {\bibinfo {title} {{Large anomalous Hall effect
  in a non-collinear antiferromagnet at room temperature}},\ }\href
  {https://doi.org/10.1038/nature15723} {\bibfield  {journal} {\bibinfo
  {journal} {Nature}\ }\textbf {\bibinfo {volume} {527}},\ \bibinfo {pages}
  {212} (\bibinfo {year} {2015})}\BibitemShut {NoStop}%
\bibitem [{\citenamefont {He}\ \emph {et~al.}(2023)\citenamefont {He},
  \citenamefont {Wang}, \citenamefont {Luo}, \citenamefont {Zeng},
  \citenamefont {Chen},\ and\ \citenamefont {Tang}}]{he2023nonrelativistic}%
  \BibitemOpen
  \bibfield  {author} {\bibinfo {author} {\bibfnamefont {R.}~\bibnamefont
  {He}}, \bibinfo {author} {\bibfnamefont {D.}~\bibnamefont {Wang}}, \bibinfo
  {author} {\bibfnamefont {N.}~\bibnamefont {Luo}}, \bibinfo {author}
  {\bibfnamefont {J.}~\bibnamefont {Zeng}}, \bibinfo {author} {\bibfnamefont
  {K.-Q.}\ \bibnamefont {Chen}},\ and\ \bibinfo {author} {\bibfnamefont
  {L.-M.}\ \bibnamefont {Tang}},\ }\bibfield  {title} {\bibinfo {title}
  {{Nonrelativistic Spin-Momentum Coupling in Antiferromagnetic Twisted
  Bilayers}},\ }\href {https://doi.org/10.1103/PhysRevLett.130.046401}
  {\bibfield  {journal} {\bibinfo  {journal} {Phys. Rev. Lett.}\ }\textbf
  {\bibinfo {volume} {130}},\ \bibinfo {pages} {046401} (\bibinfo {year}
  {2023})}\BibitemShut {NoStop}%
\bibitem [{\citenamefont {Sheoran}\ and\ \citenamefont
  {Bhattacharya}(2024)}]{sheoran2024nonrelativistic}%
  \BibitemOpen
  \bibfield  {author} {\bibinfo {author} {\bibfnamefont {S.}~\bibnamefont
  {Sheoran}}\ and\ \bibinfo {author} {\bibfnamefont {S.}~\bibnamefont
  {Bhattacharya}},\ }\bibfield  {title} {\bibinfo {title} {{Nonrelativistic
  spin splittings and altermagnetism in twisted bilayers of centrosymmetric
  antiferromagnets}},\ }\href
  {https://doi.org/10.1103/PhysRevMaterials.8.L051401} {\bibfield  {journal}
  {\bibinfo  {journal} {Phys. Rev. Mater.}\ }\textbf {\bibinfo {volume} {8}},\
  \bibinfo {pages} {L051401} (\bibinfo {year} {2024})}\BibitemShut {NoStop}%
\bibitem [{\citenamefont {Mazin}\ \emph {et~al.}(2023)\citenamefont {Mazin},
  \citenamefont {González-Hernández},\ and\ \citenamefont
  {Šmejkal}}]{mazin2023induced}%
  \BibitemOpen
  \bibfield  {author} {\bibinfo {author} {\bibfnamefont {I.}~\bibnamefont
  {Mazin}}, \bibinfo {author} {\bibfnamefont {R.}~\bibnamefont
  {González-Hernández}},\ and\ \bibinfo {author} {\bibfnamefont
  {L.}~\bibnamefont {Šmejkal}},\ }\href@noop {} {\bibinfo {title} {{Induced
  Monolayer Altermagnetism in MnP(S,Se)$_3$ and FeSe}}} (\bibinfo {year}
  {2023}),\ \Eprint {https://arxiv.org/abs/2309.02355} {arXiv:2309.02355
  [cond-mat.mes-hall]} \BibitemShut {NoStop}%
\bibitem [{\citenamefont {Zhu}\ \emph {et~al.}(2024{\natexlab{b}})\citenamefont
  {Zhu}, \citenamefont {Chen}, \citenamefont {Li}, \citenamefont {Qiao},
  \citenamefont {Ma}, \citenamefont {Liu}, \citenamefont {Hu}, \citenamefont
  {Gao},\ and\ \citenamefont {Ren}}]{zhu2024multipiezo}%
  \BibitemOpen
  \bibfield  {author} {\bibinfo {author} {\bibfnamefont {Y.}~\bibnamefont
  {Zhu}}, \bibinfo {author} {\bibfnamefont {T.}~\bibnamefont {Chen}}, \bibinfo
  {author} {\bibfnamefont {Y.}~\bibnamefont {Li}}, \bibinfo {author}
  {\bibfnamefont {L.}~\bibnamefont {Qiao}}, \bibinfo {author} {\bibfnamefont
  {X.}~\bibnamefont {Ma}}, \bibinfo {author} {\bibfnamefont {C.}~\bibnamefont
  {Liu}}, \bibinfo {author} {\bibfnamefont {T.}~\bibnamefont {Hu}}, \bibinfo
  {author} {\bibfnamefont {H.}~\bibnamefont {Gao}},\ and\ \bibinfo {author}
  {\bibfnamefont {W.}~\bibnamefont {Ren}},\ }\bibfield  {title} {\bibinfo
  {title} {{Multipiezo Effect in Altermagnetic V$_2$SeTeO Monolayer}},\ }\href
  {https://doi.org/10.1021/acs.nanolett.3c04330} {\bibfield  {journal}
  {\bibinfo  {journal} {Nano Letters}\ }\textbf {\bibinfo {volume} {24}},\
  \bibinfo {pages} {472} (\bibinfo {year} {2024}{\natexlab{b}})}\BibitemShut
  {NoStop}%
\bibitem [{\citenamefont {Jaeschke-Ubiergo}\ \emph {et~al.}(2024)\citenamefont
  {Jaeschke-Ubiergo}, \citenamefont {Bharadwaj}, \citenamefont {Jungwirth},
  \citenamefont {\ifmmode~\check{S}\else \v{S}\fi{}mejkal},\ and\ \citenamefont
  {Sinova}}]{jaeschke2024supercell}%
  \BibitemOpen
  \bibfield  {author} {\bibinfo {author} {\bibfnamefont {R.}~\bibnamefont
  {Jaeschke-Ubiergo}}, \bibinfo {author} {\bibfnamefont {V.~K.}\ \bibnamefont
  {Bharadwaj}}, \bibinfo {author} {\bibfnamefont {T.}~\bibnamefont
  {Jungwirth}}, \bibinfo {author} {\bibfnamefont {L.}~\bibnamefont
  {\ifmmode~\check{S}\else \v{S}\fi{}mejkal}},\ and\ \bibinfo {author}
  {\bibfnamefont {J.}~\bibnamefont {Sinova}},\ }\bibfield  {title} {\bibinfo
  {title} {{Supercell altermagnets}},\ }\href
  {https://doi.org/10.1103/PhysRevB.109.094425} {\bibfield  {journal} {\bibinfo
   {journal} {Phys. Rev. B}\ }\textbf {\bibinfo {volume} {109}},\ \bibinfo
  {pages} {094425} (\bibinfo {year} {2024})}\BibitemShut {NoStop}%
\bibitem [{\citenamefont {Go}\ \emph {et~al.}(2021)\citenamefont {Go},
  \citenamefont {Jo}, \citenamefont {Lee}, \citenamefont {Kläui},\ and\
  \citenamefont {Mokrousov}}]{go2021orbitronics}%
  \BibitemOpen
  \bibfield  {author} {\bibinfo {author} {\bibfnamefont {D.}~\bibnamefont
  {Go}}, \bibinfo {author} {\bibfnamefont {D.}~\bibnamefont {Jo}}, \bibinfo
  {author} {\bibfnamefont {H.-W.}\ \bibnamefont {Lee}}, \bibinfo {author}
  {\bibfnamefont {M.}~\bibnamefont {Kläui}},\ and\ \bibinfo {author}
  {\bibfnamefont {Y.}~\bibnamefont {Mokrousov}},\ }\bibfield  {title} {\bibinfo
  {title} {{Orbitronics: Orbital currents in solids}},\ }\href
  {https://doi.org/10.1209/0295-5075/ac2653} {\bibfield  {journal} {\bibinfo
  {journal} {Europhysics Letters}\ }\textbf {\bibinfo {volume} {135}},\
  \bibinfo {pages} {37001} (\bibinfo {year} {2021})}\BibitemShut {NoStop}%
\bibitem [{\citenamefont {Jo}\ \emph {et~al.}(2024)\citenamefont {Jo},
  \citenamefont {Go}, \citenamefont {Choi},\ and\ \citenamefont
  {Lee}}]{jo2024spintronics}%
  \BibitemOpen
  \bibfield  {author} {\bibinfo {author} {\bibfnamefont {D.}~\bibnamefont
  {Jo}}, \bibinfo {author} {\bibfnamefont {D.}~\bibnamefont {Go}}, \bibinfo
  {author} {\bibfnamefont {G.-M.}\ \bibnamefont {Choi}},\ and\ \bibinfo
  {author} {\bibfnamefont {H.-W.}\ \bibnamefont {Lee}},\ }\bibfield  {title}
  {\bibinfo {title} {{Spintronics meets orbitronics: Emergence of orbital
  angular momentum in solids}},\ }\href
  {https://doi.org/10.1038/s44306-024-00023-6} {\bibfield  {journal} {\bibinfo
  {journal} {npj Spintronics}\ }\textbf {\bibinfo {volume} {2}},\ \bibinfo
  {pages} {19} (\bibinfo {year} {2024})}\BibitemShut {NoStop}%
\bibitem [{\citenamefont {Slater}\ and\ \citenamefont
  {Koster}(1954)}]{slater1954simplified}%
  \BibitemOpen
  \bibfield  {author} {\bibinfo {author} {\bibfnamefont {J.~C.}\ \bibnamefont
  {Slater}}\ and\ \bibinfo {author} {\bibfnamefont {G.~F.}\ \bibnamefont
  {Koster}},\ }\bibfield  {title} {\bibinfo {title} {{Simplified LCAO Method
  for the Periodic Potential Problem}},\ }\href
  {https://doi.org/10.1103/PhysRev.94.1498} {\bibfield  {journal} {\bibinfo
  {journal} {Phys. Rev.}\ }\textbf {\bibinfo {volume} {94}},\ \bibinfo {pages}
  {1498} (\bibinfo {year} {1954})}\BibitemShut {NoStop}%
\bibitem [{fle()}]{fleurWeb}%
  \BibitemOpen
  \href@noop {} {\bibinfo {title} {{The FLEUR project}}},\ \bibinfo
  {howpublished} {\url{https://www.flapw.de/}}\BibitemShut {NoStop}%
\bibitem [{\citenamefont {Wortmann}\ \emph {et~al.}(2023)\citenamefont
  {Wortmann}, \citenamefont {Michalicek}, \citenamefont {Baadji}, \citenamefont
  {Betzinger}, \citenamefont {Bihlmayer}, \citenamefont {Br\"oder},
  \citenamefont {Burnus}, \citenamefont {Enkovaara}, \citenamefont {Freimuth},
  \citenamefont {Friedrich}, \citenamefont {Gerhorst}, \citenamefont
  {Granberg~Cauchi}, \citenamefont {Grytsiuk}, \citenamefont {Hanke},
  \citenamefont {Hanke}, \citenamefont {Heide}, \citenamefont {Heinze},
  \citenamefont {Hilgers}, \citenamefont {Janssen}, \citenamefont
  {Kl\"uppelberg}, \citenamefont {Kovacik}, \citenamefont {Kurz}, \citenamefont
  {Lezaic}, \citenamefont {Madsen}, \citenamefont {Mokrousov}, \citenamefont
  {Neukirchen}, \citenamefont {Redies}, \citenamefont {Rost}, \citenamefont
  {Schlipf}, \citenamefont {Schindlmayr}, \citenamefont {Winkelmann},\ and\
  \citenamefont {Bl\"ugel}}]{fleurCode}%
  \BibitemOpen
  \bibfield  {author} {\bibinfo {author} {\bibfnamefont {D.}~\bibnamefont
  {Wortmann}}, \bibinfo {author} {\bibfnamefont {G.}~\bibnamefont
  {Michalicek}}, \bibinfo {author} {\bibfnamefont {N.}~\bibnamefont {Baadji}},
  \bibinfo {author} {\bibfnamefont {M.}~\bibnamefont {Betzinger}}, \bibinfo
  {author} {\bibfnamefont {G.}~\bibnamefont {Bihlmayer}}, \bibinfo {author}
  {\bibfnamefont {J.}~\bibnamefont {Br\"oder}}, \bibinfo {author}
  {\bibfnamefont {T.}~\bibnamefont {Burnus}}, \bibinfo {author} {\bibfnamefont
  {J.}~\bibnamefont {Enkovaara}}, \bibinfo {author} {\bibfnamefont
  {F.}~\bibnamefont {Freimuth}}, \bibinfo {author} {\bibfnamefont
  {C.}~\bibnamefont {Friedrich}}, \bibinfo {author} {\bibfnamefont {C.-R.}\
  \bibnamefont {Gerhorst}}, \bibinfo {author} {\bibfnamefont {S.}~\bibnamefont
  {Granberg~Cauchi}}, \bibinfo {author} {\bibfnamefont {U.}~\bibnamefont
  {Grytsiuk}}, \bibinfo {author} {\bibfnamefont {A.}~\bibnamefont {Hanke}},
  \bibinfo {author} {\bibfnamefont {J.-P.}\ \bibnamefont {Hanke}}, \bibinfo
  {author} {\bibfnamefont {M.}~\bibnamefont {Heide}}, \bibinfo {author}
  {\bibfnamefont {S.}~\bibnamefont {Heinze}}, \bibinfo {author} {\bibfnamefont
  {R.}~\bibnamefont {Hilgers}}, \bibinfo {author} {\bibfnamefont
  {H.}~\bibnamefont {Janssen}}, \bibinfo {author} {\bibfnamefont {D.~A.}\
  \bibnamefont {Kl\"uppelberg}}, \bibinfo {author} {\bibfnamefont
  {R.}~\bibnamefont {Kovacik}}, \bibinfo {author} {\bibfnamefont
  {P.}~\bibnamefont {Kurz}}, \bibinfo {author} {\bibfnamefont {M.}~\bibnamefont
  {Lezaic}}, \bibinfo {author} {\bibfnamefont {G.~K.~H.}\ \bibnamefont
  {Madsen}}, \bibinfo {author} {\bibfnamefont {Y.}~\bibnamefont {Mokrousov}},
  \bibinfo {author} {\bibfnamefont {A.}~\bibnamefont {Neukirchen}}, \bibinfo
  {author} {\bibfnamefont {M.}~\bibnamefont {Redies}}, \bibinfo {author}
  {\bibfnamefont {S.}~\bibnamefont {Rost}}, \bibinfo {author} {\bibfnamefont
  {M.}~\bibnamefont {Schlipf}}, \bibinfo {author} {\bibfnamefont
  {A.}~\bibnamefont {Schindlmayr}}, \bibinfo {author} {\bibfnamefont
  {M.}~\bibnamefont {Winkelmann}},\ and\ \bibinfo {author} {\bibfnamefont
  {S.}~\bibnamefont {Bl\"ugel}},\ }\href
  {https://doi.org/10.5281/zenodo.7576163} {\bibinfo {title} {{FLEUR}}},\
  \bibinfo {howpublished} {Zenodo} (\bibinfo {year} {2023})\BibitemShut
  {NoStop}%
\bibitem [{\citenamefont {Wimmer}\ \emph {et~al.}(1981)\citenamefont {Wimmer},
  \citenamefont {Krakauer}, \citenamefont {Weinert},\ and\ \citenamefont
  {Freeman}}]{wimmer1981full}%
  \BibitemOpen
  \bibfield  {author} {\bibinfo {author} {\bibfnamefont {E.}~\bibnamefont
  {Wimmer}}, \bibinfo {author} {\bibfnamefont {H.}~\bibnamefont {Krakauer}},
  \bibinfo {author} {\bibfnamefont {M.}~\bibnamefont {Weinert}},\ and\ \bibinfo
  {author} {\bibfnamefont {A.~J.}\ \bibnamefont {Freeman}},\ }\bibfield
  {title} {\bibinfo {title} {{Full-potential self-consistent
  linearized-augmented-plane-wave method for calculating the electronic
  structure of molecules and surfaces: ${\mathrm{O}}_{2}$ molecule}},\ }\href
  {https://doi.org/10.1103/PhysRevB.24.864} {\bibfield  {journal} {\bibinfo
  {journal} {Phys. Rev. B}\ }\textbf {\bibinfo {volume} {24}},\ \bibinfo
  {pages} {864} (\bibinfo {year} {1981})}\BibitemShut {NoStop}%
\bibitem [{\citenamefont {Perdew}\ and\ \citenamefont
  {Zunger}(1981)}]{perdew1981self}%
  \BibitemOpen
  \bibfield  {author} {\bibinfo {author} {\bibfnamefont {J.~P.}\ \bibnamefont
  {Perdew}}\ and\ \bibinfo {author} {\bibfnamefont {A.}~\bibnamefont
  {Zunger}},\ }\bibfield  {title} {\bibinfo {title} {{Self-interaction
  correction to density-functional approximations for many-electron systems}},\
  }\href {https://doi.org/10.1103/PhysRevB.23.5048} {\bibfield  {journal}
  {\bibinfo  {journal} {Phys. Rev. B}\ }\textbf {\bibinfo {volume} {23}},\
  \bibinfo {pages} {5048} (\bibinfo {year} {1981})}\BibitemShut {NoStop}%
\bibitem [{\citenamefont {Li}\ \emph {et~al.}(1990)\citenamefont {Li},
  \citenamefont {Freeman}, \citenamefont {Jansen},\ and\ \citenamefont
  {Fu}}]{li1990magnetic}%
  \BibitemOpen
  \bibfield  {author} {\bibinfo {author} {\bibfnamefont {C.}~\bibnamefont
  {Li}}, \bibinfo {author} {\bibfnamefont {A.~J.}\ \bibnamefont {Freeman}},
  \bibinfo {author} {\bibfnamefont {H.~J.~F.}\ \bibnamefont {Jansen}},\ and\
  \bibinfo {author} {\bibfnamefont {C.~L.}\ \bibnamefont {Fu}},\ }\bibfield
  {title} {\bibinfo {title} {{Magnetic anisotropy in low-dimensional
  ferromagnetic systems: Fe monolayers on Ag(001), Au(001), and Pd(001)
  substrates}},\ }\href {https://doi.org/10.1103/PhysRevB.42.5433} {\bibfield
  {journal} {\bibinfo  {journal} {Phys. Rev. B}\ }\textbf {\bibinfo {volume}
  {42}},\ \bibinfo {pages} {5433} (\bibinfo {year} {1990})}\BibitemShut
  {NoStop}%
\bibitem [{\citenamefont {de~P.~R.~Moreira}\ \emph {et~al.}(2000)\citenamefont
  {de~P.~R.~Moreira}, \citenamefont {Dovesi}, \citenamefont {Roetti},
  \citenamefont {Saunders},\ and\ \citenamefont {Orlando}}]{moreira2000ab}%
  \BibitemOpen
  \bibfield  {author} {\bibinfo {author} {\bibfnamefont {I.}~\bibnamefont
  {de~P.~R.~Moreira}}, \bibinfo {author} {\bibfnamefont {R.}~\bibnamefont
  {Dovesi}}, \bibinfo {author} {\bibfnamefont {C.}~\bibnamefont {Roetti}},
  \bibinfo {author} {\bibfnamefont {V.~R.}\ \bibnamefont {Saunders}},\ and\
  \bibinfo {author} {\bibfnamefont {R.}~\bibnamefont {Orlando}},\ }\bibfield
  {title} {\bibinfo {title} {{Ab initio study of $M{\mathrm{F}}_{2}$
  $(M=\mathrm{M}\mathrm{n},\mathrm{}\mathrm{F}\mathrm{e},\mathrm{}\mathrm{C}\mathrm{o},\mathrm{}\mathrm{Ni})$
  rutile-type compounds using the periodic unrestricted Hartree-Fock
  approach}},\ }\href {https://doi.org/10.1103/PhysRevB.62.7816} {\bibfield
  {journal} {\bibinfo  {journal} {Phys. Rev. B}\ }\textbf {\bibinfo {volume}
  {62}},\ \bibinfo {pages} {7816} (\bibinfo {year} {2000})}\BibitemShut
  {NoStop}%
\bibitem [{\citenamefont {Monkhorst}\ and\ \citenamefont
  {Pack}(1976)}]{monkhorst1976special}%
  \BibitemOpen
  \bibfield  {author} {\bibinfo {author} {\bibfnamefont {H.~J.}\ \bibnamefont
  {Monkhorst}}\ and\ \bibinfo {author} {\bibfnamefont {J.~D.}\ \bibnamefont
  {Pack}},\ }\bibfield  {title} {\bibinfo {title} {{Special points for
  Brillouin-zone integrations}},\ }\href
  {https://doi.org/10.1103/PhysRevB.13.5188} {\bibfield  {journal} {\bibinfo
  {journal} {Phys. Rev. B}\ }\textbf {\bibinfo {volume} {13}},\ \bibinfo
  {pages} {5188} (\bibinfo {year} {1976})}\BibitemShut {NoStop}%
\bibitem [{\citenamefont {Pizzi}\ \emph {et~al.}(2020)\citenamefont {Pizzi},
  \citenamefont {Vitale}, \citenamefont {Arita}, \citenamefont {Blügel},
  \citenamefont {Freimuth}, \citenamefont {Géranton}, \citenamefont
  {Gibertini}, \citenamefont {Gresch}, \citenamefont {Johnson}, \citenamefont
  {Koretsune}, \citenamefont {Ibañez-Azpiroz}, \citenamefont {Lee},
  \citenamefont {Lihm}, \citenamefont {Marchand}, \citenamefont {Marrazzo},
  \citenamefont {Mokrousov}, \citenamefont {Mustafa}, \citenamefont {Nohara},
  \citenamefont {Nomura}, \citenamefont {Paulatto}, \citenamefont {Poncé},
  \citenamefont {Ponweiser}, \citenamefont {Qiao}, \citenamefont {Thöle},
  \citenamefont {Tsirkin}, \citenamefont {Wierzbowska}, \citenamefont
  {Marzari}, \citenamefont {Vanderbilt}, \citenamefont {Souza}, \citenamefont
  {Mostofi},\ and\ \citenamefont {Yates}}]{pizzi2020wannier90}%
  \BibitemOpen
  \bibfield  {author} {\bibinfo {author} {\bibfnamefont {G.}~\bibnamefont
  {Pizzi}}, \bibinfo {author} {\bibfnamefont {V.}~\bibnamefont {Vitale}},
  \bibinfo {author} {\bibfnamefont {R.}~\bibnamefont {Arita}}, \bibinfo
  {author} {\bibfnamefont {S.}~\bibnamefont {Blügel}}, \bibinfo {author}
  {\bibfnamefont {F.}~\bibnamefont {Freimuth}}, \bibinfo {author}
  {\bibfnamefont {G.}~\bibnamefont {Géranton}}, \bibinfo {author}
  {\bibfnamefont {M.}~\bibnamefont {Gibertini}}, \bibinfo {author}
  {\bibfnamefont {D.}~\bibnamefont {Gresch}}, \bibinfo {author} {\bibfnamefont
  {C.}~\bibnamefont {Johnson}}, \bibinfo {author} {\bibfnamefont
  {T.}~\bibnamefont {Koretsune}}, \bibinfo {author} {\bibfnamefont
  {J.}~\bibnamefont {Ibañez-Azpiroz}}, \bibinfo {author} {\bibfnamefont
  {H.}~\bibnamefont {Lee}}, \bibinfo {author} {\bibfnamefont {J.-M.}\
  \bibnamefont {Lihm}}, \bibinfo {author} {\bibfnamefont {D.}~\bibnamefont
  {Marchand}}, \bibinfo {author} {\bibfnamefont {A.}~\bibnamefont {Marrazzo}},
  \bibinfo {author} {\bibfnamefont {Y.}~\bibnamefont {Mokrousov}}, \bibinfo
  {author} {\bibfnamefont {J.~I.}\ \bibnamefont {Mustafa}}, \bibinfo {author}
  {\bibfnamefont {Y.}~\bibnamefont {Nohara}}, \bibinfo {author} {\bibfnamefont
  {Y.}~\bibnamefont {Nomura}}, \bibinfo {author} {\bibfnamefont
  {L.}~\bibnamefont {Paulatto}}, \bibinfo {author} {\bibfnamefont
  {S.}~\bibnamefont {Poncé}}, \bibinfo {author} {\bibfnamefont
  {T.}~\bibnamefont {Ponweiser}}, \bibinfo {author} {\bibfnamefont
  {J.}~\bibnamefont {Qiao}}, \bibinfo {author} {\bibfnamefont {F.}~\bibnamefont
  {Thöle}}, \bibinfo {author} {\bibfnamefont {S.~S.}\ \bibnamefont {Tsirkin}},
  \bibinfo {author} {\bibfnamefont {M.}~\bibnamefont {Wierzbowska}}, \bibinfo
  {author} {\bibfnamefont {N.}~\bibnamefont {Marzari}}, \bibinfo {author}
  {\bibfnamefont {D.}~\bibnamefont {Vanderbilt}}, \bibinfo {author}
  {\bibfnamefont {I.}~\bibnamefont {Souza}}, \bibinfo {author} {\bibfnamefont
  {A.~A.}\ \bibnamefont {Mostofi}},\ and\ \bibinfo {author} {\bibfnamefont
  {J.~R.}\ \bibnamefont {Yates}},\ }\bibfield  {title} {\bibinfo {title}
  {{Wannier90 as a community code: new features and applications}},\ }\href
  {https://doi.org/10.1088/1361-648X/ab51ff} {\bibfield  {journal} {\bibinfo
  {journal} {J. Phys. Condens. Matter.}\ }\textbf {\bibinfo {volume} {32}},\
  \bibinfo {pages} {165902} (\bibinfo {year} {2020})}\BibitemShut {NoStop}%
\bibitem [{\citenamefont {Dzialoshinskii}(1958)}]{dzialoshinskii1958magnetic}%
  \BibitemOpen
  \bibfield  {author} {\bibinfo {author} {\bibfnamefont {I.~E.}\ \bibnamefont
  {Dzialoshinskii}},\ }\bibfield  {title} {\bibinfo {title} {{The magnetic
  structure of fluorides of the transition metals}},\ }\href@noop {} {\bibfield
   {journal} {\bibinfo  {journal} {Sov. Phys. JETP}\ }\textbf {\bibinfo
  {volume} {6}},\ \bibinfo {pages} {1120} (\bibinfo {year} {1958})}\BibitemShut
  {NoStop}%
\end{thebibliography}%


\begin{thebibliography}{9}%
\makeatletter
\providecommand \@ifxundefined [1]{%
 \@ifx{#1\undefined}
}%
\providecommand \@ifnum [1]{%
 \ifnum #1\expandafter \@firstoftwo
 \else \expandafter \@secondoftwo
 \fi
}%
\providecommand \@ifx [1]{%
 \ifx #1\expandafter \@firstoftwo
 \else \expandafter \@secondoftwo
 \fi
}%
\providecommand \natexlab [1]{#1}%
\providecommand \enquote  [1]{``#1''}%
\providecommand \bibnamefont  [1]{#1}%
\providecommand \bibfnamefont [1]{#1}%
\providecommand \citenamefont [1]{#1}%
\providecommand \href@noop [0]{\@secondoftwo}%
\providecommand \href [0]{\begingroup \@sanitize@url \@href}%
\providecommand \@href[1]{\@@startlink{#1}\@@href}%
\providecommand \@@href[1]{\endgroup#1\@@endlink}%
\providecommand \@sanitize@url [0]{\catcode `\\12\catcode `\$12\catcode
  `\&12\catcode `\#12\catcode `\^12\catcode `\_12\catcode `\%12\relax}%
\providecommand \@@startlink[1]{}%
\providecommand \@@endlink[0]{}%
\providecommand \url  [0]{\begingroup\@sanitize@url \@url }%
\providecommand \@url [1]{\endgroup\@href {#1}{\urlprefix }}%
\providecommand \urlprefix  [0]{URL }%
\providecommand \Eprint [0]{\href }%
\providecommand \doibase [0]{https://doi.org/}%
\providecommand \selectlanguage [0]{\@gobble}%
\providecommand \bibinfo  [0]{\@secondoftwo}%
\providecommand \bibfield  [0]{\@secondoftwo}%
\providecommand \translation [1]{[#1]}%
\providecommand \BibitemOpen [0]{}%
\providecommand \bibitemStop [0]{}%
\providecommand \bibitemNoStop [0]{.\EOS\space}%
\providecommand \EOS [0]{\spacefactor3000\relax}%
\providecommand \BibitemShut  [1]{\csname bibitem#1\endcsname}%
\let\auto@bib@innerbib\@empty
\bibitem [{\citenamefont {Slater}\ and\ \citenamefont
  {Koster}(1954)}]{slater1954simplified}%
  \BibitemOpen
  \bibfield  {author} {\bibinfo {author} {\bibfnamefont {J.~C.}\ \bibnamefont
  {Slater}}\ and\ \bibinfo {author} {\bibfnamefont {G.~F.}\ \bibnamefont
  {Koster}},\ }\bibfield  {title} {\bibinfo {title} {{Simplified LCAO Method
  for the Periodic Potential Problem}},\ }\href
  {https://doi.org/10.1103/PhysRev.94.1498} {\bibfield  {journal} {\bibinfo
  {journal} {Phys. Rev.}\ }\textbf {\bibinfo {volume} {94}},\ \bibinfo {pages}
  {1498} (\bibinfo {year} {1954})}\BibitemShut {NoStop}%
\bibitem [{fle()}]{fleurWeb}%
  \BibitemOpen
  \href@noop {} {\bibinfo {title} {{The FLEUR project}}},\ \bibinfo
  {howpublished} {\url{https://www.flapw.de/}}\BibitemShut {NoStop}%
\bibitem [{\citenamefont {Wortmann}\ \emph {et~al.}(2023)\citenamefont
  {Wortmann}, \citenamefont {Michalicek}, \citenamefont {Baadji}, \citenamefont
  {Betzinger}, \citenamefont {Bihlmayer}, \citenamefont {Br\"oder},
  \citenamefont {Burnus}, \citenamefont {Enkovaara}, \citenamefont {Freimuth},
  \citenamefont {Friedrich}, \citenamefont {Gerhorst}, \citenamefont
  {Granberg~Cauchi}, \citenamefont {Grytsiuk}, \citenamefont {Hanke},
  \citenamefont {Hanke}, \citenamefont {Heide}, \citenamefont {Heinze},
  \citenamefont {Hilgers}, \citenamefont {Janssen}, \citenamefont
  {Kl\"uppelberg}, \citenamefont {Kovacik}, \citenamefont {Kurz}, \citenamefont
  {Lezaic}, \citenamefont {Madsen}, \citenamefont {Mokrousov}, \citenamefont
  {Neukirchen}, \citenamefont {Redies}, \citenamefont {Rost}, \citenamefont
  {Schlipf}, \citenamefont {Schindlmayr}, \citenamefont {Winkelmann},\ and\
  \citenamefont {Bl\"ugel}}]{fleurCode}%
  \BibitemOpen
  \bibfield  {author} {\bibinfo {author} {\bibfnamefont {D.}~\bibnamefont
  {Wortmann}}, \bibinfo {author} {\bibfnamefont {G.}~\bibnamefont
  {Michalicek}}, \bibinfo {author} {\bibfnamefont {N.}~\bibnamefont {Baadji}},
  \bibinfo {author} {\bibfnamefont {M.}~\bibnamefont {Betzinger}}, \bibinfo
  {author} {\bibfnamefont {G.}~\bibnamefont {Bihlmayer}}, \bibinfo {author}
  {\bibfnamefont {J.}~\bibnamefont {Br\"oder}}, \bibinfo {author}
  {\bibfnamefont {T.}~\bibnamefont {Burnus}}, \bibinfo {author} {\bibfnamefont
  {J.}~\bibnamefont {Enkovaara}}, \bibinfo {author} {\bibfnamefont
  {F.}~\bibnamefont {Freimuth}}, \bibinfo {author} {\bibfnamefont
  {C.}~\bibnamefont {Friedrich}}, \bibinfo {author} {\bibfnamefont {C.-R.}\
  \bibnamefont {Gerhorst}}, \bibinfo {author} {\bibfnamefont {S.}~\bibnamefont
  {Granberg~Cauchi}}, \bibinfo {author} {\bibfnamefont {U.}~\bibnamefont
  {Grytsiuk}}, \bibinfo {author} {\bibfnamefont {A.}~\bibnamefont {Hanke}},
  \bibinfo {author} {\bibfnamefont {J.-P.}\ \bibnamefont {Hanke}}, \bibinfo
  {author} {\bibfnamefont {M.}~\bibnamefont {Heide}}, \bibinfo {author}
  {\bibfnamefont {S.}~\bibnamefont {Heinze}}, \bibinfo {author} {\bibfnamefont
  {R.}~\bibnamefont {Hilgers}}, \bibinfo {author} {\bibfnamefont
  {H.}~\bibnamefont {Janssen}}, \bibinfo {author} {\bibfnamefont {D.~A.}\
  \bibnamefont {Kl\"uppelberg}}, \bibinfo {author} {\bibfnamefont
  {R.}~\bibnamefont {Kovacik}}, \bibinfo {author} {\bibfnamefont
  {P.}~\bibnamefont {Kurz}}, \bibinfo {author} {\bibfnamefont {M.}~\bibnamefont
  {Lezaic}}, \bibinfo {author} {\bibfnamefont {G.~K.~H.}\ \bibnamefont
  {Madsen}}, \bibinfo {author} {\bibfnamefont {Y.}~\bibnamefont {Mokrousov}},
  \bibinfo {author} {\bibfnamefont {A.}~\bibnamefont {Neukirchen}}, \bibinfo
  {author} {\bibfnamefont {M.}~\bibnamefont {Redies}}, \bibinfo {author}
  {\bibfnamefont {S.}~\bibnamefont {Rost}}, \bibinfo {author} {\bibfnamefont
  {M.}~\bibnamefont {Schlipf}}, \bibinfo {author} {\bibfnamefont
  {A.}~\bibnamefont {Schindlmayr}}, \bibinfo {author} {\bibfnamefont
  {M.}~\bibnamefont {Winkelmann}},\ and\ \bibinfo {author} {\bibfnamefont
  {S.}~\bibnamefont {Bl\"ugel}},\ }\href
  {https://doi.org/10.5281/zenodo.7576163} {\bibinfo {title} {{FLEUR}}},\
  \bibinfo {howpublished} {Zenodo} (\bibinfo {year} {2023})\BibitemShut
  {NoStop}%
\bibitem [{\citenamefont {Wimmer}\ \emph {et~al.}(1981)\citenamefont {Wimmer},
  \citenamefont {Krakauer}, \citenamefont {Weinert},\ and\ \citenamefont
  {Freeman}}]{wimmer1981full}%
  \BibitemOpen
  \bibfield  {author} {\bibinfo {author} {\bibfnamefont {E.}~\bibnamefont
  {Wimmer}}, \bibinfo {author} {\bibfnamefont {H.}~\bibnamefont {Krakauer}},
  \bibinfo {author} {\bibfnamefont {M.}~\bibnamefont {Weinert}},\ and\ \bibinfo
  {author} {\bibfnamefont {A.~J.}\ \bibnamefont {Freeman}},\ }\bibfield
  {title} {\bibinfo {title} {{Full-potential self-consistent
  linearized-augmented-plane-wave method for calculating the electronic
  structure of molecules and surfaces: ${\mathrm{O}}_{2}$ molecule}},\ }\href
  {https://doi.org/10.1103/PhysRevB.24.864} {\bibfield  {journal} {\bibinfo
  {journal} {Phys. Rev. B}\ }\textbf {\bibinfo {volume} {24}},\ \bibinfo
  {pages} {864} (\bibinfo {year} {1981})}\BibitemShut {NoStop}%
\bibitem [{\citenamefont {Perdew}\ \emph {et~al.}(1996)\citenamefont {Perdew},
  \citenamefont {Burke},\ and\ \citenamefont
  {Ernzerhof}}]{perdew1996generalized}%
  \BibitemOpen
  \bibfield  {author} {\bibinfo {author} {\bibfnamefont {J.~P.}\ \bibnamefont
  {Perdew}}, \bibinfo {author} {\bibfnamefont {K.}~\bibnamefont {Burke}},\ and\
  \bibinfo {author} {\bibfnamefont {M.}~\bibnamefont {Ernzerhof}},\ }\bibfield
  {title} {\bibinfo {title} {{Generalized Gradient Approximation Made
  Simple}},\ }\href {https://doi.org/10.1103/PhysRevLett.77.3865} {\bibfield
  {journal} {\bibinfo  {journal} {Phys. Rev. Lett.}\ }\textbf {\bibinfo
  {volume} {77}},\ \bibinfo {pages} {3865} (\bibinfo {year}
  {1996})}\BibitemShut {NoStop}%
\bibitem [{\citenamefont {Li}\ \emph {et~al.}(1990)\citenamefont {Li},
  \citenamefont {Freeman}, \citenamefont {Jansen},\ and\ \citenamefont
  {Fu}}]{li1990magnetic}%
  \BibitemOpen
  \bibfield  {author} {\bibinfo {author} {\bibfnamefont {C.}~\bibnamefont
  {Li}}, \bibinfo {author} {\bibfnamefont {A.~J.}\ \bibnamefont {Freeman}},
  \bibinfo {author} {\bibfnamefont {H.~J.~F.}\ \bibnamefont {Jansen}},\ and\
  \bibinfo {author} {\bibfnamefont {C.~L.}\ \bibnamefont {Fu}},\ }\bibfield
  {title} {\bibinfo {title} {{Magnetic anisotropy in low-dimensional
  ferromagnetic systems: Fe monolayers on Ag(001), Au(001), and Pd(001)
  substrates}},\ }\href {https://doi.org/10.1103/PhysRevB.42.5433} {\bibfield
  {journal} {\bibinfo  {journal} {Phys. Rev. B}\ }\textbf {\bibinfo {volume}
  {42}},\ \bibinfo {pages} {5433} (\bibinfo {year} {1990})}\BibitemShut
  {NoStop}%
\bibitem [{\citenamefont {de~P.~R.~Moreira}\ \emph {et~al.}(2000)\citenamefont
  {de~P.~R.~Moreira}, \citenamefont {Dovesi}, \citenamefont {Roetti},
  \citenamefont {Saunders},\ and\ \citenamefont {Orlando}}]{moreira2000ab}%
  \BibitemOpen
  \bibfield  {author} {\bibinfo {author} {\bibfnamefont {I.}~\bibnamefont
  {de~P.~R.~Moreira}}, \bibinfo {author} {\bibfnamefont {R.}~\bibnamefont
  {Dovesi}}, \bibinfo {author} {\bibfnamefont {C.}~\bibnamefont {Roetti}},
  \bibinfo {author} {\bibfnamefont {V.~R.}\ \bibnamefont {Saunders}},\ and\
  \bibinfo {author} {\bibfnamefont {R.}~\bibnamefont {Orlando}},\ }\bibfield
  {title} {\bibinfo {title} {{Ab initio study of $M{\mathrm{F}}_{2}$
  $(M=\mathrm{M}\mathrm{n},\mathrm{}\mathrm{F}\mathrm{e},\mathrm{}\mathrm{C}\mathrm{o},\mathrm{}\mathrm{Ni})$
  rutile-type compounds using the periodic unrestricted Hartree-Fock
  approach}},\ }\href {https://doi.org/10.1103/PhysRevB.62.7816} {\bibfield
  {journal} {\bibinfo  {journal} {Phys. Rev. B}\ }\textbf {\bibinfo {volume}
  {62}},\ \bibinfo {pages} {7816} (\bibinfo {year} {2000})}\BibitemShut
  {NoStop}%
\bibitem [{\citenamefont {Monkhorst}\ and\ \citenamefont
  {Pack}(1976)}]{monkhorst1976special}%
  \BibitemOpen
  \bibfield  {author} {\bibinfo {author} {\bibfnamefont {H.~J.}\ \bibnamefont
  {Monkhorst}}\ and\ \bibinfo {author} {\bibfnamefont {J.~D.}\ \bibnamefont
  {Pack}},\ }\bibfield  {title} {\bibinfo {title} {{Special points for
  Brillouin-zone integrations}},\ }\href
  {https://doi.org/10.1103/PhysRevB.13.5188} {\bibfield  {journal} {\bibinfo
  {journal} {Phys. Rev. B}\ }\textbf {\bibinfo {volume} {13}},\ \bibinfo
  {pages} {5188} (\bibinfo {year} {1976})}\BibitemShut {NoStop}%
\bibitem [{\citenamefont {Pizzi}\ \emph {et~al.}(2020)\citenamefont {Pizzi},
  \citenamefont {Vitale}, \citenamefont {Arita}, \citenamefont {Blügel},
  \citenamefont {Freimuth}, \citenamefont {Géranton}, \citenamefont
  {Gibertini}, \citenamefont {Gresch}, \citenamefont {Johnson}, \citenamefont
  {Koretsune}, \citenamefont {Ibañez-Azpiroz}, \citenamefont {Lee},
  \citenamefont {Lihm}, \citenamefont {Marchand}, \citenamefont {Marrazzo},
  \citenamefont {Mokrousov}, \citenamefont {Mustafa}, \citenamefont {Nohara},
  \citenamefont {Nomura}, \citenamefont {Paulatto}, \citenamefont {Poncé},
  \citenamefont {Ponweiser}, \citenamefont {Qiao}, \citenamefont {Thöle},
  \citenamefont {Tsirkin}, \citenamefont {Wierzbowska}, \citenamefont
  {Marzari}, \citenamefont {Vanderbilt}, \citenamefont {Souza}, \citenamefont
  {Mostofi},\ and\ \citenamefont {Yates}}]{pizzi2020wannier90}%
  \BibitemOpen
  \bibfield  {author} {\bibinfo {author} {\bibfnamefont {G.}~\bibnamefont
  {Pizzi}}, \bibinfo {author} {\bibfnamefont {V.}~\bibnamefont {Vitale}},
  \bibinfo {author} {\bibfnamefont {R.}~\bibnamefont {Arita}}, \bibinfo
  {author} {\bibfnamefont {S.}~\bibnamefont {Blügel}}, \bibinfo {author}
  {\bibfnamefont {F.}~\bibnamefont {Freimuth}}, \bibinfo {author}
  {\bibfnamefont {G.}~\bibnamefont {Géranton}}, \bibinfo {author}
  {\bibfnamefont {M.}~\bibnamefont {Gibertini}}, \bibinfo {author}
  {\bibfnamefont {D.}~\bibnamefont {Gresch}}, \bibinfo {author} {\bibfnamefont
  {C.}~\bibnamefont {Johnson}}, \bibinfo {author} {\bibfnamefont
  {T.}~\bibnamefont {Koretsune}}, \bibinfo {author} {\bibfnamefont
  {J.}~\bibnamefont {Ibañez-Azpiroz}}, \bibinfo {author} {\bibfnamefont
  {H.}~\bibnamefont {Lee}}, \bibinfo {author} {\bibfnamefont {J.-M.}\
  \bibnamefont {Lihm}}, \bibinfo {author} {\bibfnamefont {D.}~\bibnamefont
  {Marchand}}, \bibinfo {author} {\bibfnamefont {A.}~\bibnamefont {Marrazzo}},
  \bibinfo {author} {\bibfnamefont {Y.}~\bibnamefont {Mokrousov}}, \bibinfo
  {author} {\bibfnamefont {J.~I.}\ \bibnamefont {Mustafa}}, \bibinfo {author}
  {\bibfnamefont {Y.}~\bibnamefont {Nohara}}, \bibinfo {author} {\bibfnamefont
  {Y.}~\bibnamefont {Nomura}}, \bibinfo {author} {\bibfnamefont
  {L.}~\bibnamefont {Paulatto}}, \bibinfo {author} {\bibfnamefont
  {S.}~\bibnamefont {Poncé}}, \bibinfo {author} {\bibfnamefont
  {T.}~\bibnamefont {Ponweiser}}, \bibinfo {author} {\bibfnamefont
  {J.}~\bibnamefont {Qiao}}, \bibinfo {author} {\bibfnamefont {F.}~\bibnamefont
  {Thöle}}, \bibinfo {author} {\bibfnamefont {S.~S.}\ \bibnamefont {Tsirkin}},
  \bibinfo {author} {\bibfnamefont {M.}~\bibnamefont {Wierzbowska}}, \bibinfo
  {author} {\bibfnamefont {N.}~\bibnamefont {Marzari}}, \bibinfo {author}
  {\bibfnamefont {D.}~\bibnamefont {Vanderbilt}}, \bibinfo {author}
  {\bibfnamefont {I.}~\bibnamefont {Souza}}, \bibinfo {author} {\bibfnamefont
  {A.~A.}\ \bibnamefont {Mostofi}},\ and\ \bibinfo {author} {\bibfnamefont
  {J.~R.}\ \bibnamefont {Yates}},\ }\bibfield  {title} {\bibinfo {title}
  {{Wannier90 as a community code: new features and applications}},\ }\href
  {https://doi.org/10.1088/1361-648X/ab51ff} {\bibfield  {journal} {\bibinfo
  {journal} {J. Phys. Condens. Matter.}\ }\textbf {\bibinfo {volume} {32}},\
  \bibinfo {pages} {165902} (\bibinfo {year} {2020})}\BibitemShut {NoStop}%
\end{thebibliography}%

	\subsection{Appendix}

	\emph{Discussions on DMI}---Here, we demonstrate that the DMI does not contribute to the WFM in the models examined in this study [e.g., Figs.~\ref{fig_illustration}(c)-(e)]. The DMI, $E_\mathrm{DMI} = \mathbf{D}_{ij} \cdot (\mathbf{S}_i \times \mathbf{S}_j) $, between spins at sites $i$ and $j$, is characterized by the DMI vector $\mathbf{D}_{ij}$, whose orientation is dictated by the crystal symmetry. For our model in Fig.~\ref{fig_illustration}(c), the crystal structure has two mirror planes---one horizontal and one vertical---including sublattices $A$ and $B$. Since the DMI vector should be perpendicular to each mirror plane according to the Moriya’s rules~\cite{moriya1960anisotropic}, the presence of the two symmetry planes implies that the DMI vector should be zero. On the other hand, for the kagome lattice shown in Figs.~\ref{fig_illustration}(d) and \ref{fig_illustration}(e), there is only one mirror plane---horizontal---including all sublattices $A$, $B$, and $C$. Accordingly, the DMI vectors  $\mathbf{D}_{ij}$ are identically oriented along the out-of-plane direction for all $i$ and $j$, i.e., $\mathbf{D}_{AB} = \mathbf{D}_{BC} = \mathbf{D}_{CA} = D \hat{\mathbf{z}}$. The DMI energy is then given by $E_\mathrm{DMI} = D \hat{\mathbf{z}} \cdot (\mathbf{S}_A \times \mathbf{S}_B + \mathbf{S}_B \times \mathbf{S}_C + \mathbf{S}_C \times \mathbf{S}_A) $. Depending on the sign of $D$, the DMI energetically stabilizes either the normal ($D<0$) [Fig.~\ref{fig_illustration}(d)] or inverse ($D>0$) [Fig.~\ref{fig_illustration}(e)] triangular spin structure~\cite{tomiyoshi1982magnetic}. We emphasize that in both spin structures, spins are fully compensated. Small deviations from these fully compensated spin structures can generate a finite net spin moment but increase $E_\mathrm{DMI}$, indicating that the DMI does not produce a net magnetic moment. This consequence also shows that the $g$-tensor anisotropy mechanism can be the dominant contribution to the WFM, even in the presence of the DMI.

	Although the DMI does not contribute to the WFM in our models, the $g$-tensor anisotropy and DMI mechanisms for WFM can coexist in other systems. To illustrate this, let’s consider a fully compensated spin system. If we include SOC to this system, it can affect the compensation in two ways: (i) giving additional interaction terms to the spin Hamiltonian, and (ii) inducing a local orbital moment. The effect (i) can lead to the spin rearrangement, e.g., spin canting due to the DMI, so that the canted spin configuration with finite net spin moment minimizes the energy. On the other hand, the effect (ii) occurs independently of the spin rearrangement. Once the spin configuration is given, regardless of whether the effect (i) is present or not, the local orbital moment is induced by the spin moment through SOC, with its magnitude and direction being associated with the local $g$-tensor. The correction to the net orbital moment due to (i) is proportional to the square of the SOC magnitude. Thus, as long as the SOC is weak, the magnitude and the direction of the net orbital moment are essentially unaffected by (i) and the two processes (i) and (ii) make independent contributions to the net moment.

	\begin{figure}[b]
		\center\includegraphics[width=0.5\textwidth]{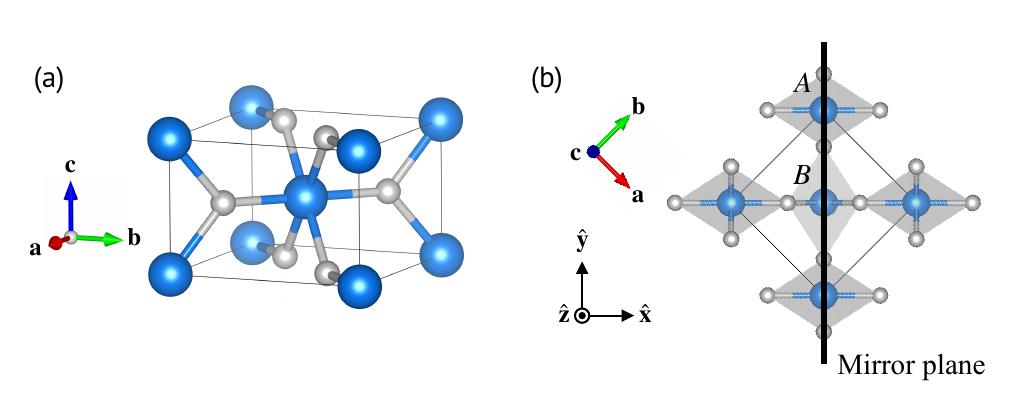}
		\caption{(a) Crystal structure of a rutile-typle AM, where blue and gray spheres represent magnetic and nonmagnetic ions, respectively. (b) Top view of the structure. The black solid line indicates the mirror plane (110) that includes two magnetic sublattices, $A$ and $B$.
		}
		\label{fig5} 
	\end{figure}
	
	\emph{WFM in rutile-type AMs}---In addition to the tight-binding models, we discuss the origin of WFM in rutile-type antiferromagnets [Fig.~\ref{fig5}(a)], such as RuO$_2$ and NiF$_2$, which are regarded as prototypical AMs. These systems can exhibit the WFM for specific directions of the Néel vector. For example, a recent first-principles calculation~\cite{smejkal2020crystal} reported that in RuO$_2$, a net magnetic moment is induced along the $b$ axis when the Néel vector is along the $a$ axis. Although this WFM was attributed to the DMI in the recent study, it cannot be explained by the conventional DMI mechanism, which involves an antisymmetric exchange interaction $\mathbf{D}_{AB} \cdot (\mathbf{S}_A \times \mathbf{S}_B)$ between spins at sublattices $A$ and $B$. For the DMI to induce spin canting in the $ab$-plane, the DMI vector $\mathbf{D}_{AB}$ must point along the $c$ axis. However, by Moriya's rules~\cite{moriya1960anisotropic}, the $c$ component of $\mathbf{D}_{AB}$ is forbidden due to the presence of a vertical mirror plane containing sublattices $A$ and $B$ [Fig.~\ref{fig5}(b)]. Indeed, an earlier study by Dzyaloshinskii~\cite{dzialoshinskii1958magnetic} demonstrated that the rutile structure does not exhibit the antisymmetric exchange interaction. Additionally, Moriya~\cite{moriya1960theory} explained the WFM in NiF$_2$ as arising from single-ion anisotropy. This explanation aligns with our $g$-tensor anisotropy mechanism from a symmetry perspective, though it did not recognize the importance of the orbital moment. Our work goes beyond the Moriya’s explanation by considering the orbital moment and shows that the net orbital moment dominates over the net spin moment in rutile-type AMs such as RuO$_2$~\cite{smejkal2020crystal} and NiF$_2$ (see below), highlighting the $g$-tensor anisotropy mechanism as the most relevant mechanism for the WFM in these materials.

	\emph{First-principles calculation for NiF$_2$}---We show that the $g$-tensor anisotropy mechanism plays a dominant role in the WFM of a rutile AM using the density functional theory code \texttt{FLEUR}~\cite{fleurWeb, fleurCode} based on the full-potential linearized augmented plane-wave method~\cite{wimmer1981full} (see Supplemental Material~\cite{supp} for details). Specifically, we calculated spin and orbital magnetic moments in insulating NiF$_2$ [Fig.~\ref{fig5}], where initial spin moments for Ni $A$ and Ni $B$ sublattices are aligned along $-\mathbf{a}$ and $+\mathbf{a}$, respectively, and then relaxed after including the atomic SOC. Table~\ref{table1} presents the calculated spin and orbital magnetic moments for each Ni atom. The resulting magnetic moment deviates from the initial direction ($-\mathbf{a}$ or $+\mathbf{a}$), leading to a net magnetic moment along the $b$ axis, where the orbital contribution ($m_b^\mathrm{orb}$) dominates over the spin contribution ($m_b^\mathrm{spin}$). This dominance of the orbital contribution and the net magnetic moment direction [Eq.~\eqref{eq:m_tot2} with $\phi = - 45^\circ$] are consistent with our mechanism based on the $g$-tensor anisotropy. We note that this result cannot be attributed to the DMI, as discussed in the previous paragraph.

\begin{figure}[b!]
		\center\includegraphics[width=0.5\textwidth]{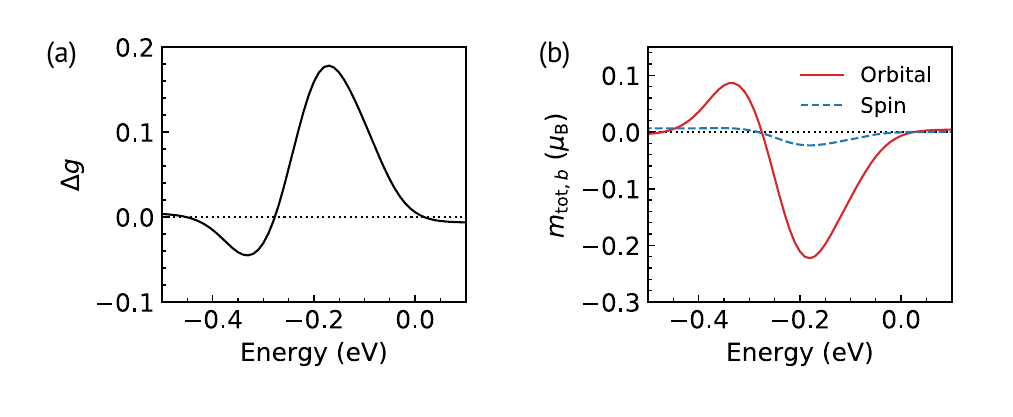}
		\caption{(a) Local $g$-tensor anisotropy ($\Delta g$) of the Ni atom in NiF$_2$ as a function of the Fermi energy. (b) Orbital and spin contributions to the net magnetic moment in NiF$_2$.
		}
		\label{fig6} 
	\end{figure}

	The net moment in NiF$_2$ is relatively small since $\Delta g$ is weak in an insulating phase, but the  $\Delta g$ and the resulting net moment can be significantly enhanced by doping. To investigate this, we evaluated  $\Delta g$ in NiF$_2$ as a function of the Fermi energy. The local $g$-tensor elements at the Ni atom can be obtained by $g_\parallel = 2 + \langle L_x^A \rangle/ \langle S_x^A \rangle $ and $g_\perp = 2 + \langle L_y^A \rangle/ \langle S_y^A \rangle $ [see Fig.~\ref{fig5}(b) for the axis convention]. The local $g$-tensor anisotropy is then given by $\Delta g = g_\parallel - g_\perp$. Figure~\ref{fig6}(a) shows that the calculated $\Delta g$ significantly increases with $p$-doping, reaching the order of 0.1. Consequently, the net magnetic moment becomes larger, with the orbital contribution dominating over the spin contribution [Fig.~\ref{fig6}(b)]. Notably, the magnitude and energy dependence of $\Delta g$ and the net moment align well with our tight-binding results [Figs.~\ref{fig_square2}(c) and \ref{fig_square2}(d)]. This large net orbital moment of the order of $0.1~\mu_\mathrm{B}$ in the metallic phase is also consistent with first-principles calculations for the altermagnetic metal RuO$_2$~\cite{smejkal2020crystal}. 
	
	\newpage
	
	\begin{center}
		\begin{table}[b]
			\caption{The $a$, $b$, and $c$ components of spin ($\mathbf{m}^\mathrm{spin}$) and orbital ($\mathbf{m}^\mathrm{orb}$) magnetic moments for Ni atoms at sublattices $A$ and $B$.}\label{table1}
			\begin{tabularx}{0.5\textwidth} { 
					>{\centering\arraybackslash}X 
					>{\centering\arraybackslash}X 
					>{\centering\arraybackslash}X 
					>{\centering\arraybackslash}X 
					>{\centering\arraybackslash}X 
					>{\centering\arraybackslash}X 
					>{\centering\arraybackslash}X  }
				\hline\hline
				Atom & $m_a^\mathrm{spin}$ ($\mu_\mathrm{B}$) & $m_b^\mathrm{spin}$ ($\mu_\mathrm{B}$) & $m_c^\mathrm{spin}$ ($\mu_\mathrm{B}$) & $m_a^\mathrm{orb}$ ($\mu_\mathrm{B}$) & $m_b^\mathrm{orb}$ ($\mu_\mathrm{B}$) & $m_c^\mathrm{orb}$ ($\mu_\mathrm{B}$) \\ 
				\hline
				Ni $A$  & $-1.72$ & $-2.3 \times 10^{-4} $ & 0 & $-0.16$ & $-2.5 \times 10^{-3} $ & 0 \\ 
				Ni $B$  & $1.72$ & $-2.3 \times 10^{-4} $ & 0 & $0.16$ & $-2.5 \times 10^{-3} $ & 0 \\
				\hline\hline
			\end{tabularx}
		\end{table}
	\end{center}

\end{document}


\title{Supplemental Material for "Weak Ferromagnetism in Altermagnets from Alternating $g$-Tensor Anisotropy"}
	
	\date{\today}

	\author{Daegeun~Jo}
	\email{daegeun.jo@physics.uu.se}
	\affiliation{Department of Physics and Astronomy, Uppsala University, P.O. Box 516, SE-75120 Uppsala, Sweden}
	\affiliation{Wallenberg Initiative Materials Science for Sustainability, Uppsala University, SE-75120 Uppsala, Sweden}
	
	\author{Dongwook~Go}
	\affiliation{Institute of Physics, Johannes Gutenberg University Mainz, 55099 Mainz, Germany}
	
	\author{Yuriy~Mokrousov}
	\affiliation{Institute of Physics, Johannes Gutenberg University Mainz, 55099 Mainz, Germany}
	\affiliation{Peter Gr\"unberg Institut and Institute for Advanced Simulation, Forschungszentrum J\"ulich and JARA, 52425 J\"ulich, Germany \looseness=-1}

	\author{Peter~M.~Oppeneer}
	\affiliation{Department of Physics and Astronomy, Uppsala University, P.O. Box 516, SE-75120 Uppsala, Sweden}
	\affiliation{Wallenberg Initiative Materials Science for Sustainability, Uppsala University, SE-75120 Uppsala, Sweden}

	\author{Sang-Wook~Cheong}
	\email{sangc@physics.rutgers.edu}
	\affiliation{Rutgers Center for Emergent Materials and Department of Physics and Astronomy, Rutgers University, Piscataway, New Jersey 08854, USA}
	
	\author{Hyun-Woo~Lee}
	\email{hwl@postech.ac.kr}
	\affiliation{Department of Physics, Pohang University of Science and Technology, Pohang 37673,Korea}
	
	\maketitle
	
	\newpage
	
	\tableofcontents

	\section{Details of tight-binding models}\label{sec1}
	
	In the main text, we investigated tight-binding models for altermagnetic systems with collinear and noncollinear spin structures. The altermagnet with collinear spins, whose results are shown in Figs.~2 and 3 of the main text, was modeled as a square lattice consisting of two magnetic atoms with local anisotropy. For the noncollinear case (Fig.~4 of the main text), we examined various systems based on the kagome lattice, where the three magnetic atoms form either a normal or inverse triangular spin structure. Specifically, these systems include pristine [Fig.~4(b)] and distorted [Fig.~4(c)] monolayer kagome lattices, and a bulk system composed of periodically repeated ABC-stacked kagome layers [Fig.~4(d)]. In this section, we provide the details of these models.
	
	\subsection{Lattice and spin structures}\label{sec1a}
	
	The two-dimensional square lattice system is defined by the following lattice vectors:
	\begin{subequations}
		\begin{align}
			\mathbf{a}_1 & = \frac{a}{\sqrt{2}} \left(1,-1,0 \right) , \\
			\mathbf{a}_2 & = \frac{a}{\sqrt{2}} \left(1,1,0 \right) .
		\end{align}
	\end{subequations}
	The unit cell consists of two magnetic sublattices, $A$ and $B$, positioned at
	\begin{subequations}
		\begin{align}
			\mathbf{r}_A & = \mathbf{0} , \\
			\mathbf{r}_B & = \frac{1}{2}\mathbf{a}_1 + \frac{1}{2} \mathbf{a}_2 ,
		\end{align}
	\end{subequations}
	where the corresponding spins point along
	\begin{subequations}
		\begin{align}
			\hat{\mathbf{s}}^A & = (\cos \phi , \sin \phi, 0) , \\
			\hat{\mathbf{s}}^B & =  (-\cos \phi , -\sin \phi, 0) ,
		\end{align}
	\end{subequations}
	respectively.

	For the two-dimensional kagome lattice, the unit cell is defined by the lattice vectors
	\begin{subequations}
		\begin{align}
			\mathbf{a}_1 & = \frac{a}{2} \left(\sqrt{3},-1,0 \right) , \\
			\mathbf{a}_2 & = \frac{a}{2} \left(\sqrt{3},1,0 \right) ,
		\end{align}
	\end{subequations}
	and contains three magnetic sublattices, denoted as $A$, $B$, and $C$, positioned at 
		\begin{equation}
			\mathbf{r}_A  = \mathbf{0} , 	\quad	\mathbf{r}_B  = \frac{1}{2}\mathbf{a}_1 , \quad
			\mathbf{r}_C  = \frac{1}{2} \mathbf{a}_2 .
		\end{equation}
	Here, we consider two types of spin configurations, where the spin direction at each site is given by:
	\begin{subequations}
		\begin{align}
			\hat{\mathbf{s}}^A & = \left(\cos \phi , \sin \phi, 0 \right) , \\
			\hat{\mathbf{s}}^B & =  \left(\cos (\phi \pm 2\pi/3) , \sin (\phi \pm 2\pi/3), 0 \right) , \\
			\hat{\mathbf{s}}^C & =  \left(\cos (\phi \pm 4\pi/3) , \sin (\phi \pm 4\pi/3), 0 \right) ,
		\end{align}
	\end{subequations}
	 with the sign distinguishing between normal $(+)$ and inverse $(-)$ triangular spin structures.

	 The three-dimensional bulk system [Fig.~4(d) of the main text] is constructed by stacking three kagome planes along the $z$-direction, with lattice vectors
	 \begin{subequations}
	 	\begin{align}
	 		\mathbf{a}_1 & = \frac{a}{2} \left(\sqrt{3},-1,0 \right) , \\
	 		\mathbf{a}_2 & = \frac{a}{2} \left(\sqrt{3},1,0 \right) , \\
	 		\mathbf{a}_3 & =  a \left(0, 0, \sqrt{\frac{3}{2}} \right) .
	 	\end{align}
	 \end{subequations}
	 The unit cell has nine sublattices, $A_\alpha$, $B_\alpha$, $C_\alpha$, corresponding to sublattices $A$, $B$, and $C$ in the $\alpha$-th layer ($\alpha=1,2,3$), positioned at	 
	 \begin{equation}
	 	\mathbf{r}_{A_\alpha} = \mathbf{r}_A + \mathbf{d}_\alpha,	\quad	
	 	\mathbf{r}_{B_\alpha} = \mathbf{r}_B+ \mathbf{d}_\alpha,    \quad
	 	\mathbf{r}_{C_\alpha} = \mathbf{r}_C+ \mathbf{d}_\alpha ,
	 \end{equation}
	 where
	 \begin{subequations}
	 	\begin{align}
	 		\mathbf{d}_1 & = \mathbf{0} , \\
	 		\mathbf{d}_2 & = \frac{1}{3}(\mathbf{a}_1  + \mathbf{a}_2 +  \mathbf{a}_3) , \\
	 		\mathbf{d}_3 & = \frac{2}{3}(\mathbf{a}_1  + \mathbf{a}_2 +  \mathbf{a}_3) ,
	 	\end{align}
	 \end{subequations}
	 are defined such that the interlayer nearest-neighbor hopping distance is equivalent to the intralayer nearest-neighbor hopping distance (e.g., $\vert \mathbf{r}_{A_2} - \mathbf{r}_{B_1} \vert = \vert \mathbf{r}_{A_1} - \mathbf{r}_{B_1} \vert $). The spin orientations are independent of the layer position, i.e., $\hat{\mathbf{s}}^{A_\alpha} = \hat{\mathbf{s}}^A$, $\hat{\mathbf{s}}^{B_\alpha} = \hat{\mathbf{s}}^B$, and $\hat{\mathbf{s}}^{C_\alpha} = \hat{\mathbf{s}}^C$.

	\subsection{Hamiltonian in real-space representation}\label{sec1b}

	The real-space representation of the Hamiltonian $\mathcal{H}$ for our models is given by 
	\begin{equation}\label{eq:h_tb} 
		\mathcal{H}  = 
		\mathcal{H}_t + \mathcal{H}_\mathrm{sd} + \mathcal{H}_\mathrm{SO} + \mathcal{H}_\mathrm{CF} ,
	\end{equation}
	including the following four terms: 
	\begin{subequations}
	\begin{align}
		\mathcal{H}_t & =	
		\sum_{\langle i,j \rangle  m n \sigma } 	t_{ij,mn} c_{im\sigma}^\dagger c_{jn\sigma}  
		\label{eq:h_t} \\
		\mathcal{H}_\mathrm{sd} & =  	
		- \frac{J_\mathrm{sd}}{\hbar}	\sum_{i  n \sigma \sigma' } 
		\hat{ \mathbf{s} }^i \cdot 	c_{in\sigma}^\dagger \mathbf{S}_{\sigma \sigma'} \; c_{in\sigma'} 
		\label{eq:h_sd} \\
		\mathcal{H}_\mathrm{SO} & =  	
		\frac{\lambda_\mathrm{SO}}{\hbar^2}	\sum_{i  m n \sigma \sigma' } 	c_{im\sigma}^\dagger 	 \mathbf{L} _{mn} \cdot \mathbf{S}_{\sigma \sigma'} \; c_{in\sigma'} 
		\label{eq:h_so} \\
		\mathcal{H}_\mathrm{CF}  & = \frac{\Delta_\mathrm{CF}}{2} \sum_{i  \sigma } 	\tau \,	(c_{i,yz,\sigma}^\dagger c_{i,yz,\sigma} - c_{i,zx,\sigma}^\dagger c_{i,zx,\sigma} ), 
		\label{eq:h_cf}
	\end{align}
	\end{subequations}
	where $c_{in\sigma}^\dagger$ ($c_{in\sigma}$) is the creation (annihilation) operator of an electron at site $i$ with orbital index $n$ and spin $\sigma = \uparrow,\downarrow$. Here, $\mathbf{S}$ and $\mathbf{L}$ are the spin and orbital angular momentum operators, respectively. The orbital basis is chosen as $\{ d_{xy}, d_{yz}, d_{zx} \}$ for each sublattice $A$ and $B$ in the square lattice, and $\{ d_{3z^2-r^2}, d_{yz}, d_{zx} \}$ for each sublattice $A$, $B$, and $C$ in the kagome lattice. The terms $\mathcal{H}_t$, $\mathcal{H}_\mathrm{sd}$, $\mathcal{H}_\mathrm{SO}$, and $\mathcal{H}_\mathrm{CF}$ describe electron hopping, $sd$-exchange interaction, spin-orbit coupling, and sublattice-dependent crystal field splitting, respectively. In Eq.~\eqref{eq:h_t}, electron hopping occurs only between nearest-neighbor site pairs $\langle i,j \rangle$, with $i$ and $j$ residing in different sublattices. The hopping parameter $t_{ij,mn}$ is determined using the Slater-Koster method~\cite{slater1954simplified} with parameters $t_\sigma$, $t_\pi$, and $t_\delta$, which represent $\sigma$-, $\pi$-, and $\delta$-bonding between $d$ orbitals, respectively (see Sec.~\ref{sec1c}). The $sd$-exchange term [Eq.~\eqref{eq:h_sd}] specifies the spin direction at each site as $\hat{\mathbf{s}}^i$,  expressed by a single parameter $\phi$, as introduced in Sec.~\ref{sec1a}. The sublattice-dependent crystal field splitting [Eq.~\eqref{eq:h_cf}] is included only in the square lattice, with the sublattice index $\tau$ taking values $+1(-1)$ for $A$($B$). A nonzero $\Delta_\mathrm{CF}$ splits $d_{yz}$ and $d_{zx}$ orbital levels in opposite directions for the two magnetic sublattices, reflecting the alternating local anisotropy induced by nonmagnetic atoms in typical rutile-type altermagnets. Thus, this term transforms the square lattice from a conventional antiferromagnet into an altermagnet. On the other hand, for the kagome lattice, we neglect $\mathcal{H}_\mathrm{CF}$ (i.e., set $\Delta_\mathrm{CF}=0$), since electron hopping in this crystal structure inherently leads to local anisotropy at each sublattice. All energy parameters used for the square and kagome lattices are listed in Table~\ref{table3}.

	\begin{table}[h]
		\caption{\label{table3}
			Parameters used in tight-binding models.}
		\begin{ruledtabular}
			\begin{tabular}{ccccccc}
				Lattice structure & $t_\sigma$ (eV) &  $t_\pi$ (eV)  & $t_\delta$ (eV) & $J_\mathrm{sd}$ (eV) & $\lambda_\mathrm{SO}$ (eV) & $\Delta_\mathrm{CF}$ (eV) \\
				\hline
				Square & 0 &  0.5  & 0 & 1.0 &  0.02 & 1.0 \\
				Kagome & -1.0 &  0.5  & -0.1 & 0.7 &  0.02 & 0 \\
			\end{tabular}
		\end{ruledtabular}
	\end{table}

	\subsection{Hamiltonian in momentum-space representation}\label{sec1c}
	
	While the real-space representation of the Hamiltonian (Sec.~\ref{sec1b}) was presented in the main text, the calculations were performed in momentum space. To obtain the momentum-space representation of the Hamiltonian, we construct the Bloch-like basis $\ket{\varphi_{n \sigma \mathbf{k}}}$ for the crystal momentum $\mathbf{k}$, as follows:
	\begin{equation}
		\ket{\varphi_{n \sigma \mathbf{k}}} = \frac{1}{\sqrt{N}} \sum_\mathbf{R} e^{i \mathbf{k} \cdot (\mathbf{R} + \mathbf{r}_n)} \ket{\phi_{n \sigma \mathbf{R} }}  ,
	\end{equation}
	where $N$ is the number of atoms in the system, $\mathbf{R}$ is the lattice vector, and $\ket{\phi_{n \sigma \mathbf{R}} }$ represents the tight-binding orbital basis centered at $\mathbf{R} + \mathbf{r}_n $. Note that the index $n$ is defined by both the orbital species and the sublattice.

	In this basis, the total Hamiltonian matrix $\mathcal{H}_\mathrm{tot}(\mathbf{k})$ at each $\mathbf{k}$ is given by
	\begin{align}\label{eq:h_k_tb}
		\mathcal{H}_\mathrm{tot}(\mathbf{k})   & = 	\mathcal{H}_t(\mathbf{k}) - \frac{J_\mathrm{sd}}{\hbar}	\sum_i
		\hat{ \mathbf{s} }^i \cdot 	\mathbf{S}^i +  \frac{\lambda_\mathrm{SO}}{\hbar^2} \mathbf{L}  \cdot \mathbf{S}  \nonumber \\
		& + \frac{\Delta_\mathrm{CF}}{2} \sum_{i\sigma} \tau
		\left( \ket{\varphi_{yz(i),\sigma \mathbf{k}}} \bra{\varphi_{yz(i), \sigma \mathbf{k}}} -  \ket{\varphi_{zx(i), \sigma \mathbf{k}}} \bra{\varphi_{zx(i), \sigma \mathbf{k}}} \right) ,
	\end{align}
	where $i$ denotes the sublattice species. The orbital and spin angular momentum operators, $\mathbf{L}$ and $\mathbf{S}$, are locally defined in the tight-binding orbital basis. Specifically, $\bra{\phi_{m\sigma ' \mathbf{R} '}} \mathbf{L} \ket{\phi_{n\sigma \mathbf{R}}} $ and $\bra{\phi_{m\sigma ' \mathbf{R} '}} \mathbf{S} \ket{\phi_{n\sigma \mathbf{R}}} $ are nonzero only for $\mathbf{R} = \mathbf{R}'$, making them constant in $\mathbf{k}$ space. In the second term on the right-hand side of Eq.~\eqref{eq:h_k_tb}, $\mathbf{S}^i = P^i  \mathbf{S}  P^i $ corresponds to the spin operator projected onto the subspace spanned by the $i$-sublattice basis, with the projection operator $P^i = \sum_{n \in i} \sum_\sigma \ket{\varphi_{n\sigma\mathbf{k}}} \bra{\varphi_{n\sigma\mathbf{k}}} $. The matrix element of the first term, describing electron hopping, is evaluated by
	\begin{equation}\label{eq:H_element}
		\bra{\varphi_{m \sigma \mathbf{k}}} \mathcal{H}_t \ket{\varphi_{n \sigma \mathbf{k}}} 
		= \sum_\mathbf{R} e^{i\mathbf{k} \cdot (\mathbf{R} + \mathbf{r}_n - \mathbf{r}_m)}
		\bra{\phi_{m \sigma \mathbf{0}} } \mathcal{H}_t \ket{\phi_{n \sigma \mathbf{R}} } ,
	\end{equation}
	which are nonzero only between the same spin $\sigma$. The summation runs over $\mathbf{R}$ such that $ \mathbf{R} + \mathbf{r}_n - \mathbf{r}_m $ connects at most nearest-neighbor sites. Within the two-center approximation, the hopping parameter on the right-hand side,  $\bra{\phi_{m \sigma \mathbf{0}} } \mathcal{H}_t \ket{\phi_{n \sigma \mathbf{R}}}$, can be expressed in terms of Slater-Koster parameters $t_\sigma$, $t_\pi$, and $t_\delta$, with coefficients given by functions of the directional cosines of $ \mathbf{R} + \mathbf{r}_n - \mathbf{r}_m $ (see Table~1 of Ref.~\cite{slater1954simplified} for the explicit formulae). 
	
	In the distorted kagome lattice [Fig.~4(c) of the main text], we assumed uniaxial strain along the $x$ axis, which modifies the nearest-neighbor distances between sublattices $A$ and $B$ and between $A$ and $C$ as
	\begin{equation}
		\vert \mathbf{R} + \mathbf{r}_n - \mathbf{r}_m \vert \rightarrow 
		(1+\epsilon) \vert \mathbf{R} + \mathbf{r}_n - \mathbf{r}_m \vert ,
	\end{equation}
	while leaving the nearest-neighbor distance between sublattices $B$ and $C$ unchanged. Here, we did not explicitly consider the change of atomic positions. Instead, the effect of this distortion was incorporated solely through the modified hopping parameter due to the changed bond length, approximately given by:
	\begin{equation}\label{eq:hopping_strain}
		\bra{\phi_{m \sigma \mathbf{0}} } \mathcal{H}_t \ket{\phi_{n \sigma \mathbf{R}} }
		\rightarrow
		\frac{1}{(1+\varepsilon)^2} 
		\bra{\phi_{m \sigma \mathbf{0}} } \mathcal{H}_t \ket{\phi_{n \sigma \mathbf{R}} } ,
	\end{equation}
	between sublattices $A$ and $B$ and between $A$ and $C$. Consequently, the hopping term in the Hamiltonian for the distorted kagome lattice is obtained from Eq.~\eqref{eq:H_element}, where $\mathbf{R} + \mathbf{r}_n - \mathbf{r}_m$ remains the same as in the pristine kagome lattice, while the hopping parameters between the relevant sublattices are scaled according to Eq.~\eqref{eq:hopping_strain}.

	\section{Details of first-principles calculation}\label{sec2}

	We calculated spin and orbital magnetic moments in an insulating rutile altermagnet NiF$_2$ using the density functional theory code \texttt{FLEUR}~\cite{fleurWeb, fleurCode} based on the full-potential linearized augmented plane-wave method~\cite{wimmer1981full}. The Perdew-Burke-Ernzerhof functional~\cite{perdew1996generalized} within the local-spin-density approximation was used and the spin-orbit coupling was included within the second variation scheme~\cite{li1990magnetic}. The rutile-type NiF$_2$ has a tetragonal unit cell (space group $P4_2/mnm$) (see Fig.~5 in End Matter of the main text), containing two Ni atoms at $(0,0,0)$ and $(\frac{1}{2},\frac{1}{2},\frac{1}{2})$, and four F atoms at $(1 \pm u, 1 \pm u, 0)$ and $(\frac{1}{2} \pm u, \frac{1}{2} \mp u, \frac{1}{2})$, where $u=0.3040$~\cite{moreira2000ab}. The unit cell parameters $a=4.742$~\AA~and $c=3.161$~\AA~were used~\cite{moreira2000ab}. The muffin-tin radius was set to 2.39 $a_0$ for Ni and 1.35 $a_0$ for F, and the plane-wave cutoff was set to 4.4 $a_0^{-1}$, where $a_0$ is the Bohr radius. The Brillouin zone was sampled by a $16\times16\times16$ Monkhorst-Pack $\mathbf{k}$-mesh~\cite{monkhorst1976special}. 	An antiferromagnetic spin ordering was assumed, with the initial spin moments for Ni $A$ at $(0,0,0)$ and Ni $B$ at $(\frac{1}{2},\frac{1}{2},\frac{1}{2})$ sublattices aligned along  $-\mathbf{a}$ and $+\mathbf{a}$, respectively.

\begin{figure}[h]
		\center\includegraphics[width=0.8\textwidth]{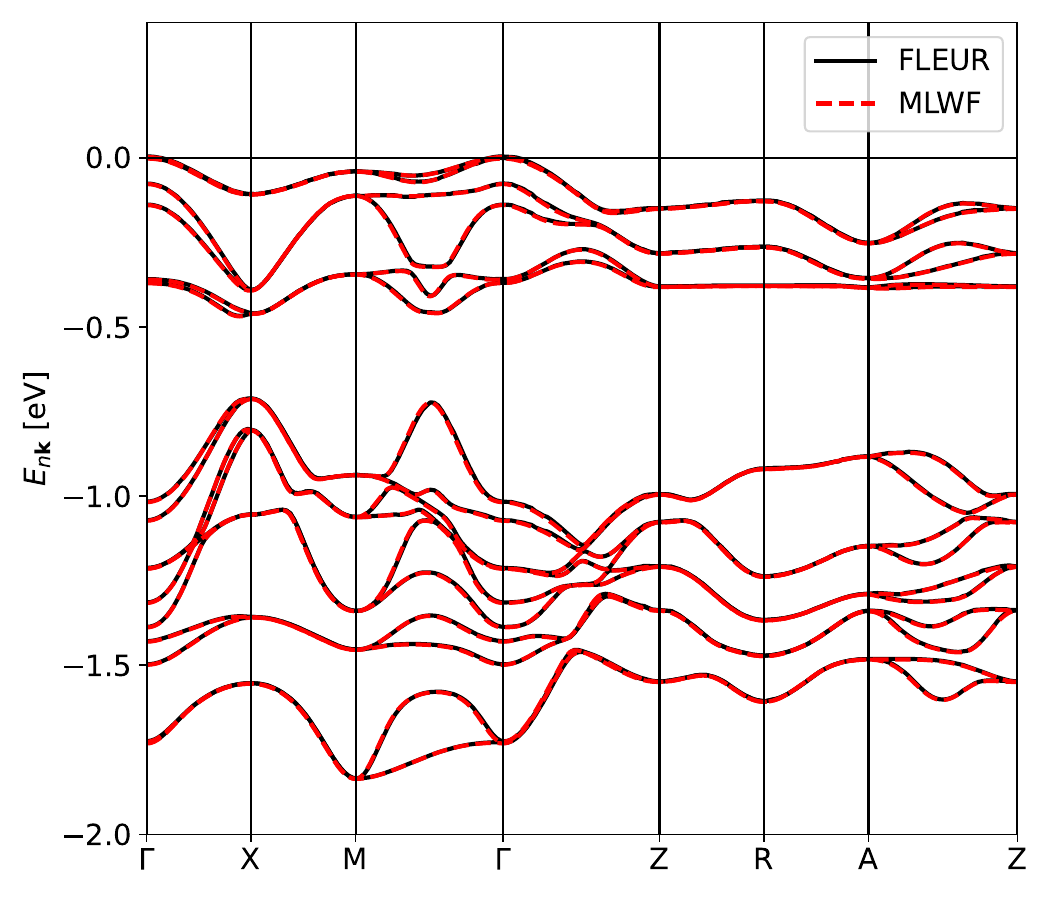}
		\caption{Band structures of NiF$_2$ calculated using Bloch functions from the \texttt{FLEUR} code (black solid lines) and MLWFs (red dashed lines). The valence-band maximum was set to 0~eV. 
		}
		\label{figS2} 
	\end{figure}

	After obtaining the self-consistent charge density, we extracted maximally localized Wannier functions (MLWFs) using the \texttt{WANNIER90} code~\cite{pizzi2020wannier90}. Although NiF$_2$ is an insulator, we disentangled 48 bands from a larger subspace spanned by 96 Bloch bands. The frozen window energy ranged from $-7.0$~eV to $5.7$~eV, with the valence-band maximum set to 0~eV. As initial functions, we chose $s$, $d_{xy}$, $d_{yz}$, $d_{zx}$, $d_{x^2-y^2}$, and $d_{3z^2-r^2}$ orbitals for each Ni atom, and $p_x$, $p_y$, and $p_z$ orbitals for for each F atom, resulting in a total of 48 MLWFs. The band structure calculated from the MLWFs agrees well with that calculated using the \texttt{FLEUR} code, as shown in Fig.~\ref{figS2}. Using the obtained MLWFs, the spin ($\mathbf{S}$) and orbital ($\mathbf{L}$) angular momentum operators were interpolated on a $150 \times 150 \times 150$ $\mathbf{k}$-mesh. The orbital angular momentum was treated within the atom-centered approximation. The spin and orbital angular momenta for atom $i$ were calculated as follows:
	\begin{subequations}
		\begin{align}
			\langle \mathbf{S}^i \rangle  & =  \frac{1}{N_\mathbf{k}} \sum_{n\mathbf{k}} f_{n\mathbf{k}} \bra{\psi_{n\mathbf{k}}} \frac{1}{2} (P^i \mathbf{S} +  \mathbf{S} P^i )   \ket{\psi_{n\mathbf{k}}} , \label{eq:spin_i} \\
			\langle \mathbf{L}^i \rangle  & =  \frac{1}{N_\mathbf{k}} \sum_{n\mathbf{k}} f_{n\mathbf{k}} \bra{\psi_{n\mathbf{k}}} \frac{1}{2} (P^i \mathbf{L} +  \mathbf{L} P^i ) \ket{\psi_{n\mathbf{k}}} , \label{eq:orbital_i}
		\end{align}
	\end{subequations}
	where $N_\mathbf{k}$ is the number of $\mathbf{k}$ points, $\ket{\psi_{n\mathbf{k}}}$ is the eigenstate of the $n$-th band, $f_{n\mathbf{k}}$ is the Fermi-Dirac distribution function, and $P^i$ is the projection operator onto the states of atom $i$. The spin and orbital magnetic moments for atom $i$ are then given by $\mathbf{m}^{\mathrm{spin},i} = -\frac{2 \mu_\mathrm{B}}{\hbar} \langle \mathbf{S}^i \rangle$ and $\mathbf{m}^{\mathrm{orb},i} = -\frac{ \mu_\mathrm{B}}{\hbar} \langle \mathbf{L}^i \rangle$, respectively. The summation over $i$ gives the total spin and orbital magnetic moments.

\clearpage

	\bibliography{ref.bib}

	\clearpage